\documentclass{jfm}
\usepackage{graphicx}
\usepackage{epstopdf, epsfig}

\usepackage{amsmath}
\usepackage{xcolor}
\usepackage{subfigure}
\usepackage{dcolumn}
\usepackage{tabularx}

\shorttitle{Minimal seeds for turbulent bands}
\shortauthor{E. Parente, J-Ch. Robinet, P. De Palma and S. Cherubini}

\title{Minimal seeds for turbulent bands in channel flow }

\author{E. Parente\aff{1,2} \corresp{\email{enzaparente@gmail.com}},
\and J.-Ch. Robinet \aff{1}, P. De Palma\aff{2}
\and S. Cherubini\aff{2}
}

\affiliation{\aff{1} DynFluid - Arts et M\'{e}tiers Paris, 151 Bd de l'H\^{o}pital, 75013 Paris, France
\aff{2} Dipartimento di Meccanica, Matematica e Management (DMMM), Politecnico di Bari, Via Re David 200, 70126 Bari, Italy}

\begin{document}

\maketitle

\begin{abstract}
In this work, nonlinear variational optimization is used for obtaining minimal seeds for the formation of turbulent bands in channel flow. Using nonlinear optimization together with energy bisection, we have found that the minimal energy threshold for obtaining spatially-patterned turbulence  scales with $Re^{-8.5}$ for $Re>1000$. 
 The minimal seed is constituted by a localized spot-like structure surrounded by a low-amplitude large-scale quadrupolar structure filling the whole domain. This minimal-energy perturbation of the laminar flow has dominant wavenumbers equal to $0.15$ and $4$ in the streamwise and spanwise directions, respectively, and is characterized by a more marked spatial localization when the Reynolds number increases. At $Re \lesssim 1200$, the minimal seed  evolves in time creating an isolated oblique band. Whereas, for $Re\gtrsim 1200$, an almost spanwise-symmetric evolution is observed, giving rise to two distinct bands. A similar evolution is found also at low $Re$ for non-minimal optimal perturbations. This highlights two different  mechanisms of formation of turbulent bands in  channel flow, depending on the Reynolds number and initial energy of the perturbation. The selection of one of these two mechanisms appears to be affected by the probability of decay of the newly-created stripe, which increases with time, but decreases with the Reynolds number.
\end{abstract}



 \section{Introduction}

Since the experimental work of \cite{reynolds1883}, who first observed subcritical transition to turbulence in a wall bounded shear flow, the dynamics of transition  has remained an open problem of fluid dynamics. One of the most triggering features of transitional flows is that turbulence does not arise at the same time in the whole domain, being preceded by the formation of localised flow structures that grow in amplitude and spread in space. The first of these localized flow features to be observed were turbulent spots, which have been investigated by many researchers in the past. \cite{emmons1951} was the first to experimentally show that turbulent spots may trigger turbulence; later this was confirmed experimentally and numerically in plane channel flow analysing the spots characteristics \citep{carlson1982,henningson1991,klingmann1992,aida2010,aida2011,lemoult2013,lemoult2014}.
In pipe flows, other localized flow structures dubbed \textit{puffs} are observed both numerically and experimentally \citep{eckhardt2007,avila2011}. Puffs are localised, downstream-travelling flow structures within a laminar field, sustained by the energy provided by the neighbouring laminar motion at the upstream end of the puff. They can decay, split or merge, filling the laminar flow with turbulent fluctuations.   
 \\
More recently, \cite{duguet2010} have shown that, locally perturbing the plane Couette flow, the fully turbulent state is preceded by the formation of turbulent bands. The same behaviour was reported by \cite{tao2013} and \cite{xiong2015}, where it was observed that, after injecting a localised perturbation, turbulent bands grow obliquely in the domain until decay or breakdown to turbulence. In the latter case, it was assessed that the turbulent bands lifetime is longer than that of turbulent spots. Besides having a crucial role in turbulent transition, oblique turbulent bands characterise also the (transiently) turbulent state in shear flows at low Reynolds numbers, as shown by \cite{tsukahara2005}. In fact, in the pipe, plane Couette and plane Poiseuille flow, there is a range of moderate Reynolds numbers for which experiments and numerical simulations have shown that turbulence is not homogeneous in space, since spatially-localized regions characterised by laminar and turbulent behaviour coexist when the statistically-steady turbulent state is reached. Indeed, elongated oblique turbulent bands plunged in a laminar flow appear in plane Couette flow at $Re>290$ \citep{prigent2002,barkley2005,duguet2010,tuckerman2011} and plane Poiseuille flow \citep{tsukahara2014,tuckerman2014,tao2018,shimizu2019,kashyap2020a}). In the channel flow, turbulent-laminar oblique patterns were observed experimentally for $Re = 1000$ by \cite{carlson1982}. Numerical simulations reproduced these patterns also at lower Reynolds numbers (\cite{xiong2015}, \cite{tao2017}, \cite{kashyap2020a}), estimating a threshold value for observing turbulent bands at $Re \approx 660$, according to \cite{tao2018}. Moreover, \cite{song2020} recently reported the onset of turbulent bands even at Reynolds number as low as $Re = 500$, generated by forcing the flow with a local perturbation with a sufficiently strong spanwise inflection. They showed that this forcing method permits to generate bands at very low values of the Reynolds number, for which bands previously appeared to be not sustained. This procedure was motivated by the work of \cite{xiao2020}, where the authors performed a 
linear stability analysis of the mean velocity profile extracted in a small domain at the head of the turbulent band. By means of stability analysis, it was suggested that spanwise inflectional instability may be the mechanism involved in the growth and self-sustaining process of turbulent bands. 
 \\
In addition to the Reynolds number, the domain size plays an important role in the growth and  self-interaction of turbulent bands. When the considered domain is sufficiently large, turbulent bands can grow for longer times, avoiding the probability of interaction with themselves or other bands. For this reason, several works were performed to study the influence of the domain in the onset of turbulent bands. In order to reduce the computational cost, some numerical studies cleverly considered computational domains tilted in the direction of the bands, as done for the plane Couette flow by \cite{barkley2005} and for the plane Poiseuille flow by \cite{tuckerman2014}. Very recently, \cite{ParenteRapid2021} carried out an optimal growth analysis in a tilted domain with given angle. They showed that, although linear optimization is able to recover the main wavenumbers of the DNS, nonlinear effects are necessary for providing large-scale flow and spatial localization able to generate a turbulent band. Although very interesting for studying the dynamics of a single band, using the tilted domain constrains the turbulent bands to develop at a fixed angle and avoids (or reduce) the interactions with other bands, resulting in a consequently less rich dynamics.  A more detailed analysis about turbulent-laminar patterns in shear flows is tackled in the review of \cite{tuckerman2020}.
\\
In the literature, two strategies are typically used to obtain laminar-turbulent bands. The first one consists in starting from a spatially-homogeneous statistically-steady turbulent state, and slowly decreasing the Reynolds number until patterned laminar regions  plunged in the initially turbulent flow begin to be observed (\cite{tsukahara2005}, \cite{tsukahara2005}, \cite{kashyap2020a}). This method usually leads to the formation of  statistically-steady laminar-turbulent patterns. 
Another possible method consists in perturbing the laminar flow with localised solutions having enough energy to trigger localized regions of turbulence eventually evolving into oblique stripes (\cite{duguet2010}, \cite{aida2010}, \cite{aida2011}, \cite{tao2013}, \cite{xiong2015}).
When focusing on the second method, the amplitude, the shape and the localization of these initial perturbations should be carefully chosen to ensure their growth towards oblique bands.  
Depending on their shape, perturbations with higher amplitude may decay, while weaker perturbations may lead the flow to transition. 
\\\
From a phase-space point of view, the problem of finding perturbations eventually leading to these laminar-turbulent patterns consists in placing the starting point of the trajectory outside the boundary of the basin of attraction of the laminar solution. 
Notably, the most relevant point of this boundary is its energy minimum, since it represents the minimal (in energy norm) perturbation of the laminar state that can lead the flow to transition. This point in the phase space has been dubbed, by \cite{rabin2012triggering},  \textit{minimal seed} for turbulent transition. This energy minimum has been assessed for several shear flows, allowing at finding minimal energy thresholds for transition. In particular, in plane Couette flows, a minimal energy threshold varying with the Reynolds number as $Re^{-2.7}$ has been found, in quantitative agreement with experimental estimates for pipe flows. Whereas, for the asymptotic-suction boundary layer \cite{CherubiniPoF2015} found a scaling law of this  energy threshold of $Re^{-2}$. \cite{vavaliaris2020} recently performed  similar computations for a non-parallel boundary-layer flow.
The determination of these energy thresholds is of primary importance for control purposes, since passive or active control methods such as boundary manipulation \citep{Rabin2014} or profile flattening \citep{marensi2019} able to increase this minimal energy would render these flows less prone to transition.  
However, to the authors knowledge, the minimal seed computation has never been carried out for the channel flow.
Moreover, all the minimal seed computations reported in the literature have been carried out in small domains, not allowing to observe laminar-turbulent patterned flow states. In large domains where laminar-turbulent oblique bands exist, the minimal transition thresholds can be potentially different from that obtained in small domains. Moreover, the analysis of these minimal perturbations and of their evolution towards oblique bands can potentially unveil the main mechanisms leading to the formation and sustainment of these laminar-turbulent patterns.
\\
This work aims at finding minimal seeds for the generation of turbulent bands in channel flow.
The methodology to find these perturbations is based on the nonlinear variational optimization first used in the boundary layer flow by \cite{cherubini2010,cherubini2011} and in pipe flow by \cite{pringle2010,pringle2012}, and then coupled with energy bisection by \cite{rabin2012triggering} for the plane Couette flow and by other authors for other shear flows \citep{duguet2013,CherubiniPoF2015,vavaliaris2020}.
In all these studies, the kinetic energy was considered as objective function of the optimization and the optimal perturbations associated to the minimal input energy were spatially localised. Similar behaviour was found in Couette flow by \cite{monokrousos2011} maximizing the time integral of the entropy production. A detailed review of these methods is reported in \cite{Kerswell2014,KerswellARFM}. 
In the present work, we carry out for the first time the computation of minimal seeds in channel flow, for very large domains  allowing the creation of laminar-turbulent patterned final states such as oblique stripes. The perturbation kinetic energy is used as objective function of the nonlinear optimization, which appears an appropriate choice since many literature studies report peaks of kinetic energy during the development of turbulent bands \citep{tao2018}. The minimal seeds obtained using this optimization-and-bisect method are found to change with the Reynolds number, presenting a power-law scaling of the initial energy $E_{0_{min}} \propto Re^{-\gamma}$, with $\gamma$ being approximately four times larger than the values reported in other shear flows in smaller domains. Moreover, two distinct mechanisms of creation of turbulent bands are reported and discussed in detail. 
\\
The paper is organised in the following way: in Section \ref{sec:problem} the used methodology is presented. In Section \ref{sec:results} the main results of the optimization are reported and the nonlinear evolution in time of the optimal perturbation is shown. In section \ref{sec:conclusions} conclusions are drawn.

\section{Problem formulation}\label{sec:problem}
The considered flow is a plane channel flow at Reynolds number $Re = U_{c} H / \nu$, with $U_c$ being the centreline velocity of the laminar Poiseuille flow, $H$ being the half width of the channel and $\nu$ the kinematic viscosity. The Reynolds number is varied by changing the pressure gradient, while the volume flux remains constant, with bulk velocity $U_b = 3/2$. The dynamics of this flow is studied by decomposing the instantaneous velocity field,  $\textbf{u}= [u,v,w]^T$, into the laminar base flow $\textbf{U} = [U(y), 0, 0]$ and a perturbation $\textbf{u}' = [u',v',w']^T$. The dynamics of the perturbations of the laminar base flow is computed by solving the perturbative nonlinear incompressible Navier-Stokes equations:

\begin{equation}
\begin{cases}
\displaystyle \frac{\partial u_i'}{\partial x_i} = 0,\\
\displaystyle \frac{\partial u_i'}{\partial t} + u_j'\frac{\partial u_i'}{\partial x_j} + u_j' \frac{\partial U_i}{\partial x_j} + U_j \frac{\partial u_i'}{\partial x_j} = - \frac{\partial p'}{\partial x_i} + \frac{1}{Re} \frac{\partial^2 u_i'}{\partial x_j},
\end{cases}
\end{equation}
with $\mathbf{U}=[U(y),0,0]$, $U(y) = 1 - y^2$ being the laminar streamwise velocity profile, $p'$ the pressure perturbation, while $x_i$ 
is the Cartesian reference frame having components $x, y, z$, for the streamwise, wall-normal and spanwise directions, respectively. No-slip boundary conditions are imposed at the walls for the three velocity components, while periodicity is fixed in the streamwise and spanwise directions.\\
In order to find the minimal seed for the considered flow, we first search for the \textit{optimal} perturbation $\textbf{u}'$ at $t = 0$, providing the maximum value of the objective function at target time $T$. Following previous works \citep{cherubini2010,pringle2010}, we choose as objective function the energy gain $G(T) = E(T)/E(0)$, where:
\begin{equation}
E(t) = \frac{1}{2V} \int_V (u'(t)^2 + v'(t)^2 + w'(t)^2) dV
\end{equation} 
is the kinetic energy at time $t$ and $V$ is the volume of the computational domain. In order to find the initial perturbation $\textbf{u}'(0)$ having given initial energy $E(0) = E_0$, providing the largest possible energy $E(T)$ at the target time, an optimization loop is set using the Lagrange multiplier technique (\cite{zuccher2004}, \cite{pringle2010}, \cite{cherubini2011}). 
A Lagrangian functional is defined by augmenting the objective function with the following constraints: i) the optimal perturbation $\mathbf{u}'(t)$ must be solution of the Navier-Stokes equations at all times $t\in]0,T[$;  ii) it must be divergence free at all times $t\in[0,T]$; and iii) it must have energy norm equal to a given value $E_0$ at $t=0$.
With these constraints, the Lagrangian functional reads:
\begin{equation}
\begin{split}
\mathcal{L}(u_k', p', u_k^{\dagger}, p^{\dagger}, u_k'(0), u_k'(T), \lambda) = \frac{E(T)}{E(0)}\\ - \int_0^T \int_V u_i^{\dagger} \left( \frac{\partial u_i'}{\partial t} + \frac{\partial (u_i' u_j')}{\partial x_j} + \frac{\partial (U_i u_j')}{\partial x_j} + \frac{\partial (u_i' U_j)}{\partial x_j} + \frac{\partial p}{\partial x_i} - \frac{1}{Re} \frac{\partial^2 u_i'}{\partial x_j} \right) dV dt \\- \int_0^T \int_V p^{\dagger} \frac{\partial u_i'}{\partial x_i} dV dt - E^\dagger \left( \frac{E_0}{E(0)} -1 \right).
\end{split}
\end{equation}
with $\mathbf{u}^{\dagger}$, $p^{\dagger}$ and $E^\dagger$ being the Lagrangian multipliers (or adjoint variables).\\
To maximise the augmented functional $\mathcal{L}$ we evaluate its variation with respect to the direct and adjoint variables and nullify it. The variation of the Lagrangian functional with respect to the direct variables $\textbf{u}'$, $p'$, provides the following adjoint equations:

\begin{equation}
\begin{cases}
\displaystyle \frac{\partial u_i^{\dagger}}{\partial x_i} = 0,\\
\displaystyle \frac{\partial u_i^{\dagger}}{\partial t} - u_j^{\dagger}\frac{\partial u_i'}{\partial x_j} + u_j'\frac{\partial u_i^{\dagger}}{\partial x_j} - u_j^{\dagger} \frac{\partial U_i}{\partial x_j} + U_j \frac{\partial u_i^{\dagger}}{\partial x_j} = - \frac{\partial p^{\dagger}}{\partial x_i} + \frac{1}{Re} \frac{\partial^2 u_i^{\dagger}}{\partial x_j}.
\end{cases}
\end{equation}
The optimization problem is then solved using a direct-adjoint looping algorithm (as done in previous works by \cite{pringle2010}, \cite{cherubini2010}, among others), which consists in integrating iteratively in time the direct and adjoint equations between $0$ and $T$ to evaluate the gradient of $\mathcal{L}$ with respect to $\mathbf{u}'(0)$, which is then used to update the initial perturbation using a gradient rotation algorithm \citep{Foures2013,farano2016,farano2017}.  Convergence is attained when the variation of the gain between two successively iterations is smaller then a chosen threshold, $\epsilon = 5 \times 10^{-8}$.\\
For computing the minimal seed, the variational optimization is coupled with an energy-bisection procedure.
Two different optimizations are initialised with a random divergence-free perturbation for low $T$, with a value of $E_0$ sufficiently high (low) to induce transition (relaminarization) at longer times. The energy is bisected and the optimization  repeated, using the results of the previous optimisation as initial guess. Energy-bisection is at first carried out for a low ($\mathcal{O}(10)$) target time in order to have a very rough (but computationally cheap) upper estimate of the initial minimal energy threshold $E_{0_{min}}$. Then, the target time is increased at $\mathcal{O}(100)$  and the energy bisection is repeated, allowing to converge towards $E_{0_{min}}$. \\
The optimization algorithm is implemented within the open source code $Channelflow$ (channelflow.ch) (\cite{channelflow}). Both direct and adjoint equations are solved using a Fourier-Chebychev discretization in space and a third-order semi-implicit backward difference scheme for the time integration. An influence matrix method with Chebyshev tau correction was used to enforce no-slip boundary condition at the walls.\\
The domain size in the streamwise, normal and spanwise directions is $L_x \times L_y \times L_z = 250 \times 2 \times 125$, while the number of grid points in the same directions are $N_x \times N_y \times N_z = 1024 \times 65 \times 1024$. This results in a numerical resolution comparable with those used by \cite{shimizu2019} and  \cite{kashyap2020a} for the same range of Reynolds numbers.
Finally, it should be noticed that the fact that optimizations are performed in a large domain involves challenging computations in terms of memory, especially for the highest considered target time, $T = 150$. In fact, the direct-adjoint algorithm requires storage of the flow field snapshots $\textbf{u}'$ at each time step, in order to evaluate  the term coupling direct and adjoint variables in the adjoint equations. For alleviating the storage requirement, a checkpointing technique is used, similar to that used in \cite{griewank2000} and \cite{hinze2006}.
\\

\section{Results}\label{sec:results}
\begin{figure}
        \centering
     
        \subfigure[]{\includegraphics[width=.48\columnwidth]{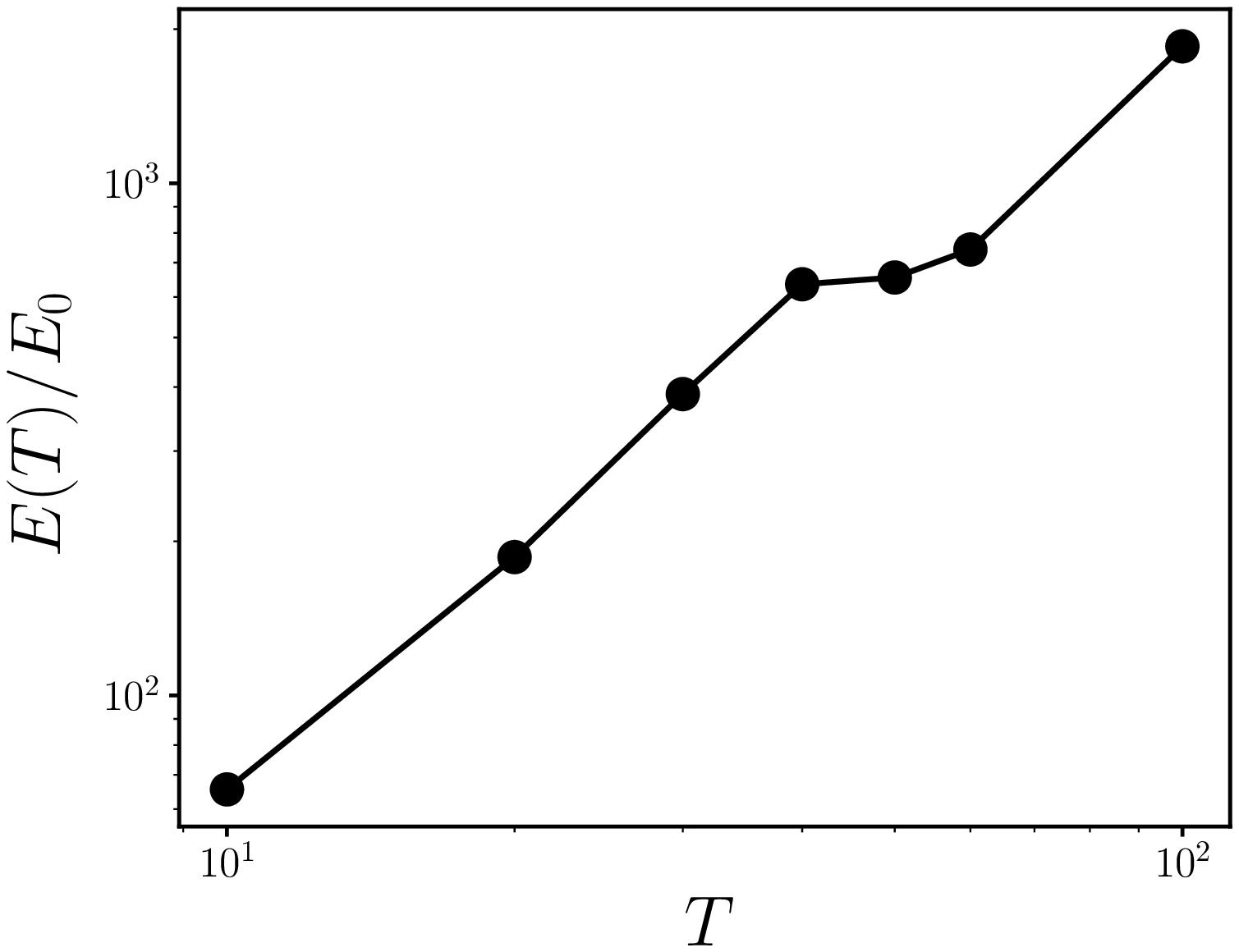}}
             \subfigure[]{\includegraphics[width=.48\columnwidth]{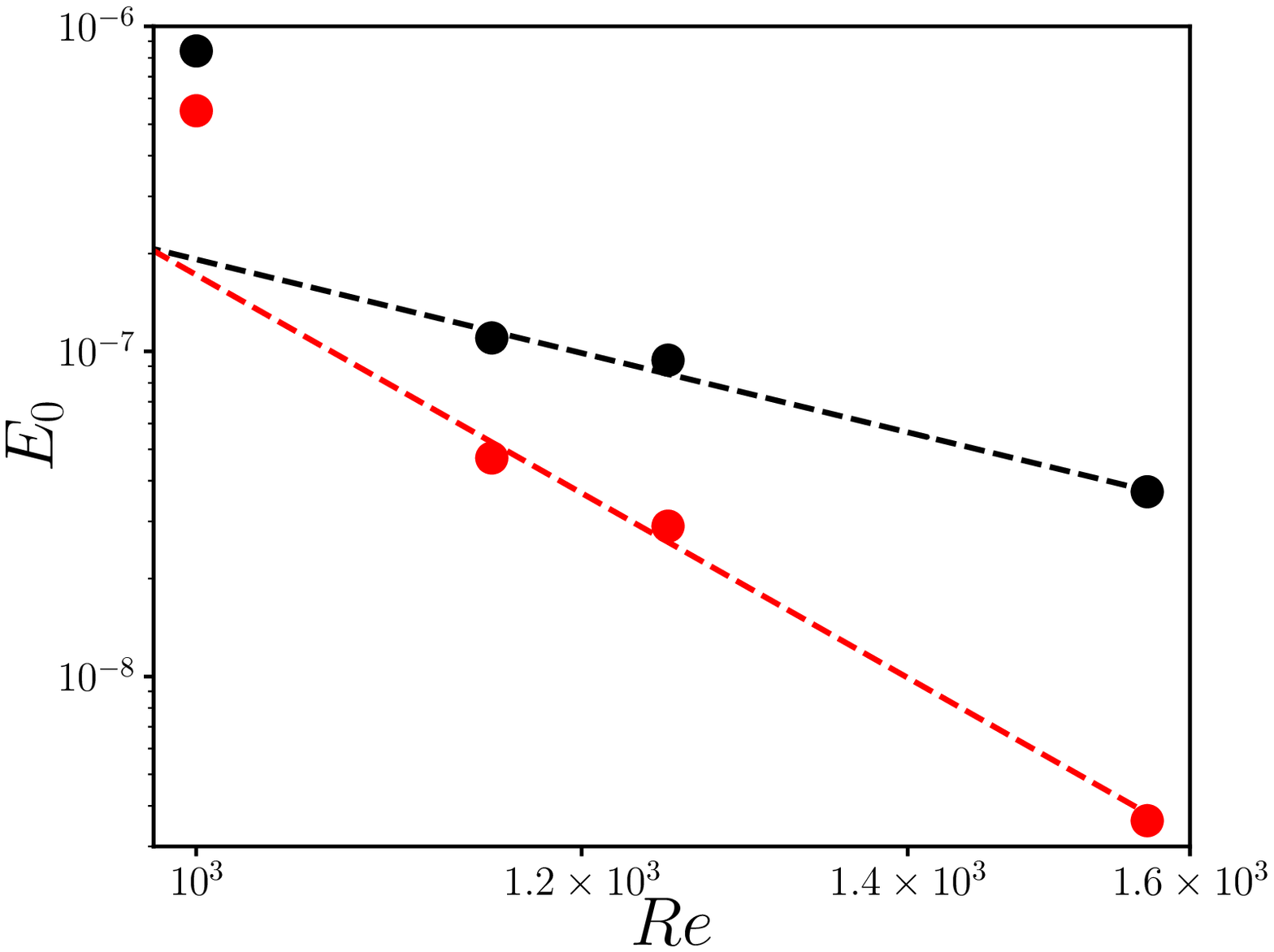}}
        \caption[]{(a) Optimal energy gain versus target time $T$ for $Re= 1150$ and $E_0=1.1\times 10^{-7}$. (b) Minimal energy threshold ${E_0}_{min}$ for transition to turbulence (red dots) and upper estimate obtained for $T = 10$ (black dots). The black and red lines represent the power-law for low ($\mathcal{O}(10)$) and high ($\mathcal{O}(100)$) target times, respectively, for $Re>1000$. 
        } 
                \label{min_energy_input}
\end{figure}

Nonlinear optimizations 
are performed for four values of $Re = 1000, 1150, 1250, 1568$, 
chosen in the range of Reynolds numbers for which the plane Poiseuille flow is linearly stable, while the turbulent state appears in the form of turbulent-laminar patterns (\cite{tao2018}, \cite{kashyap2020a}).
As reported by \cite{xiong2015}, the onset of bands does not lead to sustained turbulence for $Re<1000$. Whereas, for $Re\ge1000$, the bands can split, providing a mechanism for turbulence spreading leading to the coexistence of laminar regions with inclined turbulent bands, which persist up to $Re>3900$, for which only featureless turbulence is present.
Figure \ref{min_energy_input} (a) provides the influence of the target time on the optimal energy gain for  $Re = 1150$ and given input energy $E_{0} = 1.1 \times 10^{-7}$. 
The energy gain grows in time with an almost exponential trend for  $T < 40$ and for $T>50$, reaching an amplification of three orders of magnitude for $T>80$. No saturation of the final energy is observed for $T\le 100$, since very long (i.e., $\mathcal{O}(10^3)$) times would be needed to fill the entire domain with turbulent bands, as it will be shown in subsection \ref{sec:evolution}. 
As mentioned before, the procedure of bisection of the initial energy is at first carried out for $T=10$, in order to have a   computationally cheap upper bound for the computation of ${E_0}_{min}$, which will be then carried out for $T=100$ or $T=150$, depending on the Reynolds number.  For $Re<1568$, the estimate of ${E_0}_{min}$ was obtained for $T=100$, since we have verified that increasing the target time from $T=100$ to $T=150$ leads to slight changes of the minimal energy, namely, from $5.5 \times10^{-7}$ to $5.6 \times10^{-7}$ for $Re=1000$. Whereas, at $Re=1568$ it has been necessary to increase the target time to $T=150$ to achieve a sufficiently good approximation of the threshold energy for generating bands. 
Figure \ref{min_energy_input} (b) provides the minimal input energy able to induce transition towards the turbulent bands, ${E_0}_{min}$ (red dots), together with its upper estimate obtained for $T=10$ (black dots), for the four considered values of the Reynolds number. In the range of $Re$ analysed, we tried to fit  the minimal input energy with a power law  of the type $E_{0_{min}} \propto Re^{-\gamma}$, 
but we obtained a satisfying fit only by restraining the power law to the minimal seeds in the range $Re>1000$ (red dashed line, obtained for $\gamma \approx 8.5$).
A similar behaviour is observed for the upper estimate of $E_{0_{min}}$ obtained for $T=10$ (black dashed line), although the associated value of $\gamma$ is much lower. 
Incidentally, we observe that as expected, larger threshold initial energies are obtained for the lower target time $T = 10$ (black line). 
The fact that the minimal-seed energy obtained for $Re=1000$ appears not to be aligned with the fitting line recovered for larger values of $Re$ might have been anticipated. As reported by  \cite{xiong2015}, $Re=1000$ is exactly the limit value of the Reynolds number for which band splitting begins to be observed. Thus, at this threshold value of $Re$ the flow dynamics may present a transitional behaviour between two different regimes, not fitting with that observed at larger values of $Re$. 
Finally, it is also worth to notice that the exponent of the power law approximating the minimal-energy threshold is much higher than those reported in previous works. For the plane Couette flow, 
a minimal seed energy varying as $Re^{-2.7}$ was reported by \cite{duguet2013}; 
for the asymptotic suction boundary layer, a  scaling of $Re^{-2}$ was found by \cite{CherubiniPoF2015}. 
This discrepancy may be more likely linked to the much larger domain  considered in the present study, rather than to the different type of flow. 
\\
\begin{figure}
        \centering
        \subfigure[$Re = 1000$, $E_0 = 8.4 \times 10^{-7}$]{\includegraphics[width=.45\columnwidth]{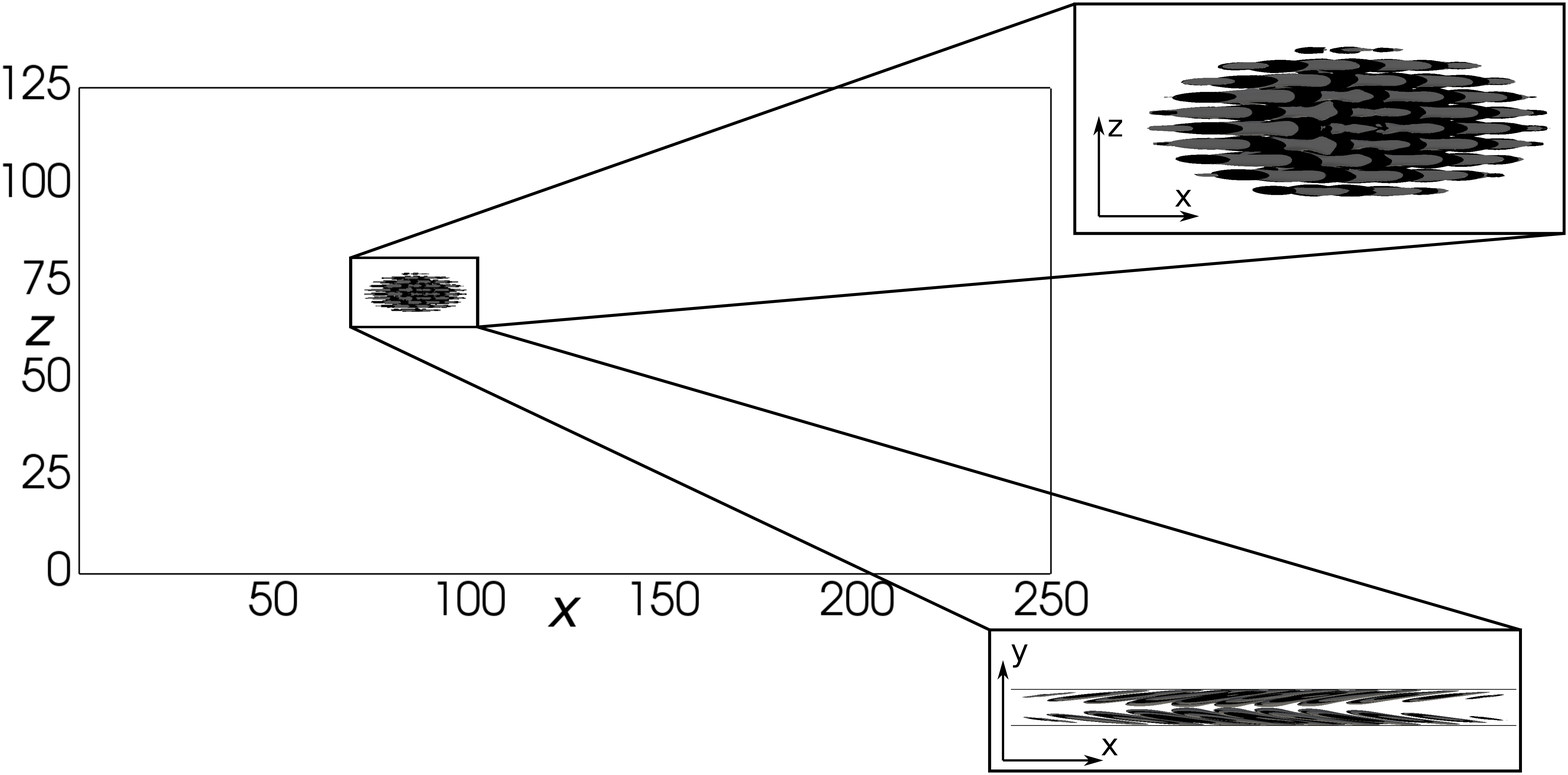}}
        \subfigure[$Re = 1150$, $E_0 = 1.1 \times 10^{-7}$]{\includegraphics[width=.45\columnwidth]{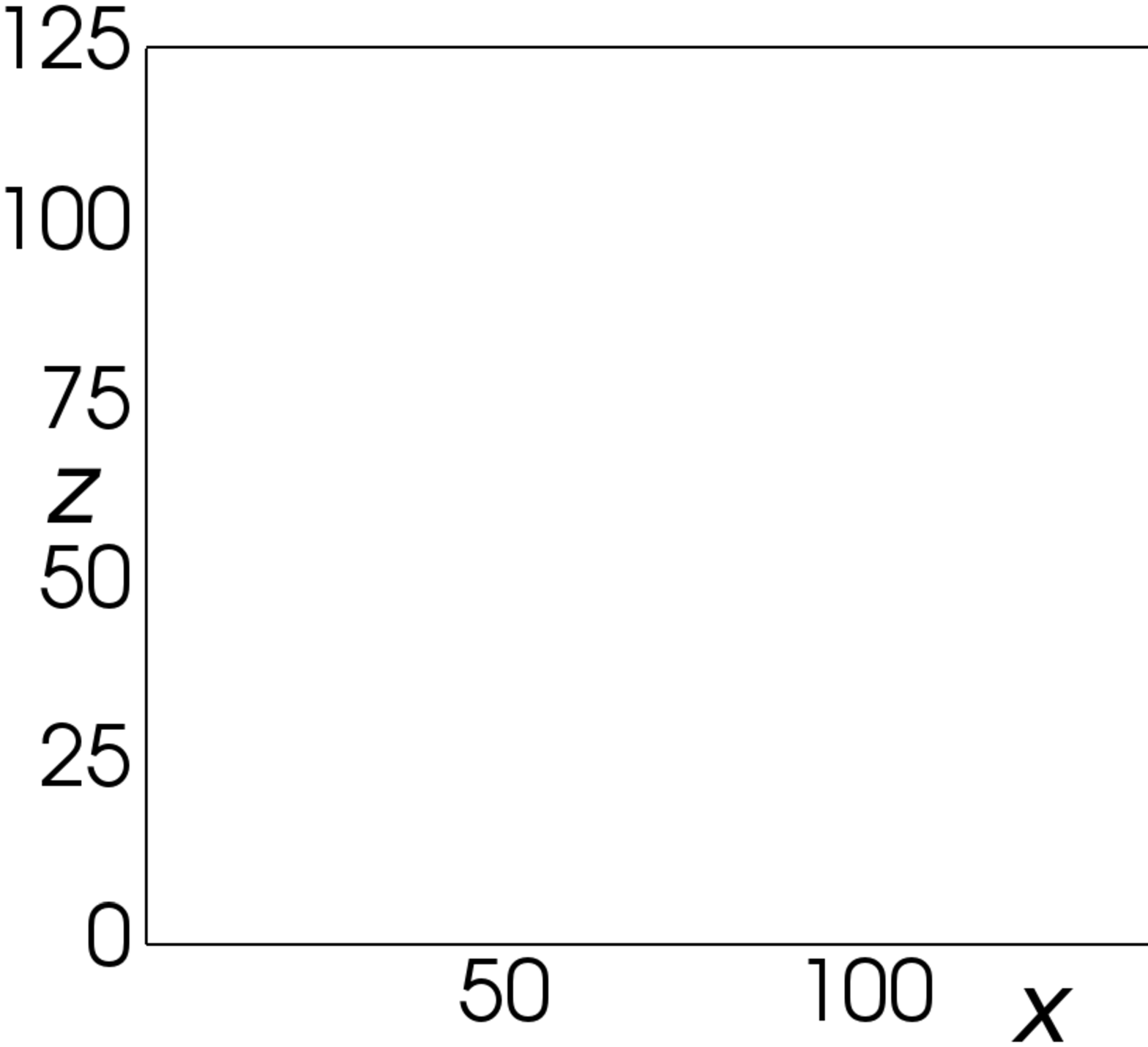}}
        \subfigure[$Re = 1250$, $E_0 = 9.4 \times 10^{-8}$]{\includegraphics[width=.45\columnwidth]{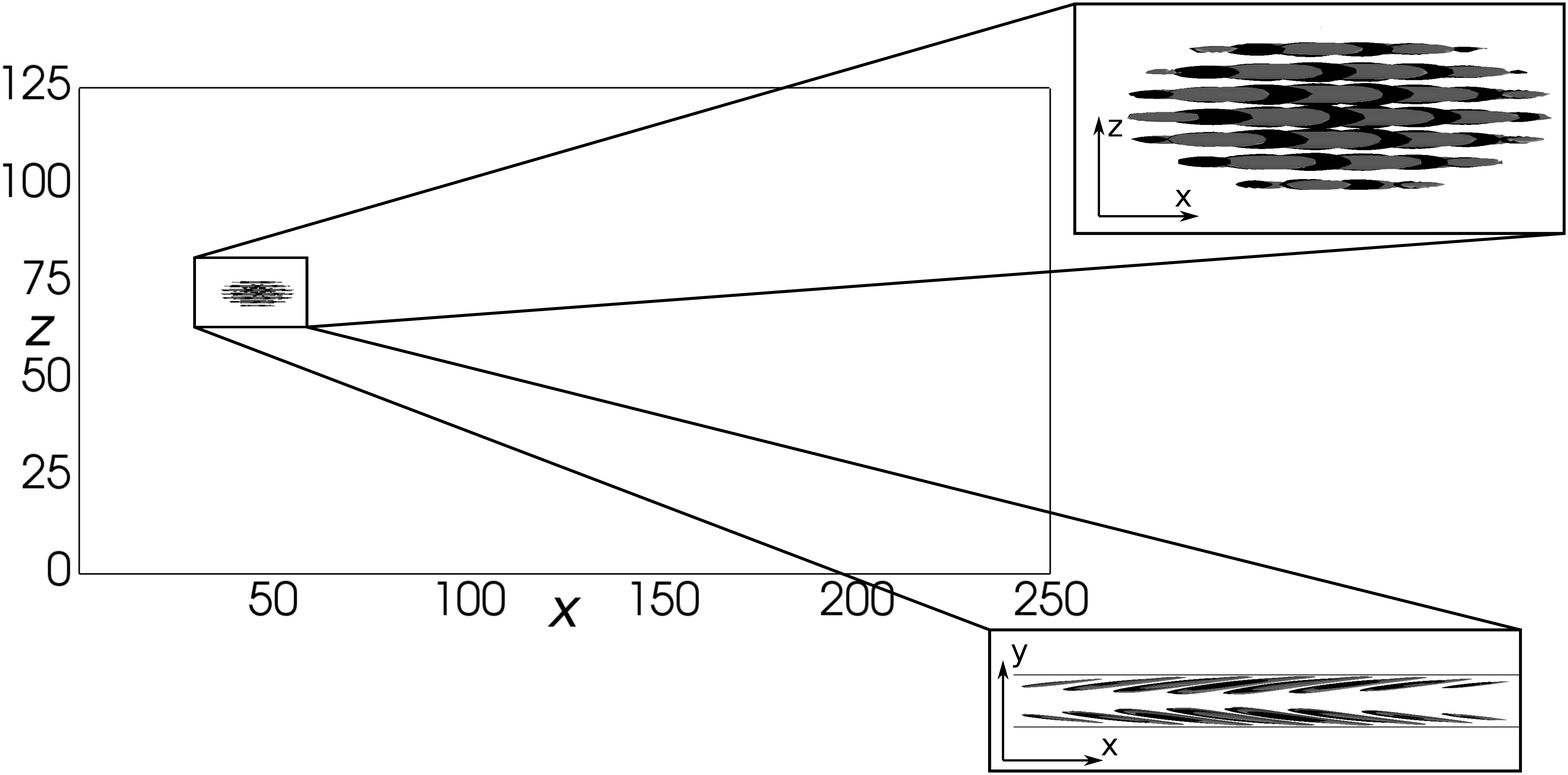}}
        \subfigure[$Re = 1568$, $E_0 = 3.7 \times 10^{-8}$]{\includegraphics[width=.45\columnwidth]{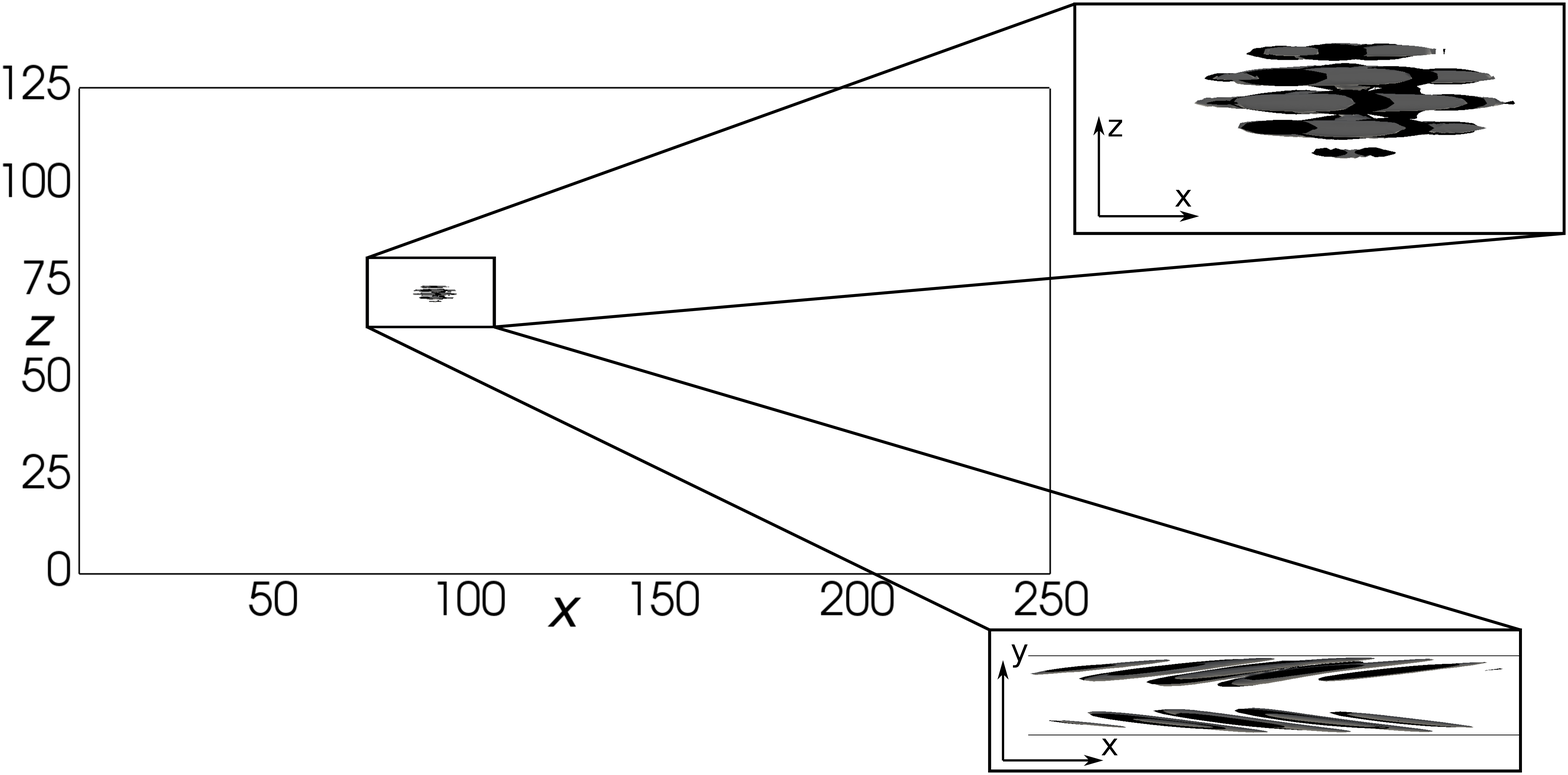}}
        \caption[]{Nonlinear optimal perturbation at time $t = 0$ for different Reynolds numbers with $U_{bulk} = 3/2$ obtained at target time $T = 10$ with the input energy reported in figure \ref{min_energy_input} by the black dots. Isosurface of the streamwise velocity (light grey for positive and black for negative values, $u = \pm 0.003$).} 
        \label{opt_sol_t0_diffRe}
\end{figure}
\begin{figure}
        \centering
        \subfigure[$Re = 1000$, $E_0 = 8.4 \times 10^{-7}$]{\includegraphics[width=.45\columnwidth]{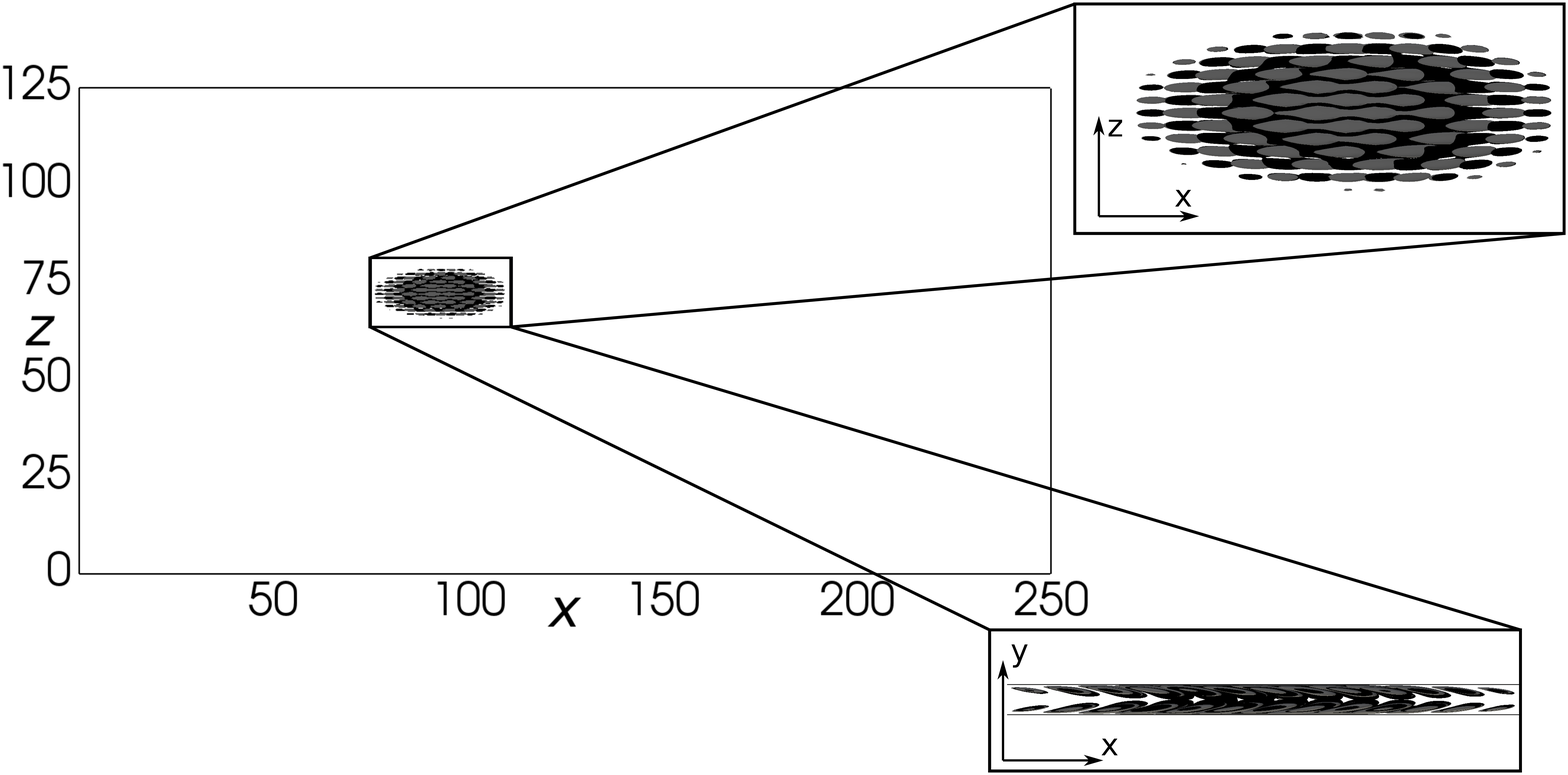}}
        \subfigure[$Re = 1150$, $E_0 = 1.1 \times 10^{-7}$]{\includegraphics[width=.45\columnwidth]{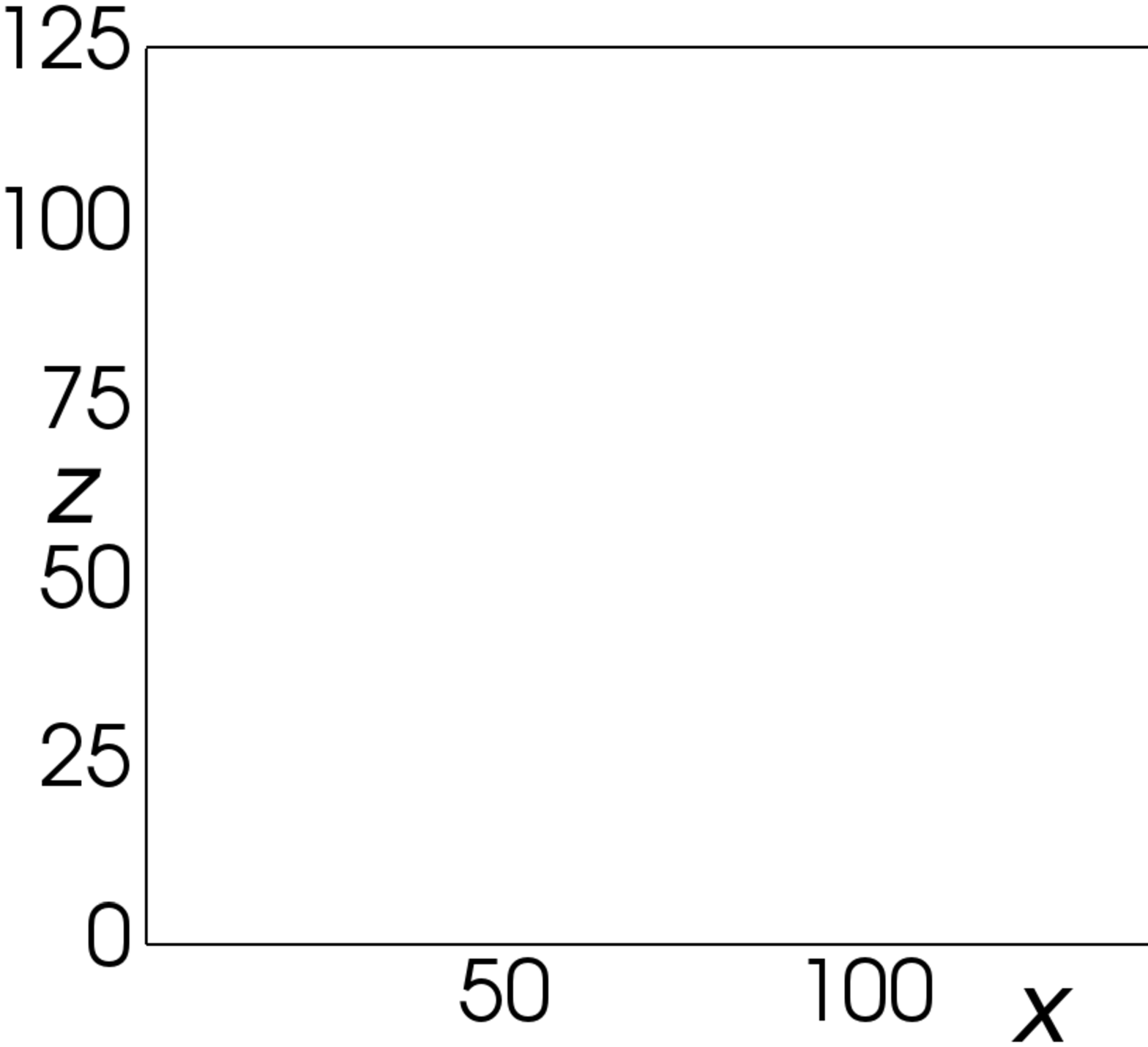}}
        \subfigure[$Re = 1250$, $E_0 = 9.4 \times 10^{-8}$]{\includegraphics[width=.45\columnwidth]{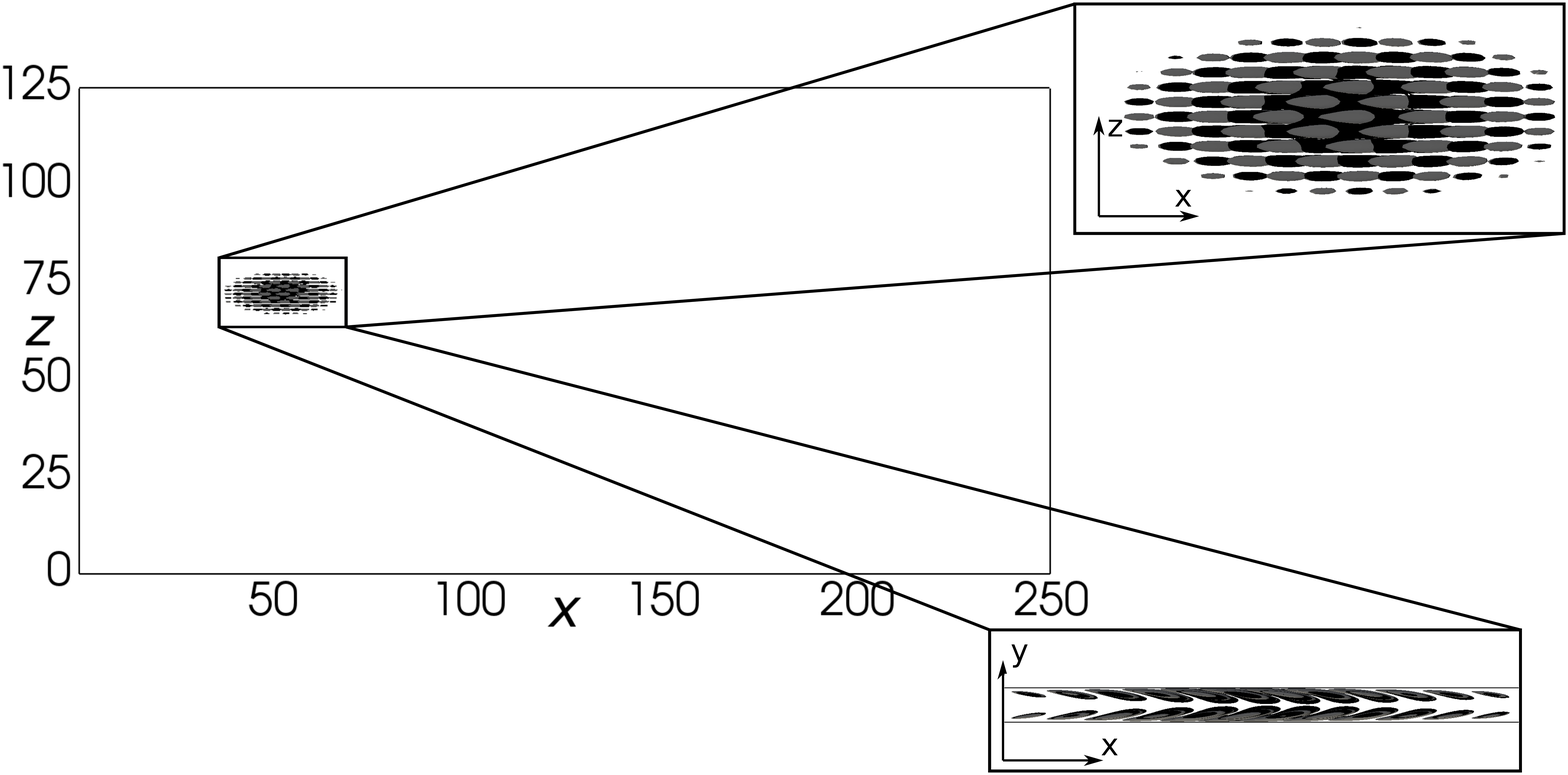}}
        \subfigure[$Re = 1568$, $E_0 = 3.7 \times 10^{-8}$]{\includegraphics[width=.45\columnwidth]{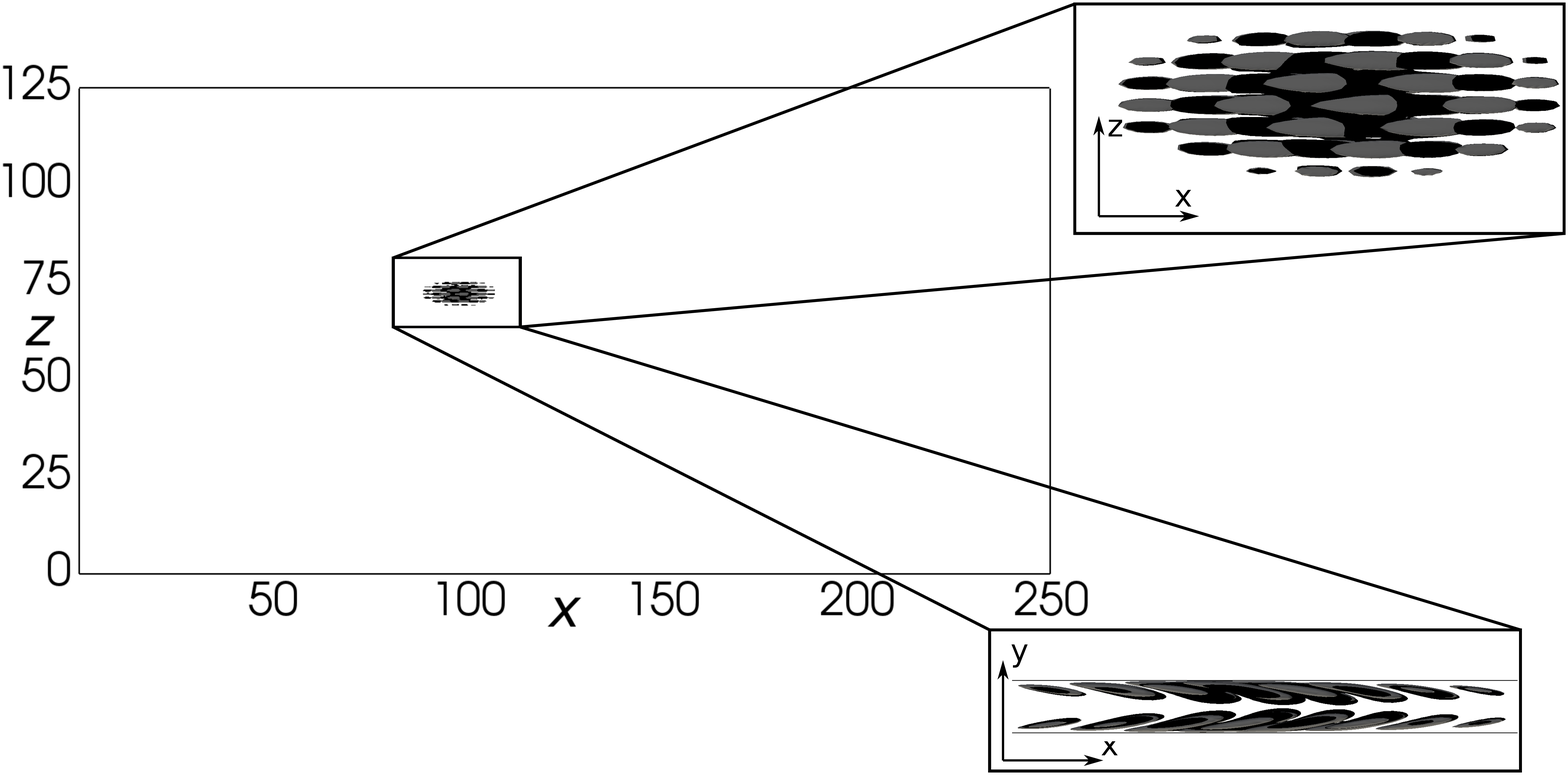}}
        \caption[]{Nonlinear optimal perturbations at time $t = T$ for different Reynolds numbers with $U_{bulk} = 3/2$ obtained at target time $T = 10$ with the input energy  reported in figure \ref{min_energy_input} by the black dots. Isosurface of the streamwise velocity (light grey for positive and black for negative values $u = \pm 0.02$).} 
        \label{opt_sol_tT_diffRe}
\end{figure}
\subsection{Nonlinear optimal perturbations}
\noindent The initial optimal solutions obtained for the target time $T = 10$ at the  threshold energy able to trigger turbulence are shown in figure \ref{opt_sol_t0_diffRe} for the four $Re$ considered. In all cases, the optimal perturbation is  spatially localised in a small region of the domain, having shape similar to that of a spot, being composed by alternating positive and negative finite-size streamwise streaks. Despite spatial localization has been already observed in nonlinear optimizations (for instance, see \cite{cherubini2011}, \cite{monokrousos2011}, \cite{pringle2012}),  the structure of this optimal perturbation 
is rather different to that of previously computed nonlinear optimal perturbations.
Whereas, it closely resembles the optimal wave packet recovered by linear optimization and windowing, obtained for the boundary layer flow by \cite{cherubiniJFMlin2010}. This is probably due to the low target time and initial energy used for this computations, which partially hinders the development of nonlinear effects. 
At $t=0$, all the optimal perturbations shown in figure \ref{opt_sol_t0_diffRe} present vortical structures pointing in the opposite direction to the shear (see the $x-y$ planes in the right bottom of each frame). Whereas, at the target time, the perturbations reverse their inclination,  pointing in the same direction of the shear flow, as shown in figure \ref{opt_sol_tT_diffRe}. This streamwise tilting is a common feature of optimal perturbations in shear flows \citep{cherubini2010,pringle2010,duguet2013}, and  suggests that the Orr mechanism is a fundamental mechanism involved in the early stage of transition to turbulence, with  characteristic time approximately equal to $10$ \citep{orr1907}.
One can also notice that, for $T=10$,  the shape of the optimal perturbations does not change much while evolving from $t=0$ to the target time, as shown in figure \ref{opt_sol_tT_diffRe}, remaining characterized by alternating positive and negative streaks localised in a spot-like region.
\\
\begin{figure}
        \centering
        \subfigure[$t = 0$]{\includegraphics[width=.495\columnwidth]{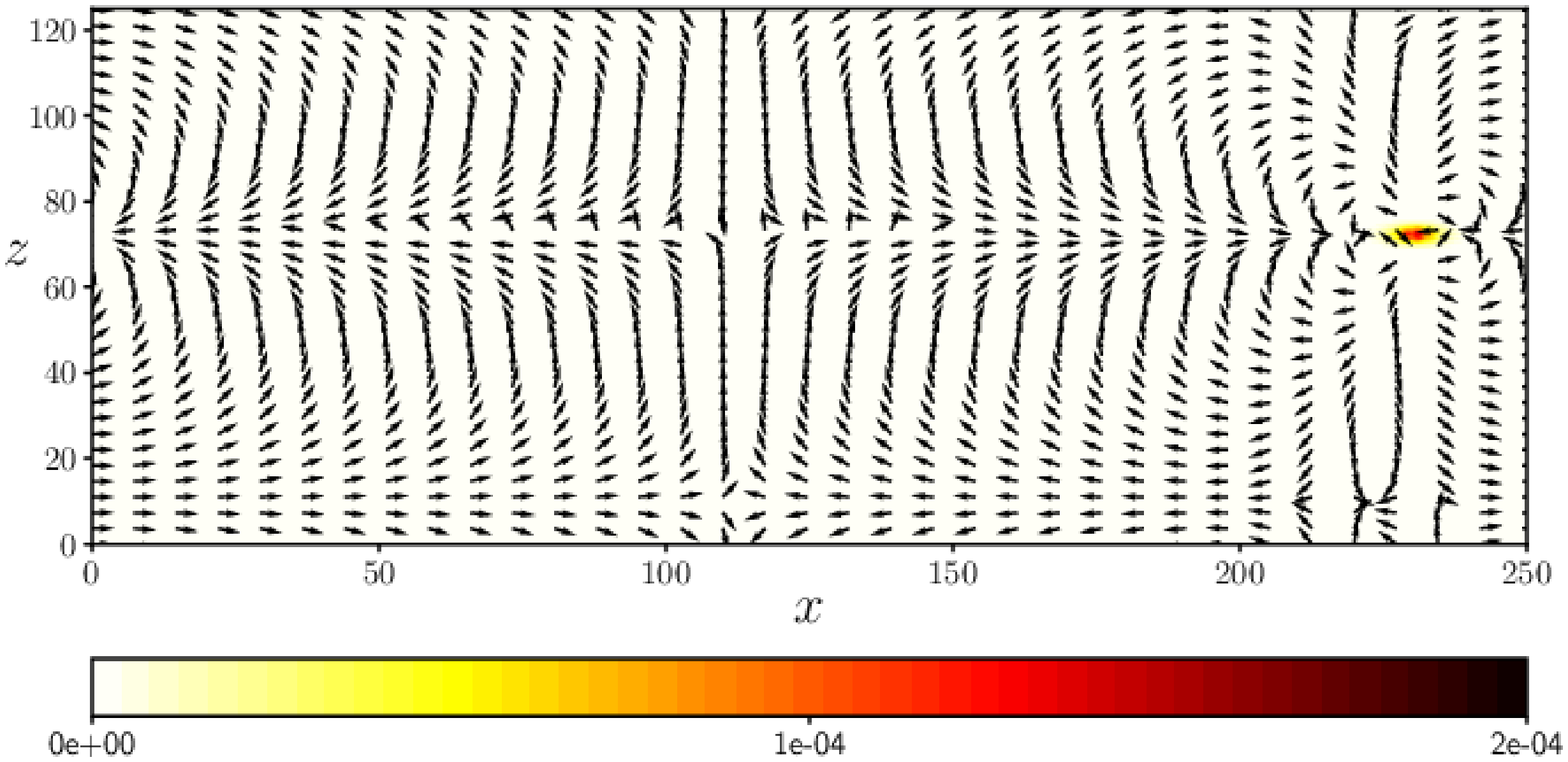}}
        \subfigure[$t = T$]{\includegraphics[width=.495\columnwidth]{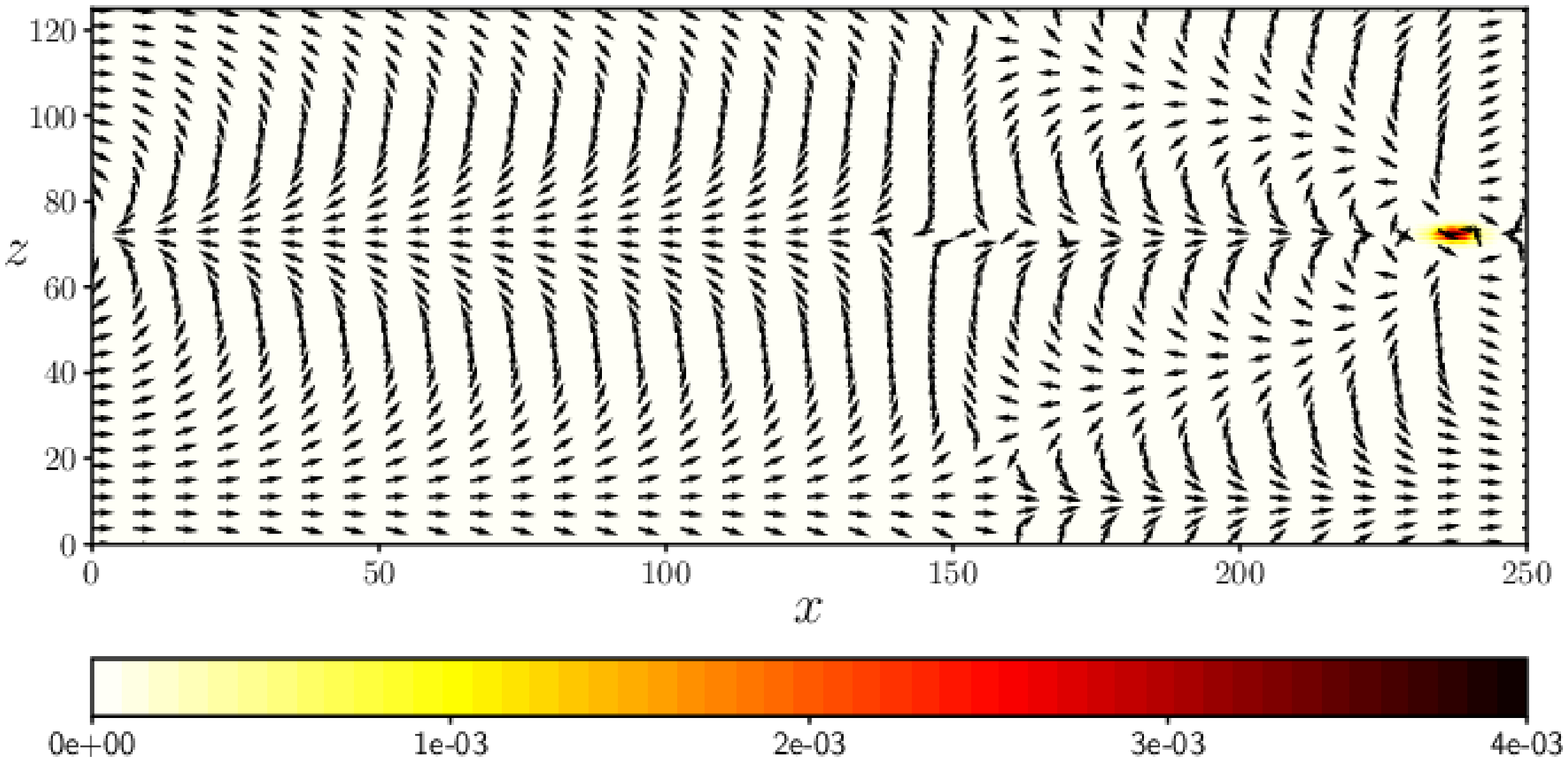}}
        \caption[]{Shaded isocontours of the crossflow energy $E_{cf}$ together with the normalized $y$-integrated large-scale flow (vectors) characterizing the nonlinear optimal for $Re = 1150$, $E_0 = 1.1 \times 10^{-7}$, $T = 10$: (a) initial optimal solution, (b) optimal solution at the target time.} 
        \label{2d_opt_sol_Re1150_E01p1e-7}
\end{figure}
In \cite{DuguetPRL2013}, it is discussed that oblique bands arise as a result of  advection of newly nucleated streaks in
the direction of a large-scale flow, which is oblique with
respect to the $x$ direction. The local orientation of the large-scale flow 
is thus responsible for the obliqueness of the
laminar-turbulent interface of growing incipient spots as
well as for maintaining turbulent stripes \citep{DuguetPRL2013}. To ascertain whether the computed nonlinear optimal perturbations contain the seed for the development of turbulent bands, we compute the large-scale flow related to the optimal disturbances by averaging the instantaneous velocity field  in the wall normal direction as $\overline{u_i} = \int_{-1}^1 u_i dy$. 
 Notice that $\overline{u}$ is zero where the flow is laminar, close to
zero where the flow is turbulent, but it is non zero on laminar-turbulent interfaces, due to a mismatch of the streamwise flow
rates across them, linked to  the presence of overhang regions. 
  In figure \ref{2d_opt_sol_Re1150_E01p1e-7}, we have reported the isocontours of the cross-flow energy $E_{cf} = (1/2) \int (v^2 +w^2) dy$, surrounded by the large-scale field $(\overline{u}, \overline{w})$, for the initial optimal perturbation at $Re=1150$, $E_0=1.1 \times 10^{-7}$. The optimal flow field is characterized by two different scales: a small-scale flow embedded within the spot-like structure and a large-scale flow in the form of large vortices filling the whole domain. The latter is characterised by a streamwise flow entering the spot and a spanwise flow exiting it, constituting a quadrupolar structure. Quadrupolar large-scale flow around spots or laminar-turbulent bands has been observed in plane Couette flow (\cite{schumacher2001,lagha2007}) and  plane Poiseuille flow \citep{lemoult2014}, due to the shearing of the streamwise velocity and the breaking of the spanwise homogeneity  \citep{wang_duguet2020}. 
  Notice that this large-scale flow, which is not associated with the presence of overhang regions, remains almost unchanged  at $t=0,T$ (compare figures \ref{2d_opt_sol_Re1150_E01p1e-7} (a) and (b)).  
\\
\begin{figure}
        \centering
        \subfigure[$t = 0$]{\includegraphics[width=.495\columnwidth]{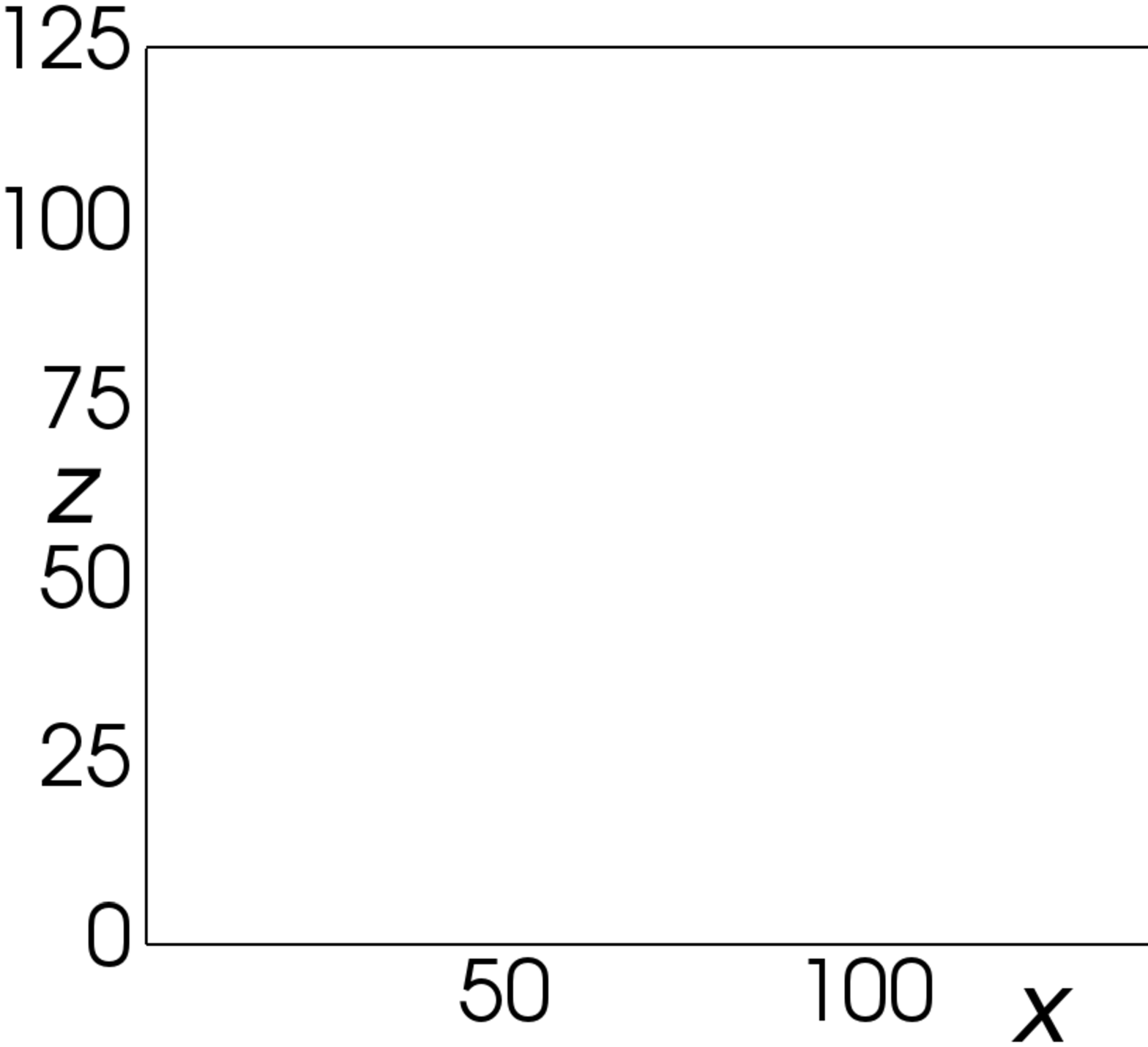}}
        \subfigure[$t = T$]{\includegraphics[width=.495\columnwidth]{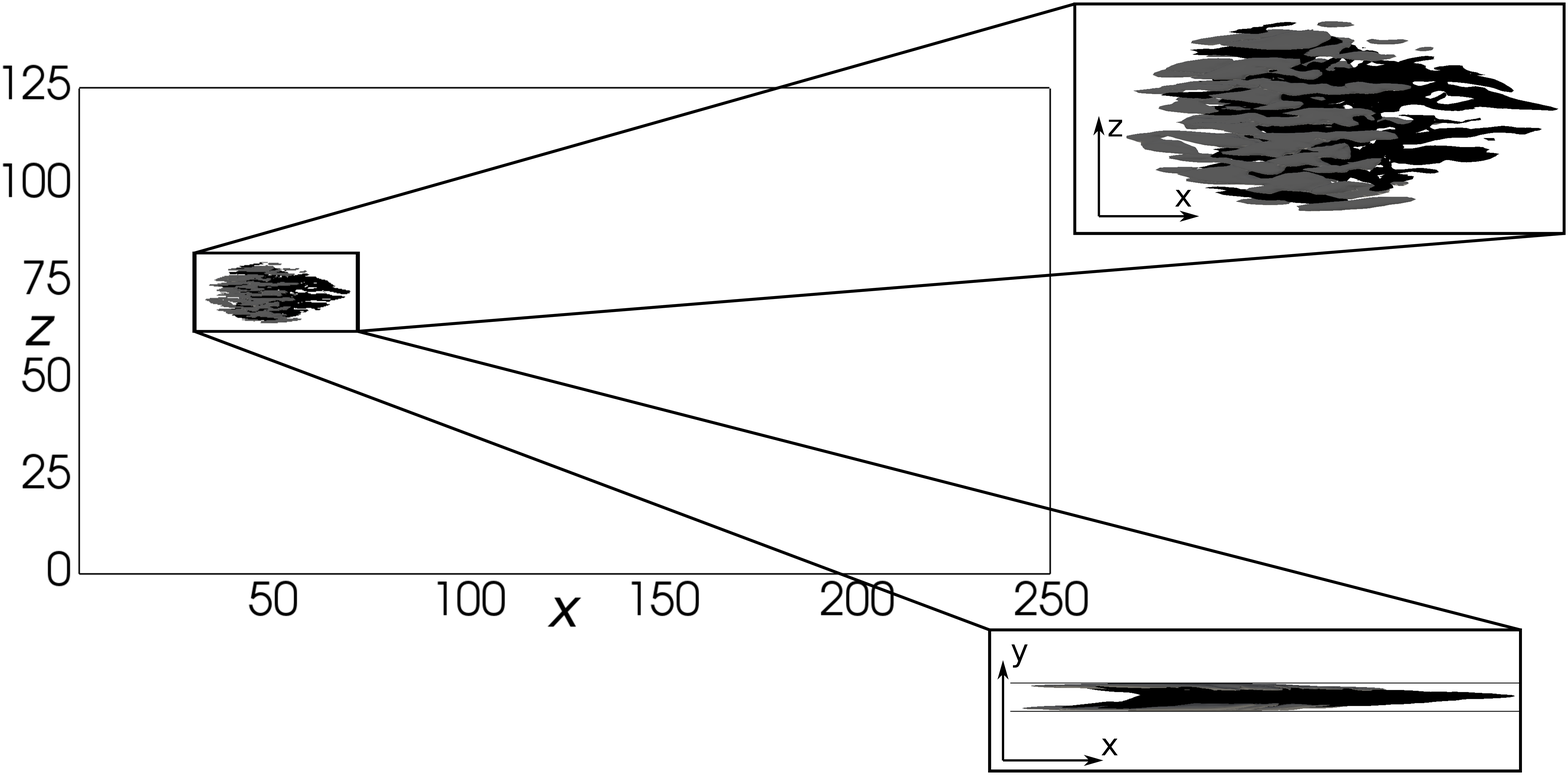}}
        \caption[]{Optimal perturbation for $Re = 1150$, $U_{bulk} = 3/2$, $T = 100$ and $E_0 = 1.1 \times 10^{-7}$ at (a) $t = 0$ and (b) $t = T$: isosurface of the streamwise velocity (light grey for positive and black for negative values, $u = \pm 0.003$ (a) and $u = \pm 0.12$ (b)).}
        \label{opt_sol_Re1150_E01p1e-7_T100} 
\end{figure}
\noindent Increasing the target time to $T=100$ for $Re=1150$, while keeping $E_0=1.1 \times 10^{-7}$, we obtain the optimal solution shown in figure \ref{opt_sol_Re1150_E01p1e-7_T100} (a), which presents a localised structure similar to that computed for low target times, despite being not perfectly symmetric 
and having thicker streaks compared to those at lower target time. The  optimal solution at the target time, shown in figure \ref{opt_sol_Re1150_E01p1e-7_T100}(b), is very similar to that of a turbulent spot, now clearly presenting an overhang region. 
\noindent Moreover, 
it is characterized by a quadrupolar large-scale flow surrounding the spot-like small-scale disturbances (not shown). 
Starting from this optimal (but yet not minimal) solution, the optimize-and-bisect procedure is carried out at $T=100$ for $Re=1150$. The same procedure is carried out for all the considered values of $Re$, for obtaining the minimal seeds whose energy is reported in figure \ref{min_energy_input} (b). 
\begin{figure}
        \centering
        \subfigure[$t = 0$]{\includegraphics[width=.495\columnwidth]{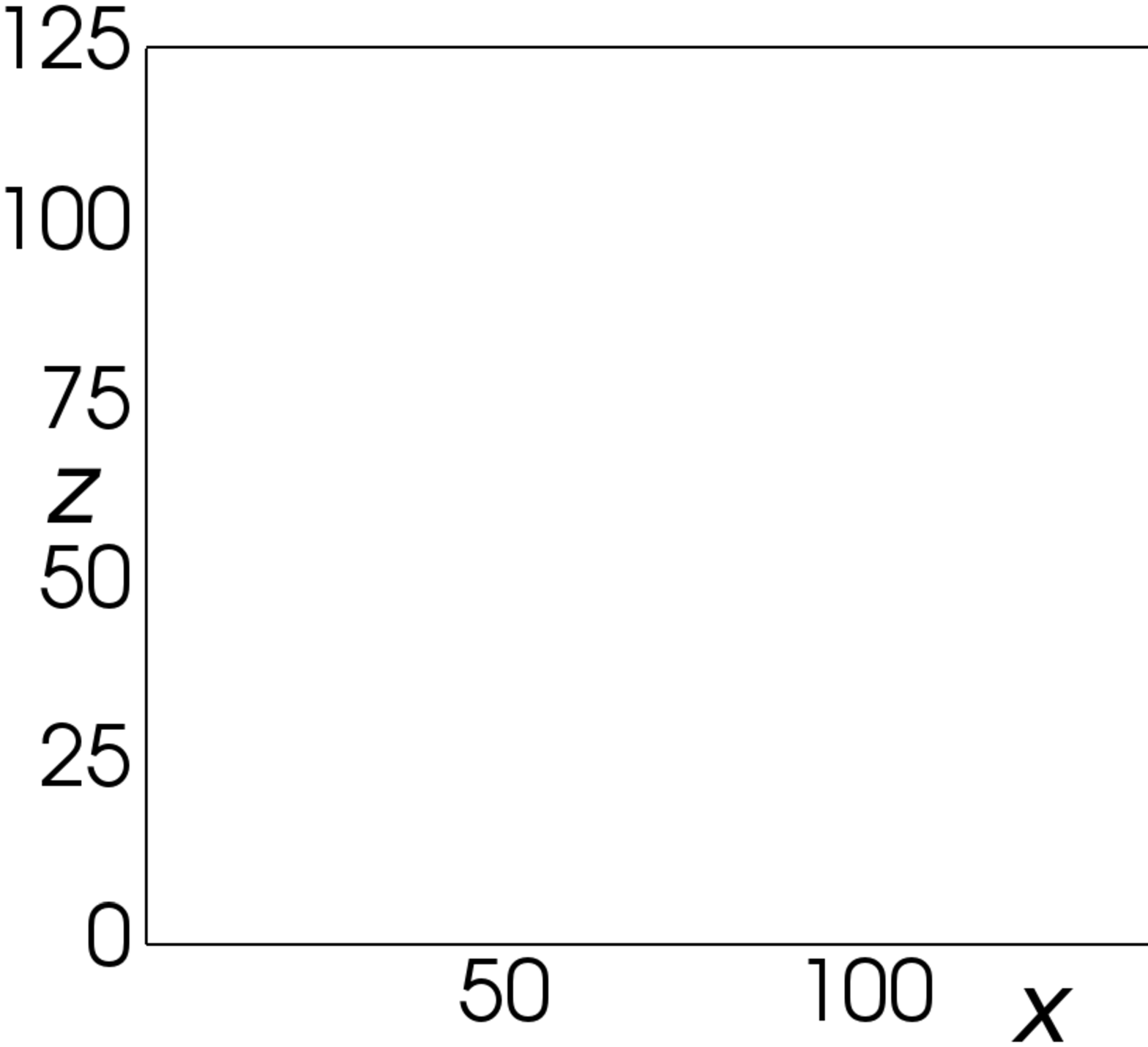}}
        \subfigure[$t = 25$]{\includegraphics[width=.495\columnwidth]{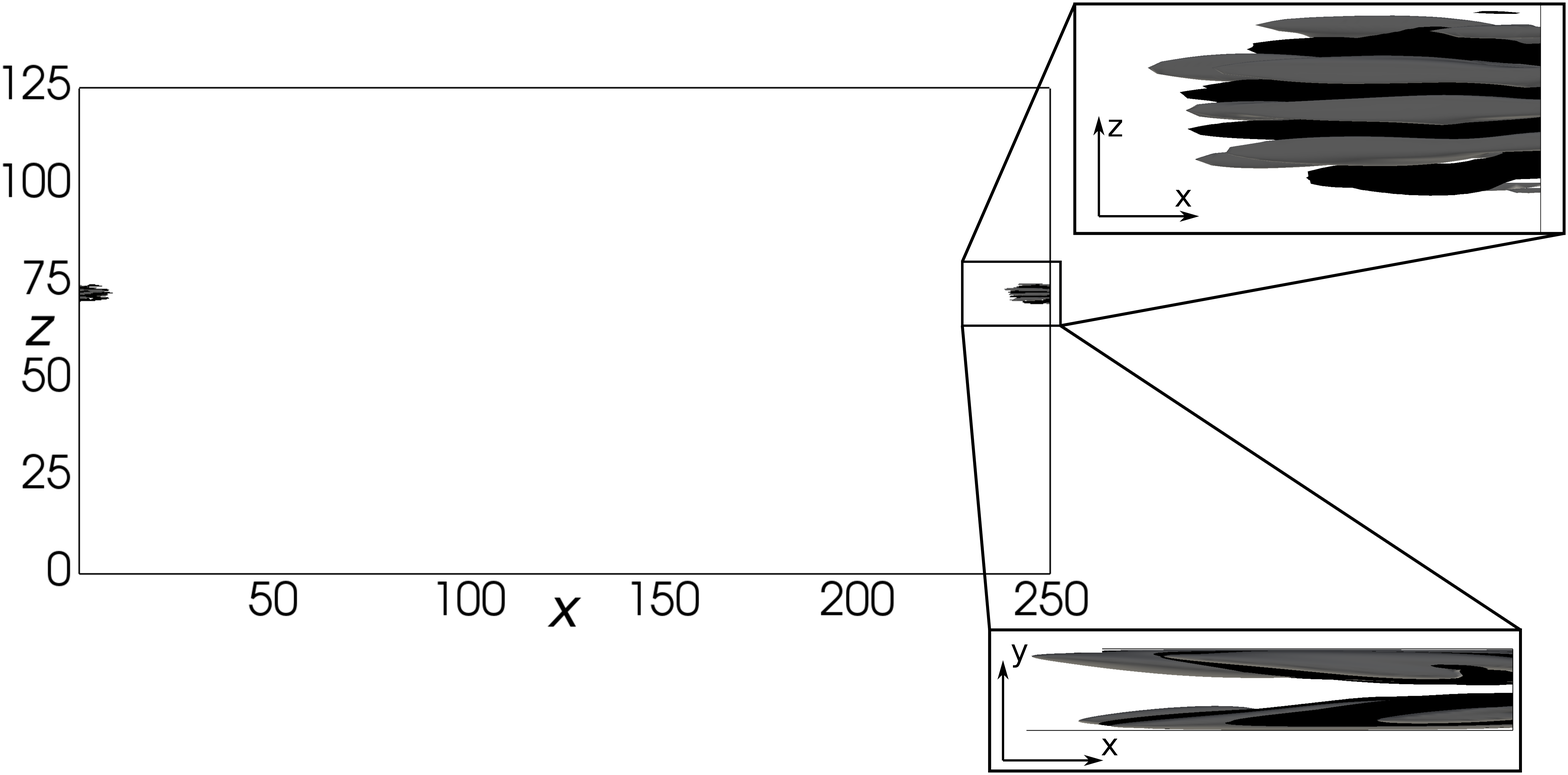}}
        \subfigure[$t = 50$]{\includegraphics[width=.495\columnwidth]{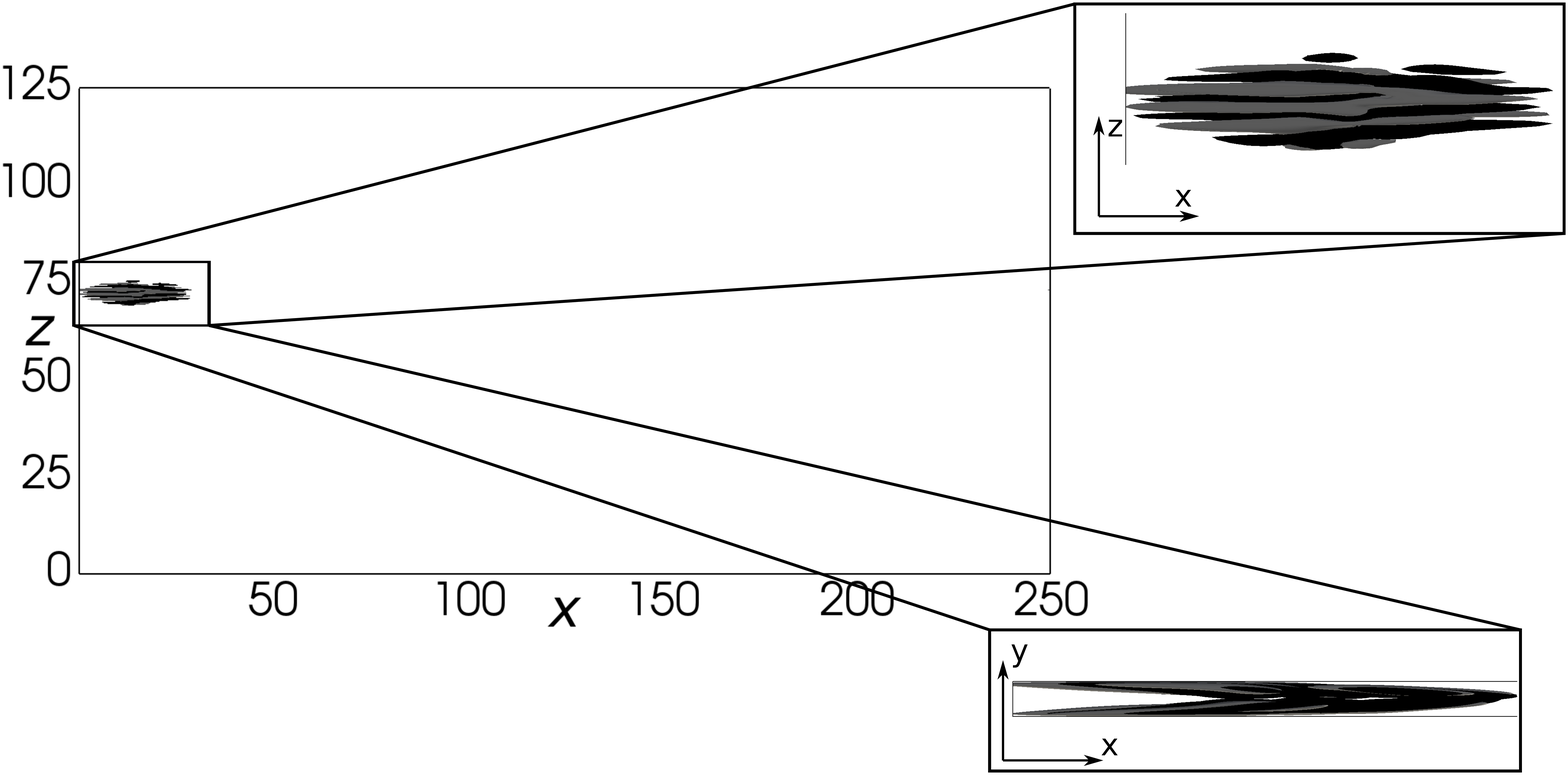}}
        \subfigure[$t = T$]{\includegraphics[width=.495\columnwidth]{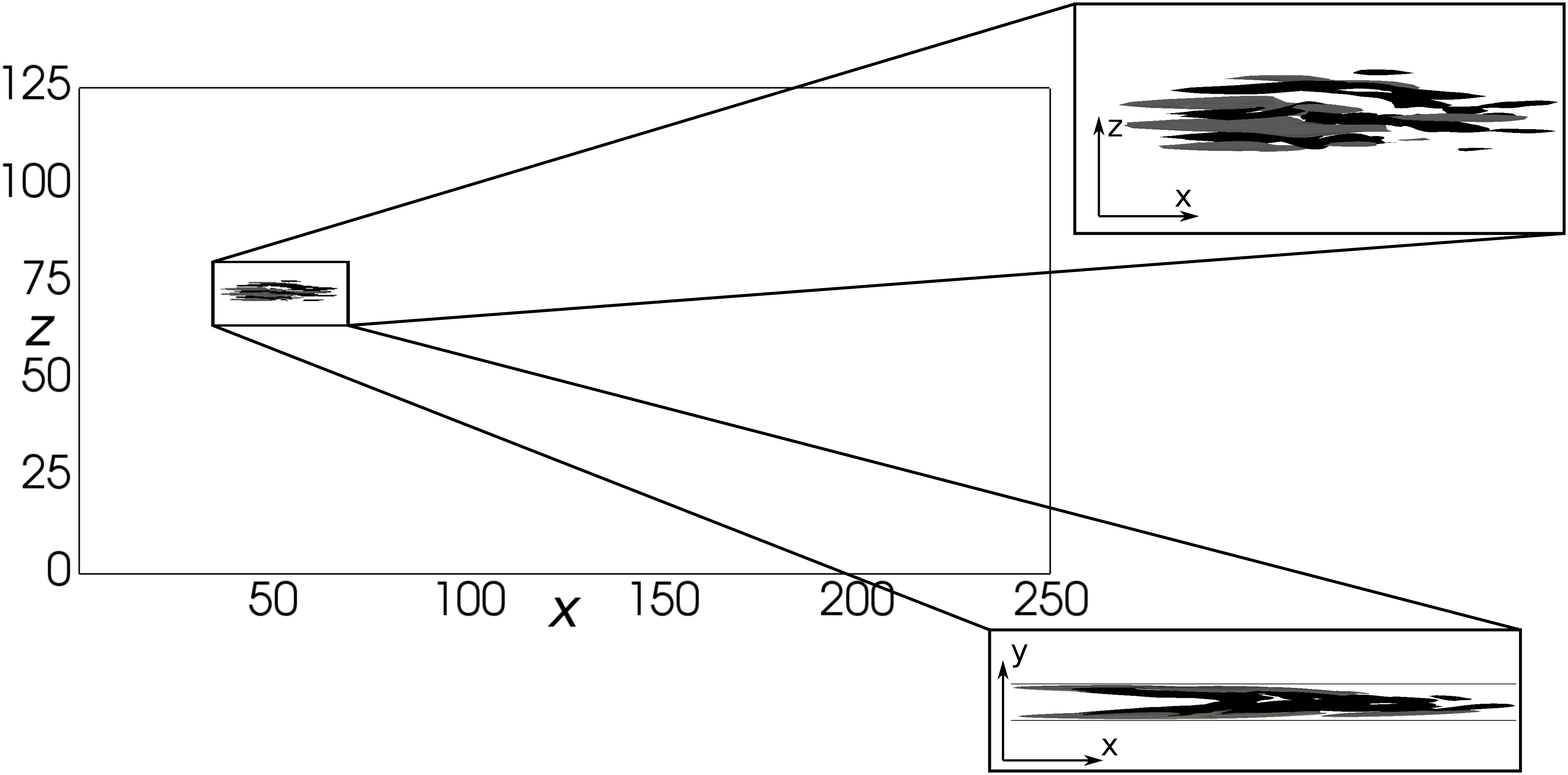}}
        \caption[]{Minimal seed for $Re = 1150$, $T = 100$ and $E_0 = 4.7 \times 10^{-8}$ at (a) $t = 0$, (b) $t = 25$, (c) $t = 50$, (d) 
        $t = T$: isosurface of the streamwise velocity (light grey for positive and black for negative values, $u = \pm 0.003$ (a), $u = \pm 0.02$ (b-c) and $u = \pm 0.08$ (d).}
        \label{opt_sol_Re1150_E04p7e-8_T100} 
\end{figure}
\subsection{Minimal seed at different Reynolds}

\begin{figure}
        \centering
        \subfigure[$Re = 1000$, $E_0 = 5.5 \times 10^{-7}$ ]{\includegraphics[width=.35\columnwidth]{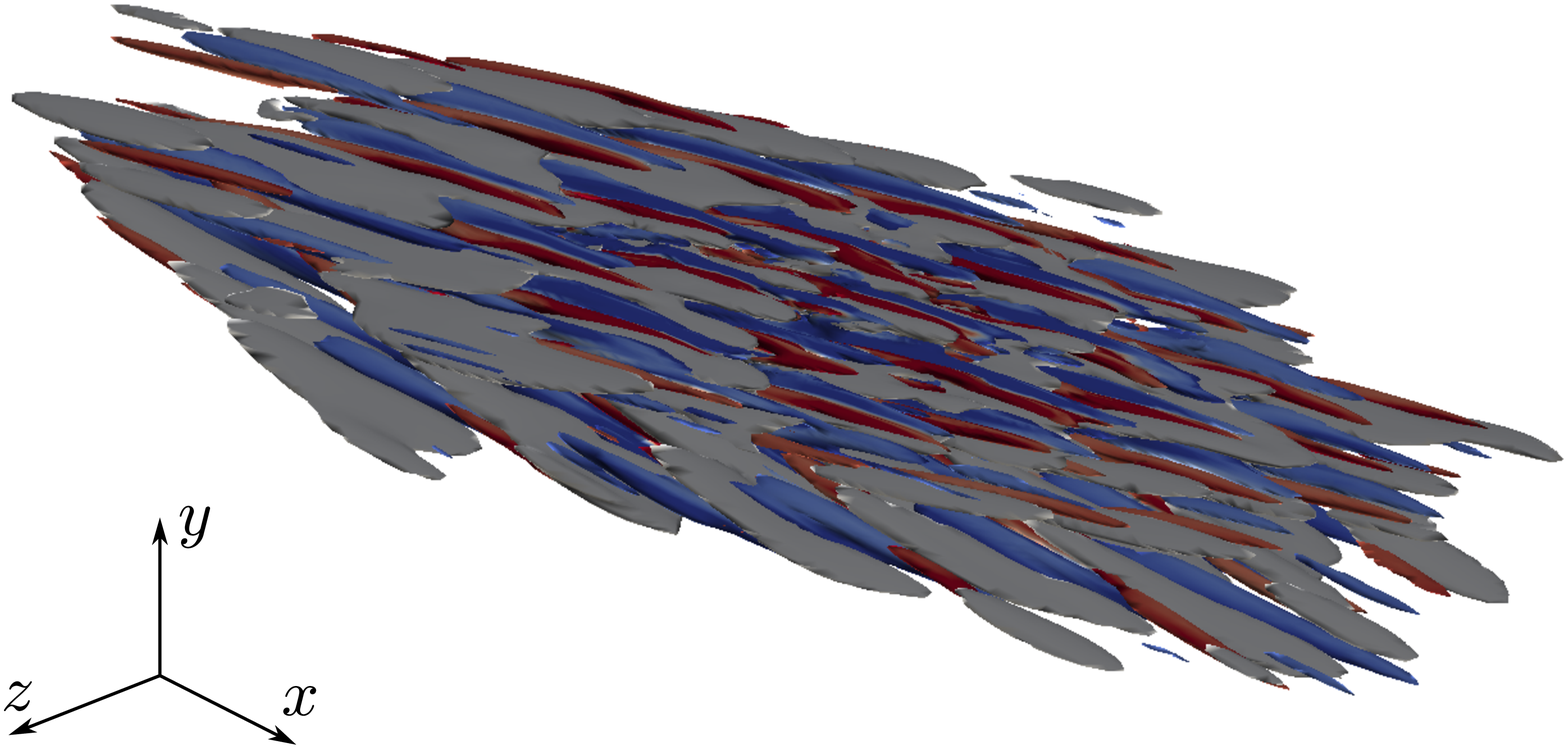}}
        \subfigure[$Re = 1150$, $E_0 = 4.7 \times 10^{-8}$ ]{\includegraphics[width=.35\columnwidth]{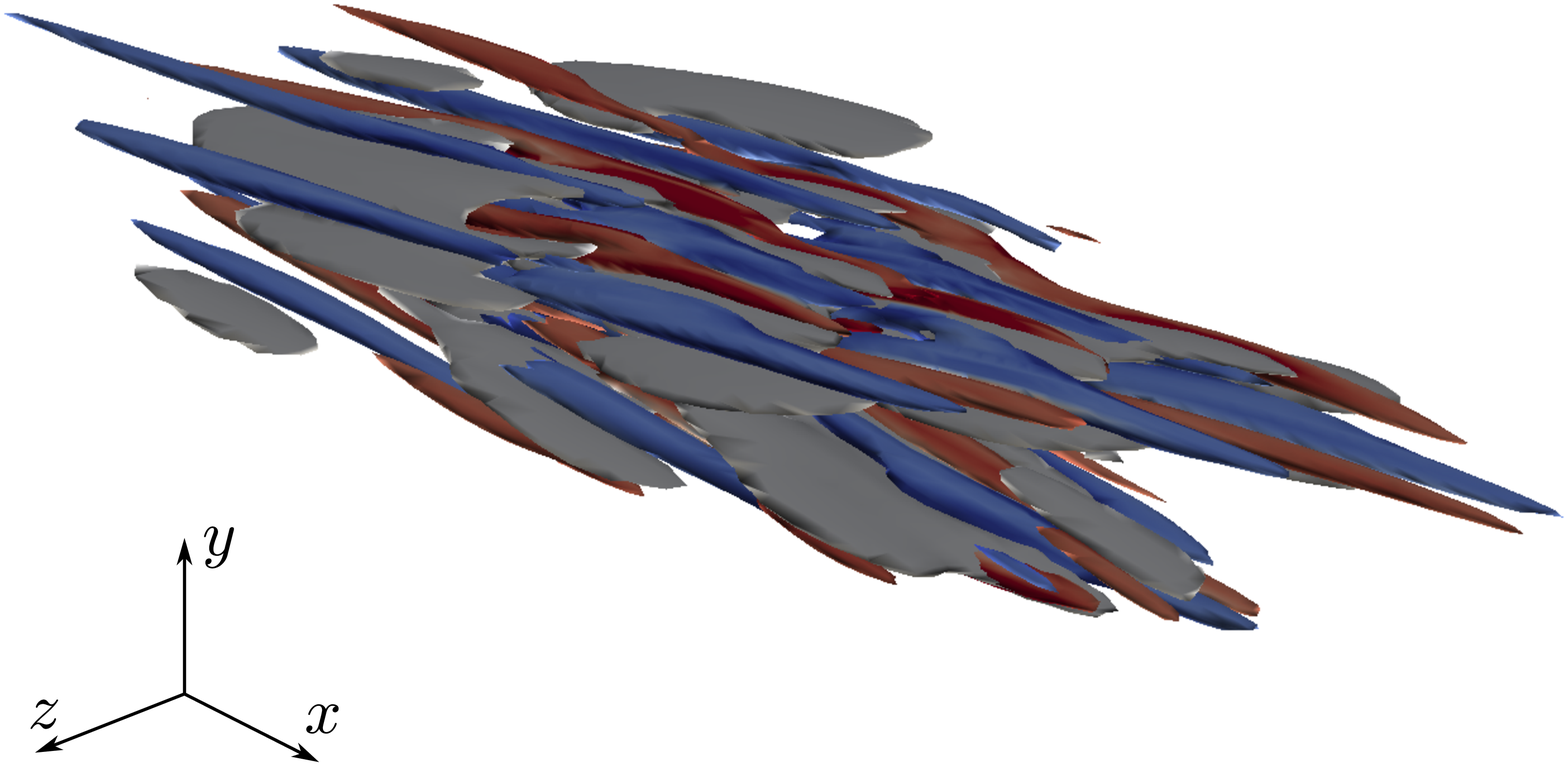}}
        \subfigure[$Re = 1250$, $E_0 = 2.9 \times 10^{-8}$]{\includegraphics[width=.35\columnwidth]{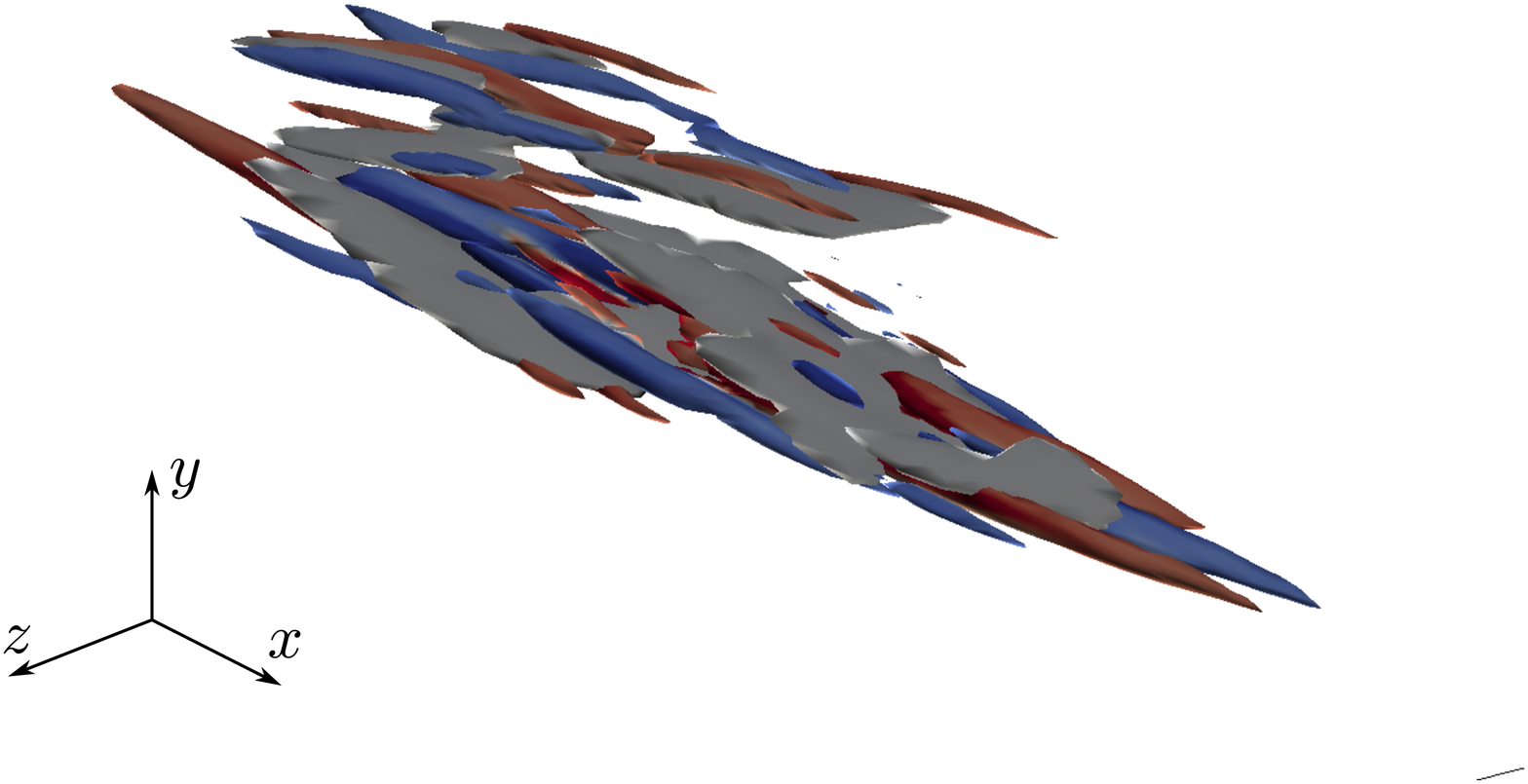}}
        \subfigure[$Re = 1568$, $E_0 = 3.6 \times 10^{-9}$]{\includegraphics[width=.35\columnwidth]{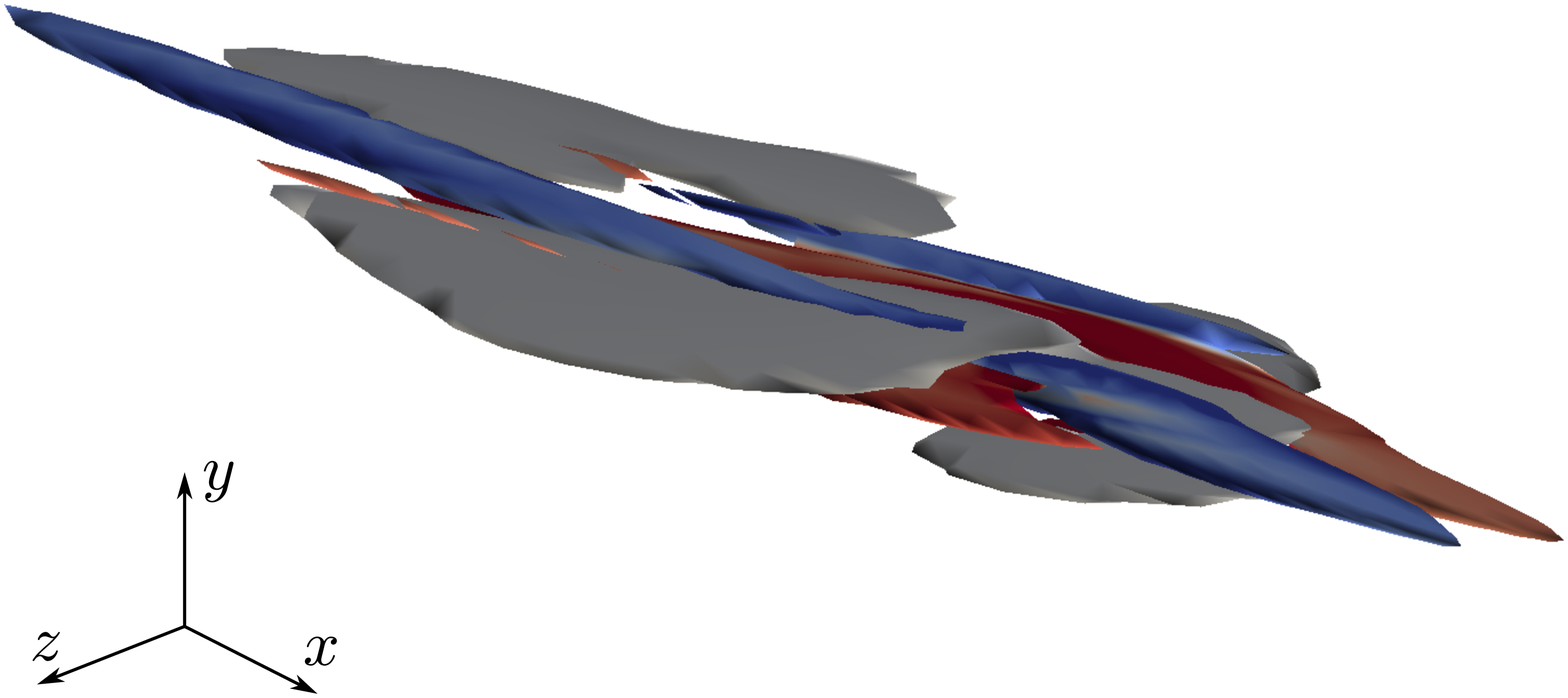}}
        \caption[]{Minimal seed at different Reynolds numbers: isosurface of negative streamwise velocity ($u = -0.0025$, light grey) and Q-criterion ($Q = 0.003$) coloured by the streamwise vorticity (positive red, negative blue).}
        \label{fig:3D-minRe1150} 
\end{figure}

\begin{figure}
        \centering
        \subfigure[$Re = 1000$, $E_0 = 5.5 \times 10^{-7}$ ]{\includegraphics[width=.35\columnwidth]{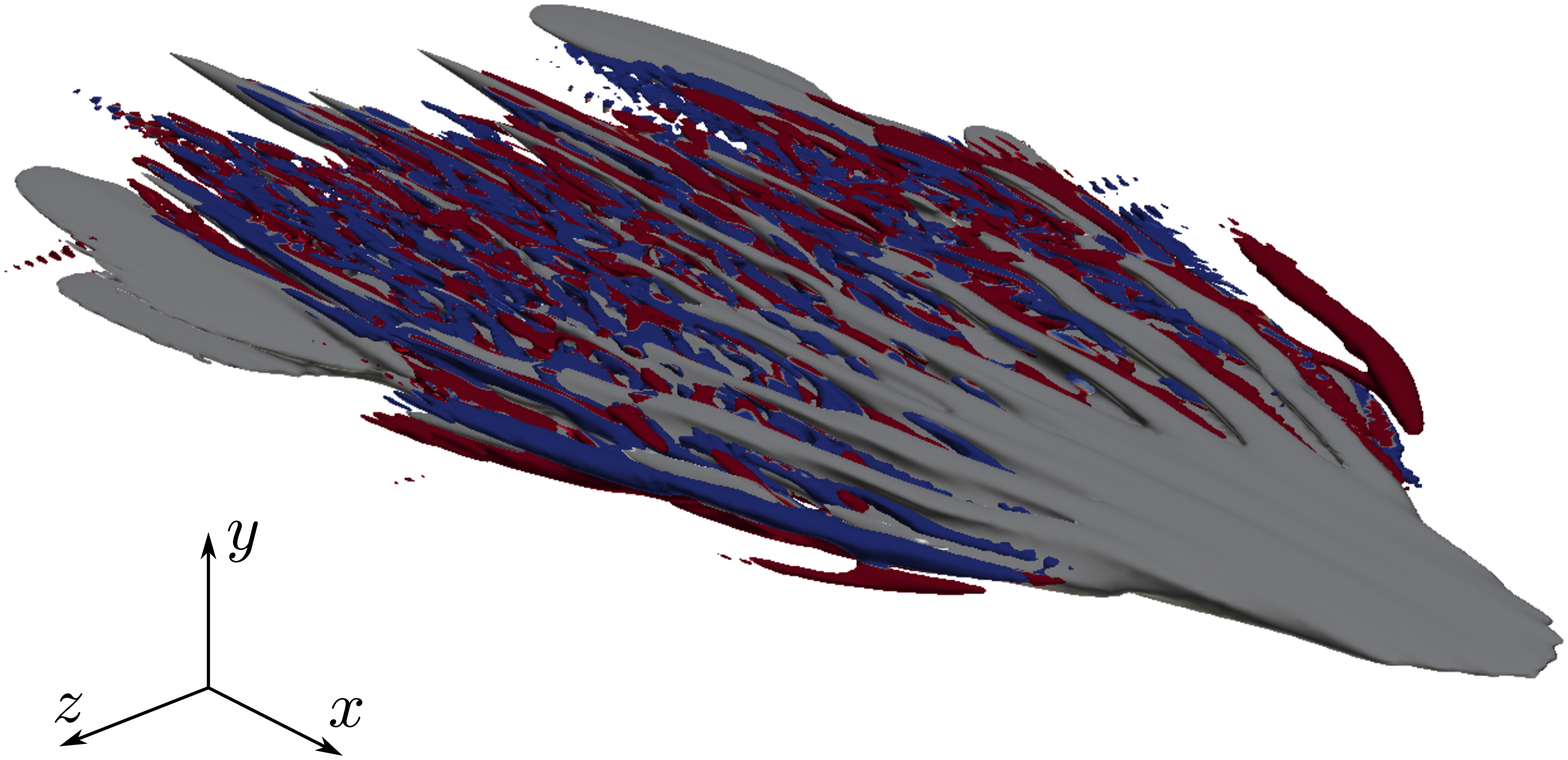}}
        \subfigure[$Re = 1150$, $E_0 = 4.7 \times 10^{-8}$ ]{\includegraphics[width=.35\columnwidth]{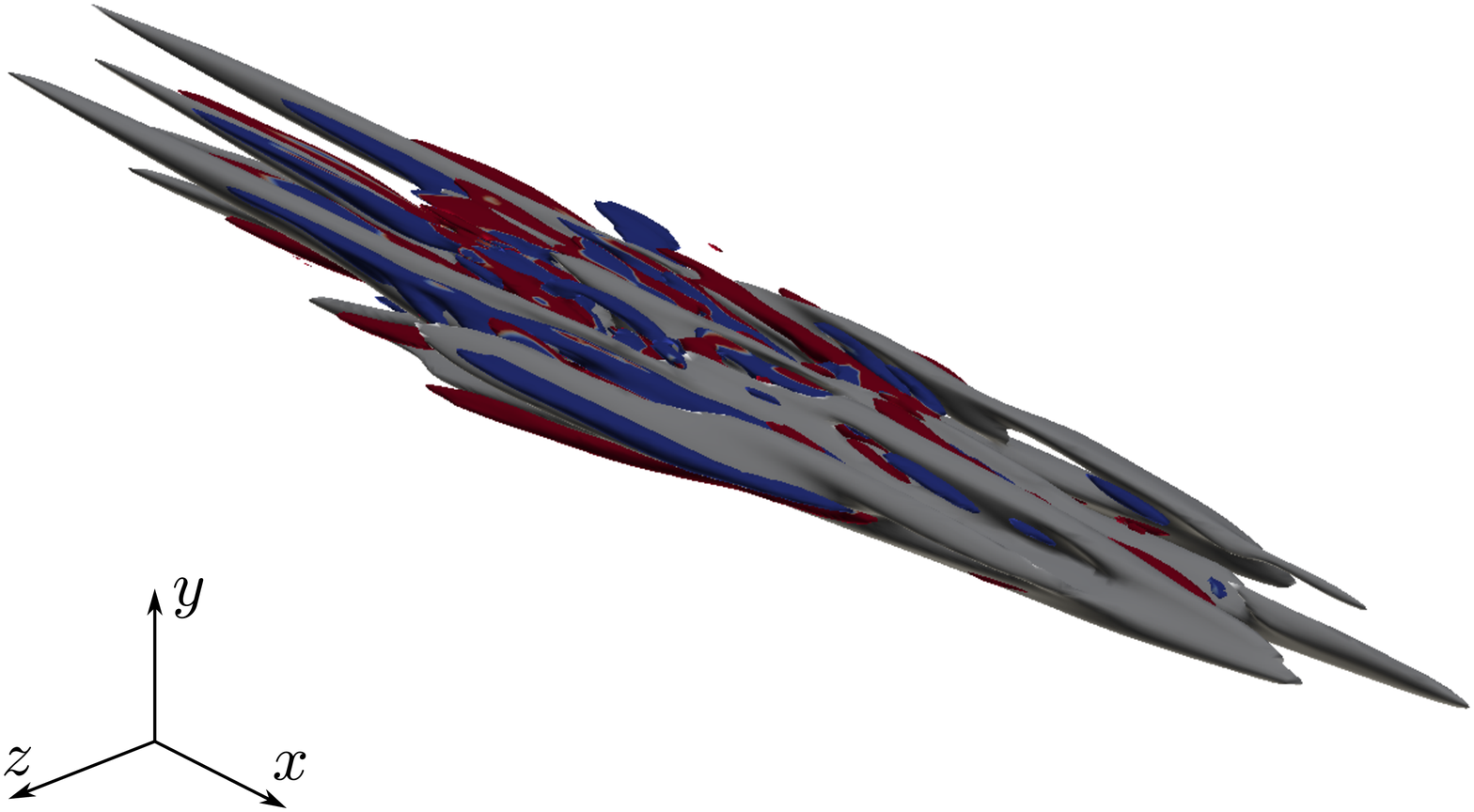}}
        \subfigure[$Re = 1250$, $E_0 = 2.9 \times 10^{-8}$]{\includegraphics[width=.35\columnwidth]{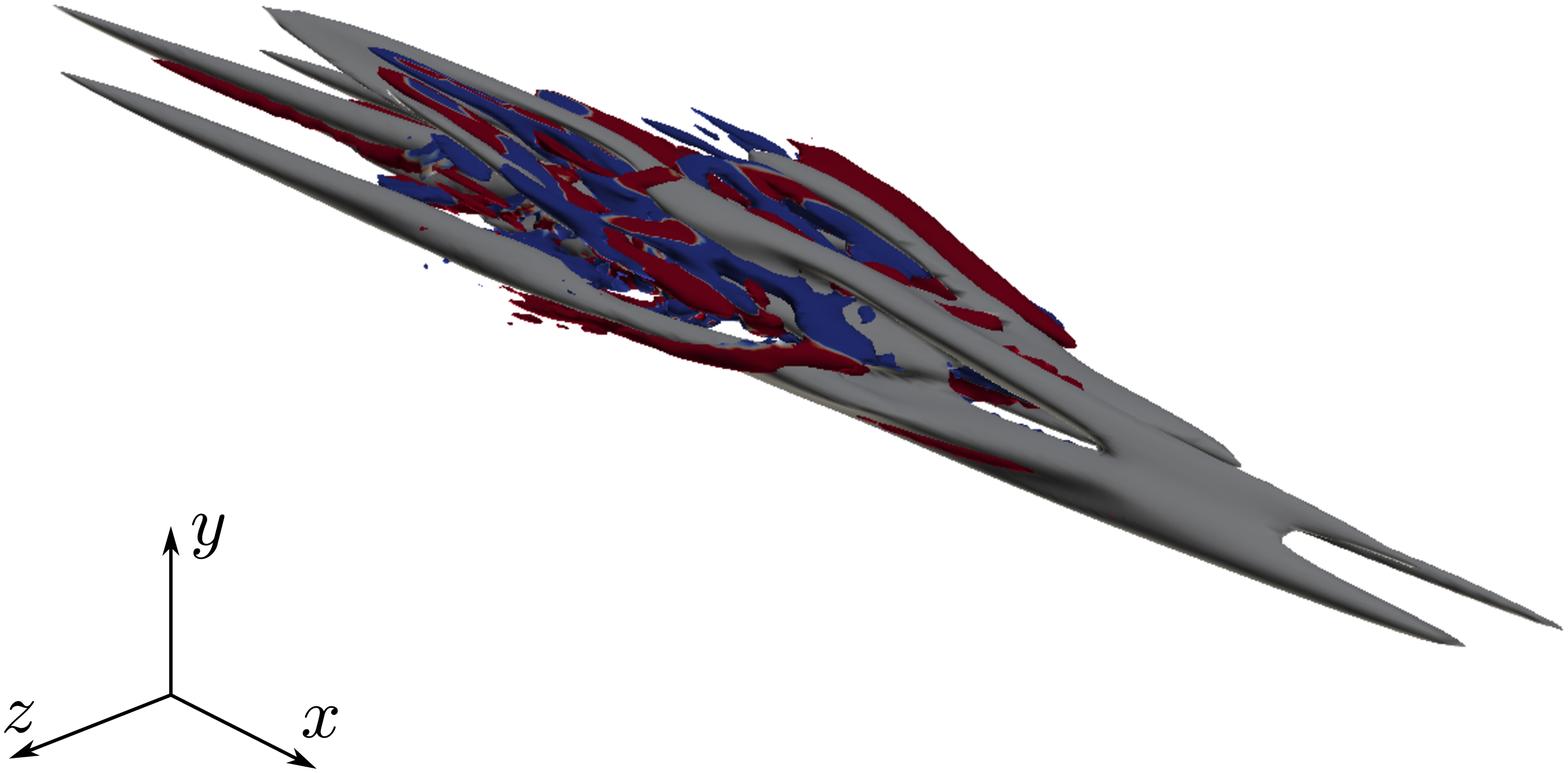}}
        \subfigure[$Re = 1568$, $E_0 = 3.6 \times 10^{-9}$]{\includegraphics[width=.35\columnwidth]{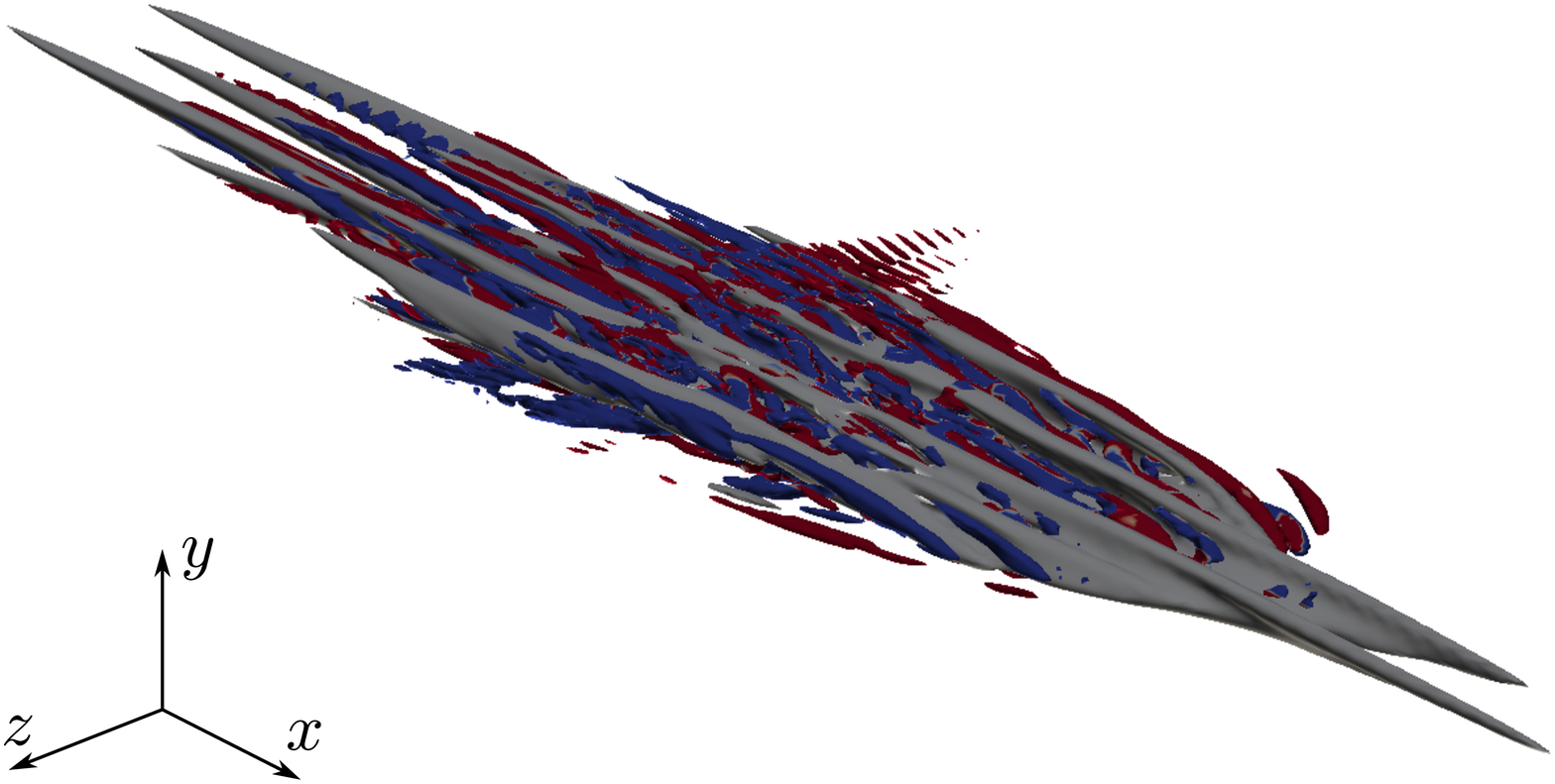}}
        \caption[]{Minimal seed at target time $t = T$ for different Reynolds numbers: isosurface of negative streamwise velocity ($u = -0.03$, light grey) and Q-criterion ($Q = 0.005$) coloured by the streamwise vorticity (positive red, negative blue).}
        \label{fig:3D-minimalseed-T} 
\end{figure}

\begin{figure}
        \centering
        \subfigure[$t = 0$]{\includegraphics[width=.495\columnwidth]{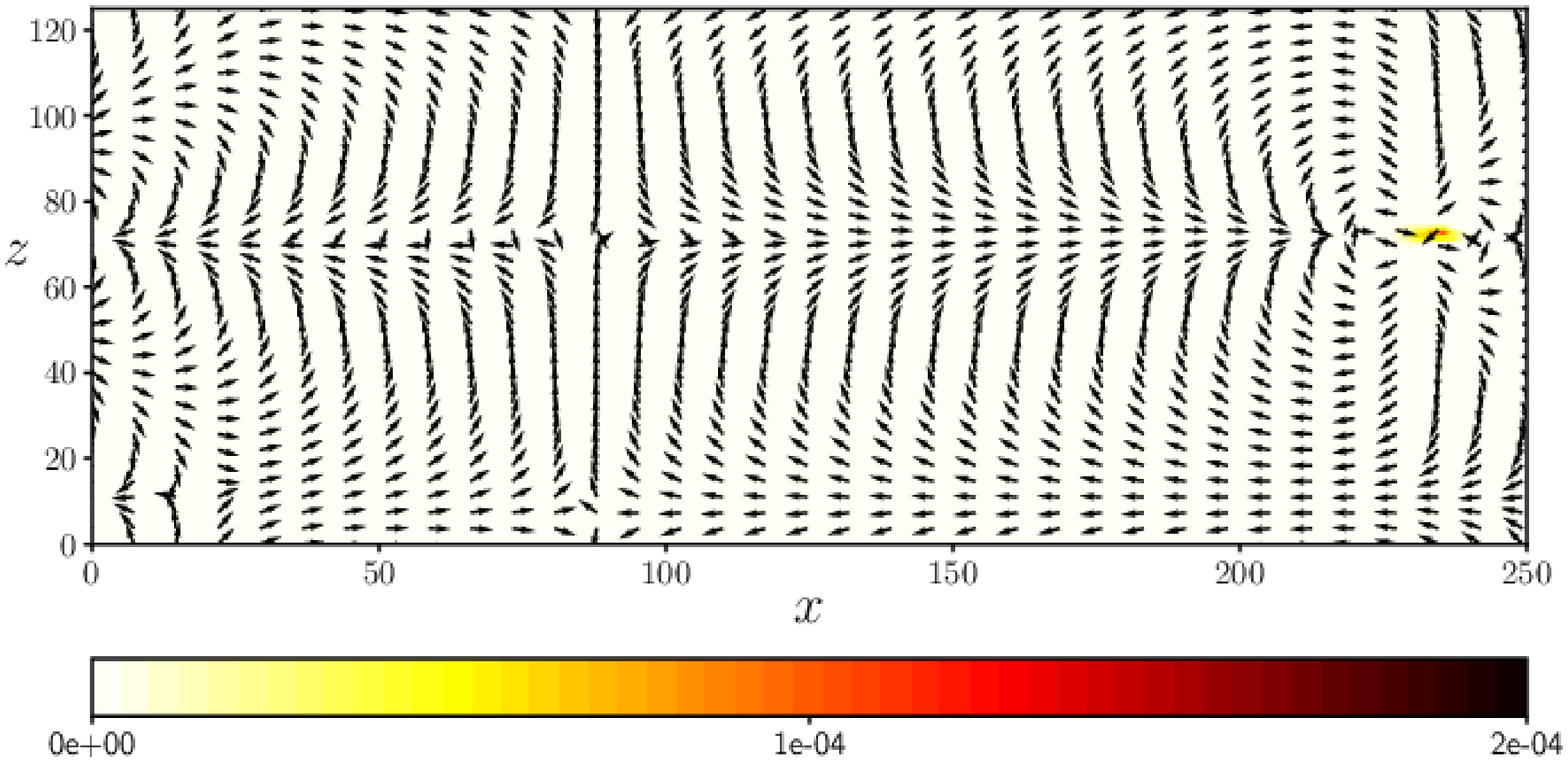}}
        \subfigure[$t = T$]{\includegraphics[width=.495\columnwidth]{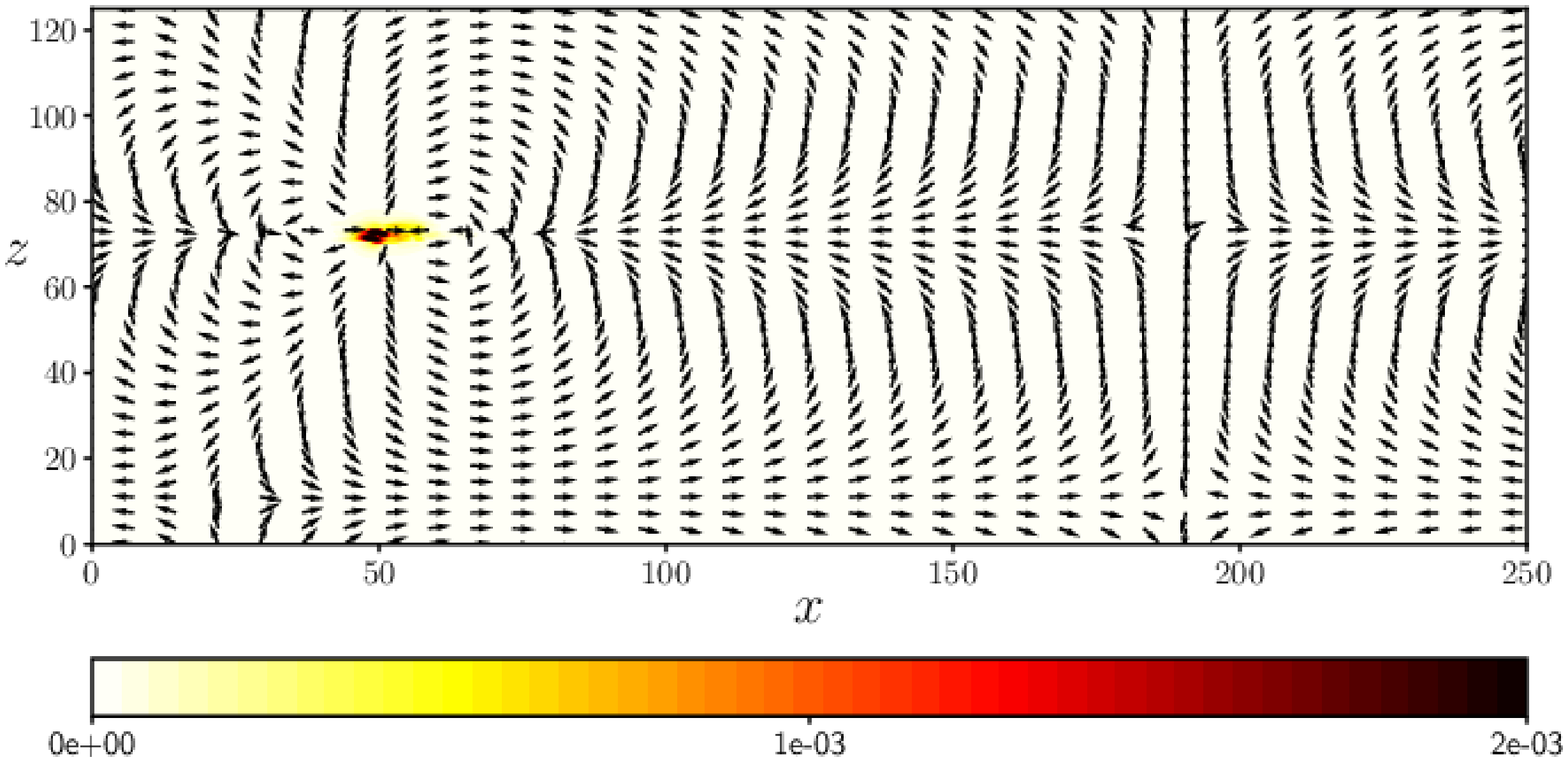}}
        \caption[]{Isocontours of the crossflow energy $E_{cf}$ with the normalized $y$-integrated large scale flow (vectors) for $Re = 1150$, $E_0 = 4.7 \times 10^{-8}$, $T = 100$: (a) initial optimal solution, (b) optimal solution at the target time.} 
        \label{2d_opt_sol_Re1150_E04p7e-8_T100}
\end{figure}

\begin{figure}
         \centering
         \subfigure[$k_x E_{uu}(k_x)$]{\includegraphics[width=.325\columnwidth]{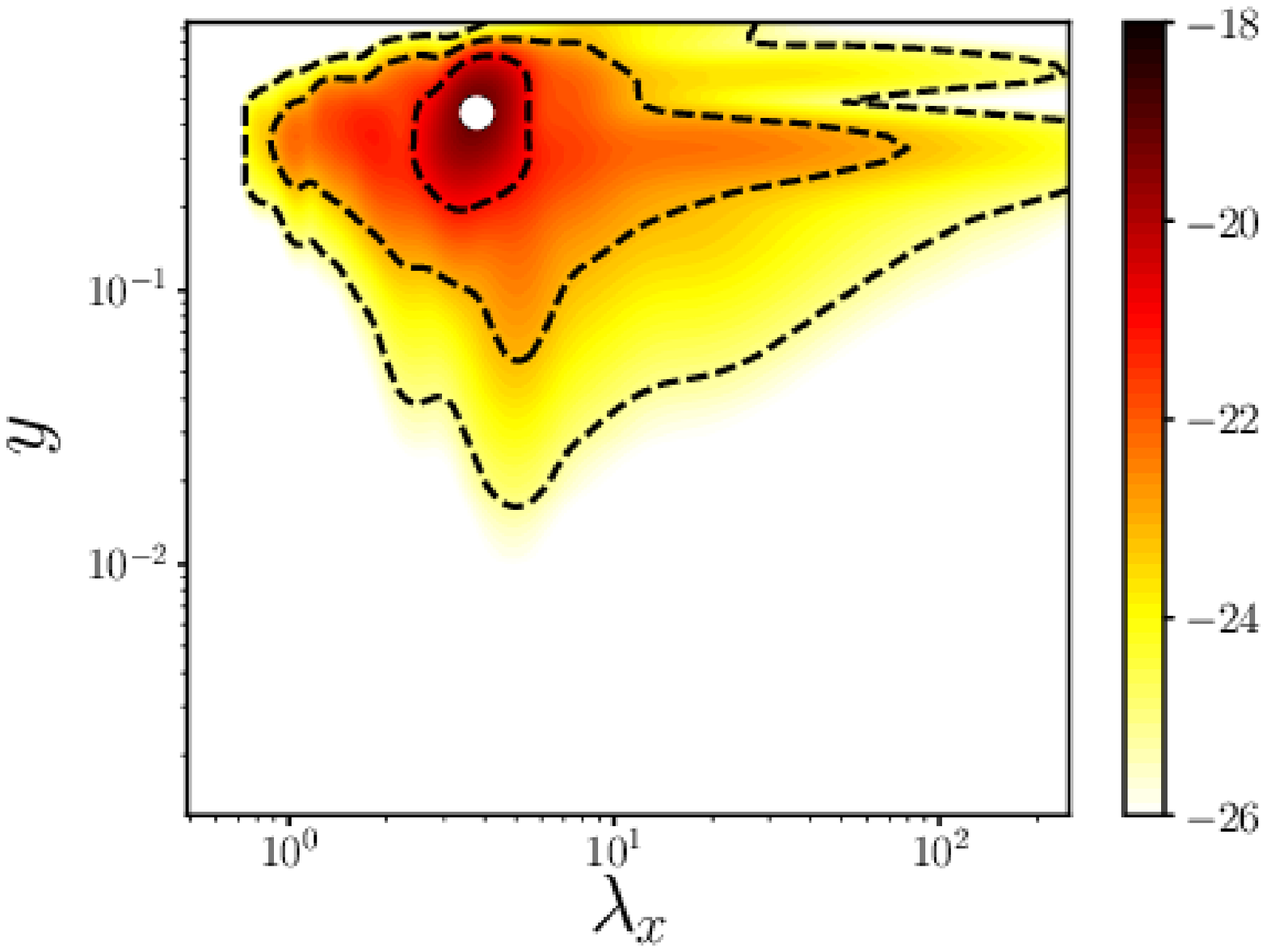}}
         \subfigure[$k_x E_{vv}(k_x)$]{\includegraphics[width=.325\columnwidth]{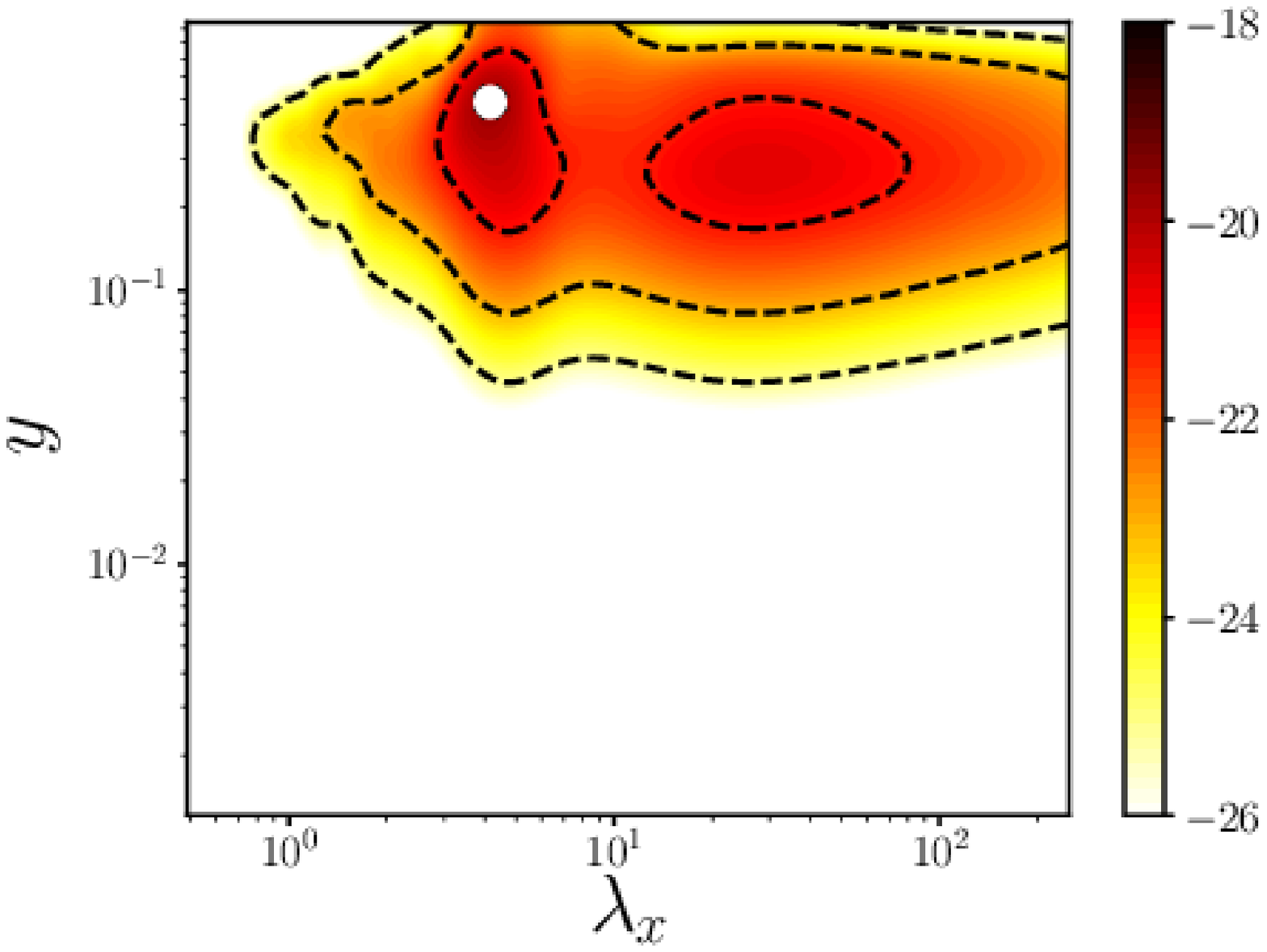}}
         \subfigure[$k_x E_{ww}(k_x)$]{\includegraphics[width=.325\columnwidth]{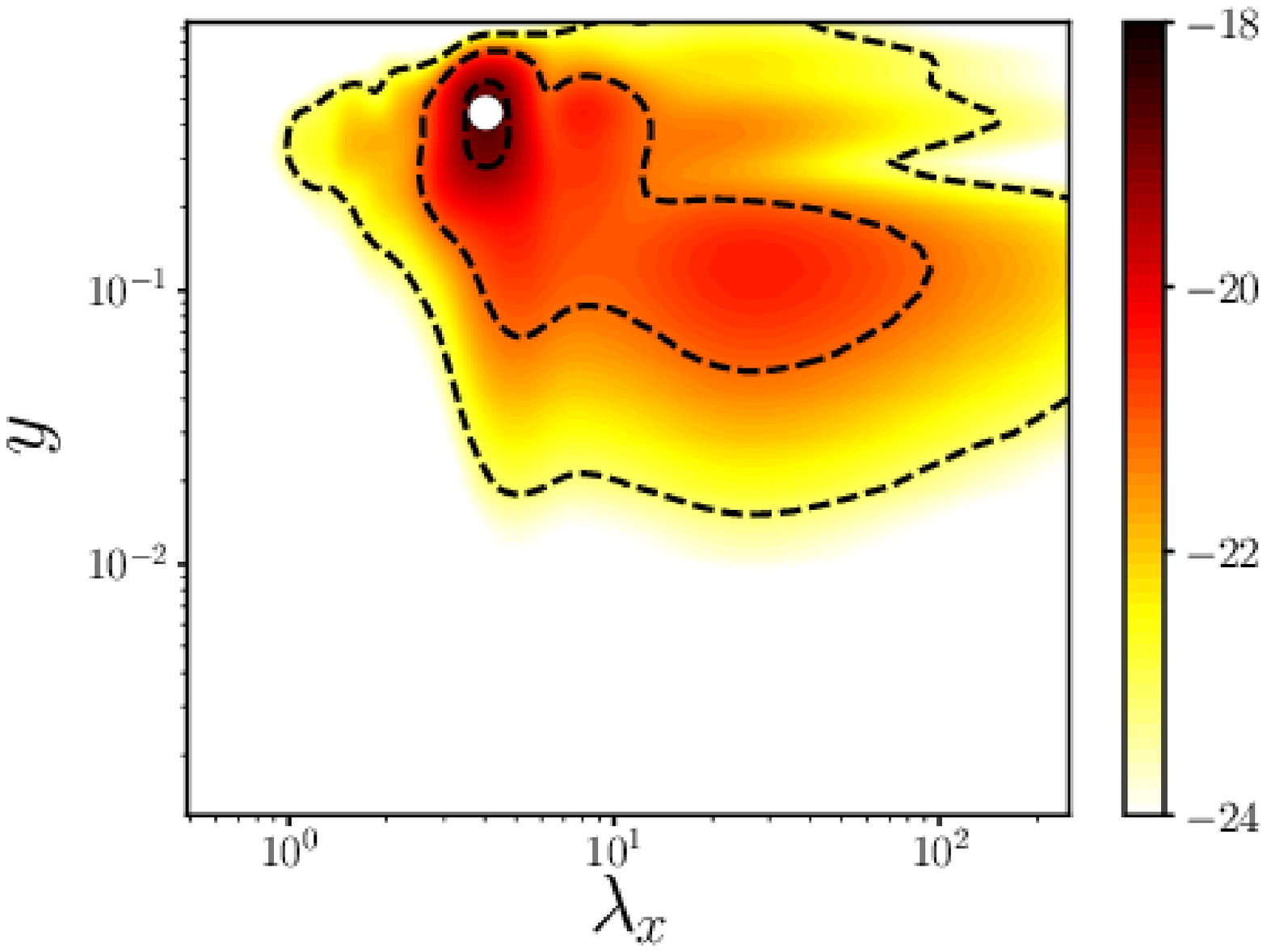}}
         \subfigure[$k_z E_{uu}(k_z)$]{\includegraphics[width=.325\columnwidth]{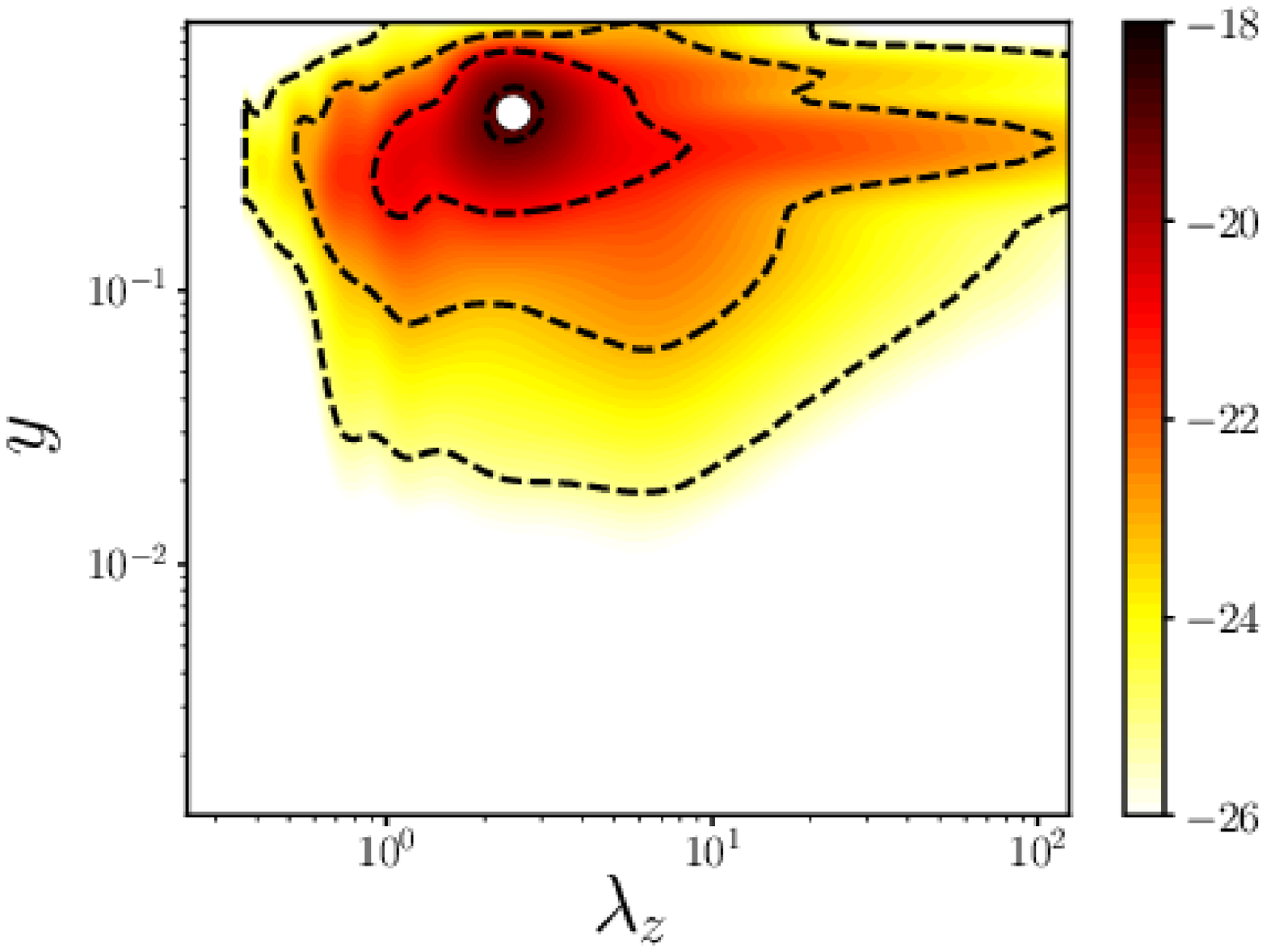}}
         \subfigure[$k_z E_{vv}(k_z)$]{\includegraphics[width=.325\columnwidth]{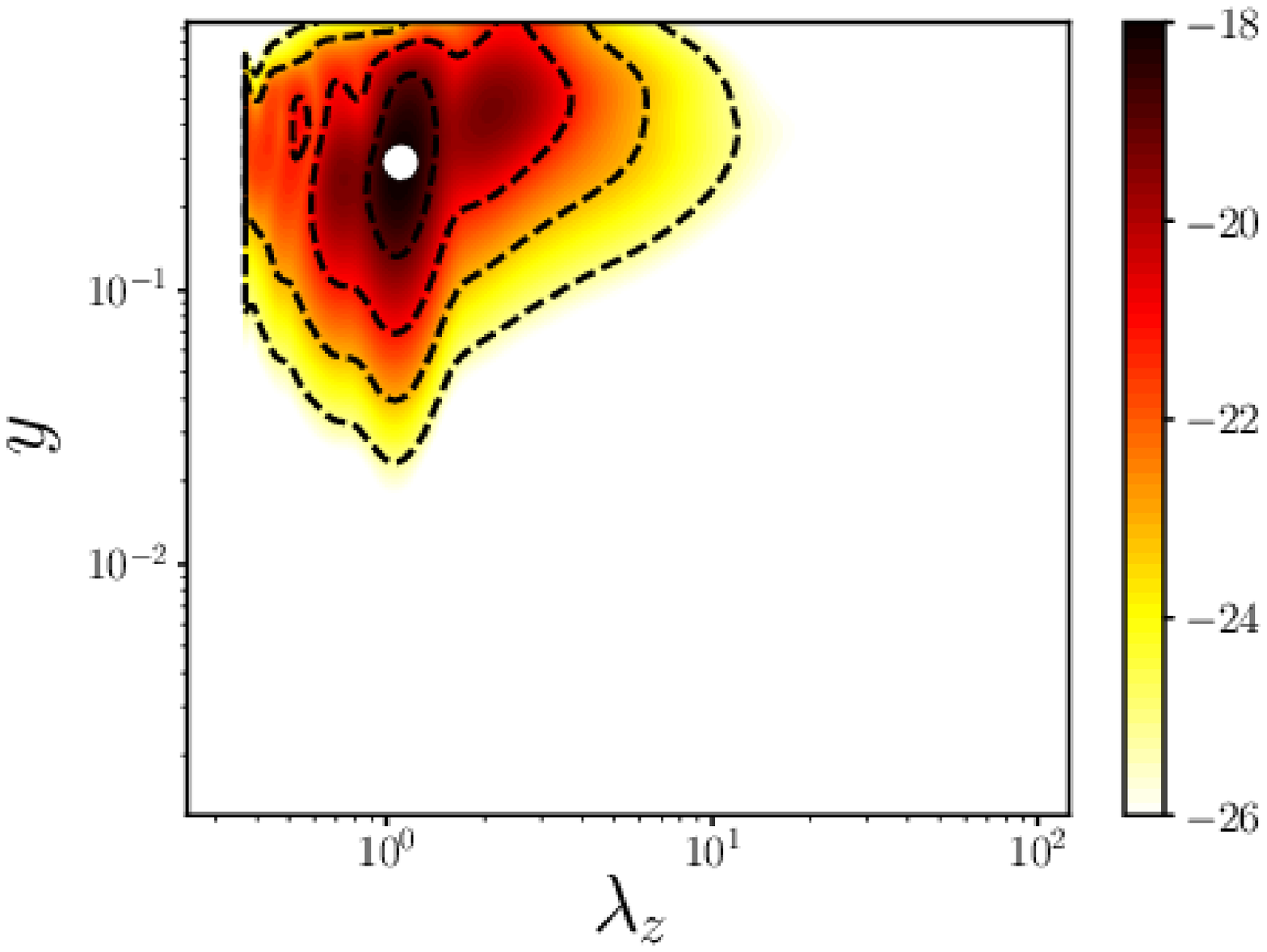}}
         \subfigure[$k_z E_{ww}(k_z)$]{\includegraphics[width=.325\columnwidth]{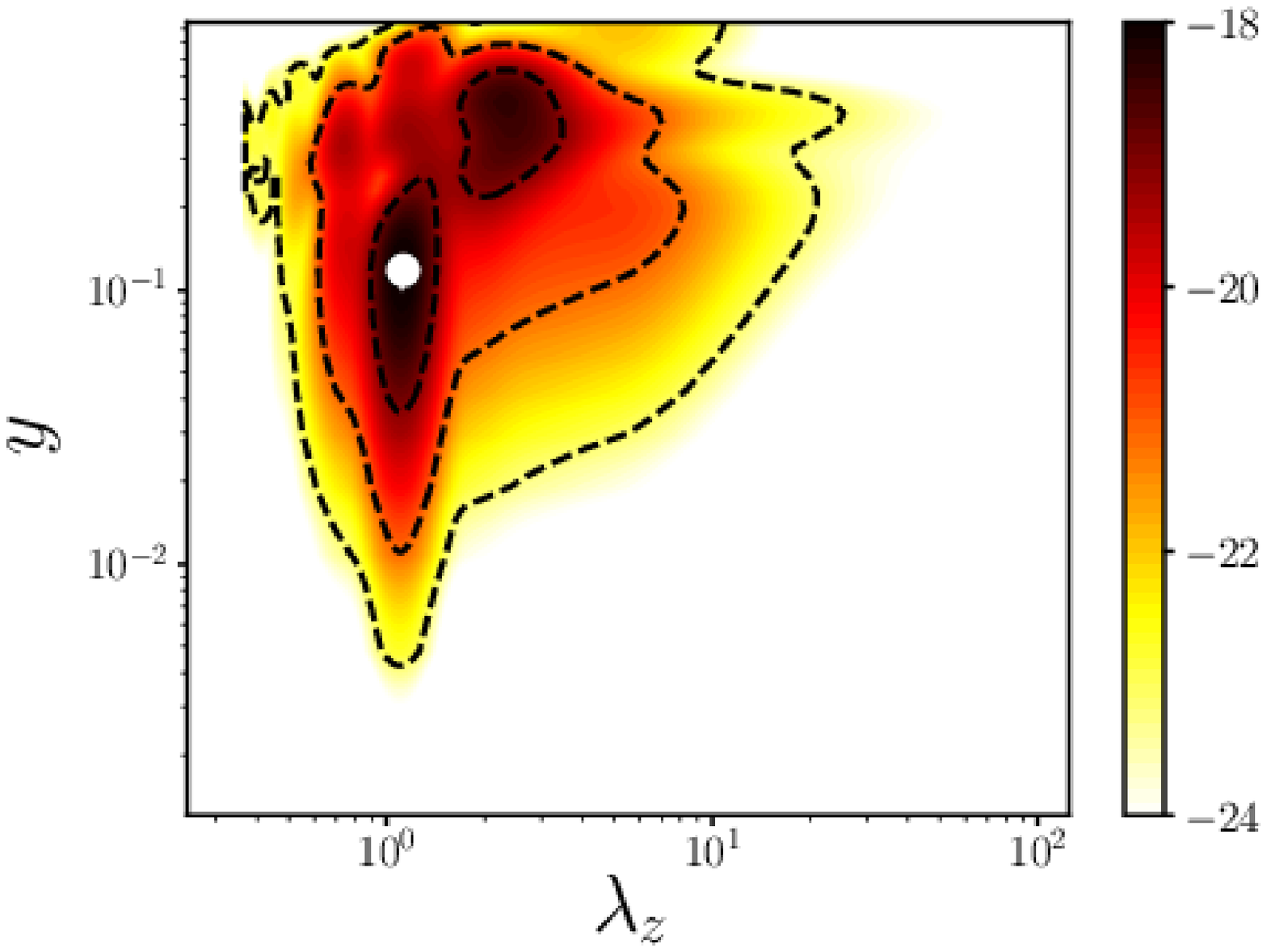}}
         \caption[]{Logarithm of the premultiplied spectral energy versus the wall-normal distance $y^+$ for the initial optimal solution at $Re = 1150$ and $T = 100$ for $E_0 = 1.1 \times 10^{-7}$ (coloured contours) and $E_0 = 4.7 \times 10^{-8}$ (black contours). The white dots indicate the energy peaks.}\label{energy_spectra_Re1150}
\end{figure}

\begin{figure}
         \centering
         \subfigure[$k_x E_{uu}(k_x)$]{\includegraphics[width=.325\columnwidth]{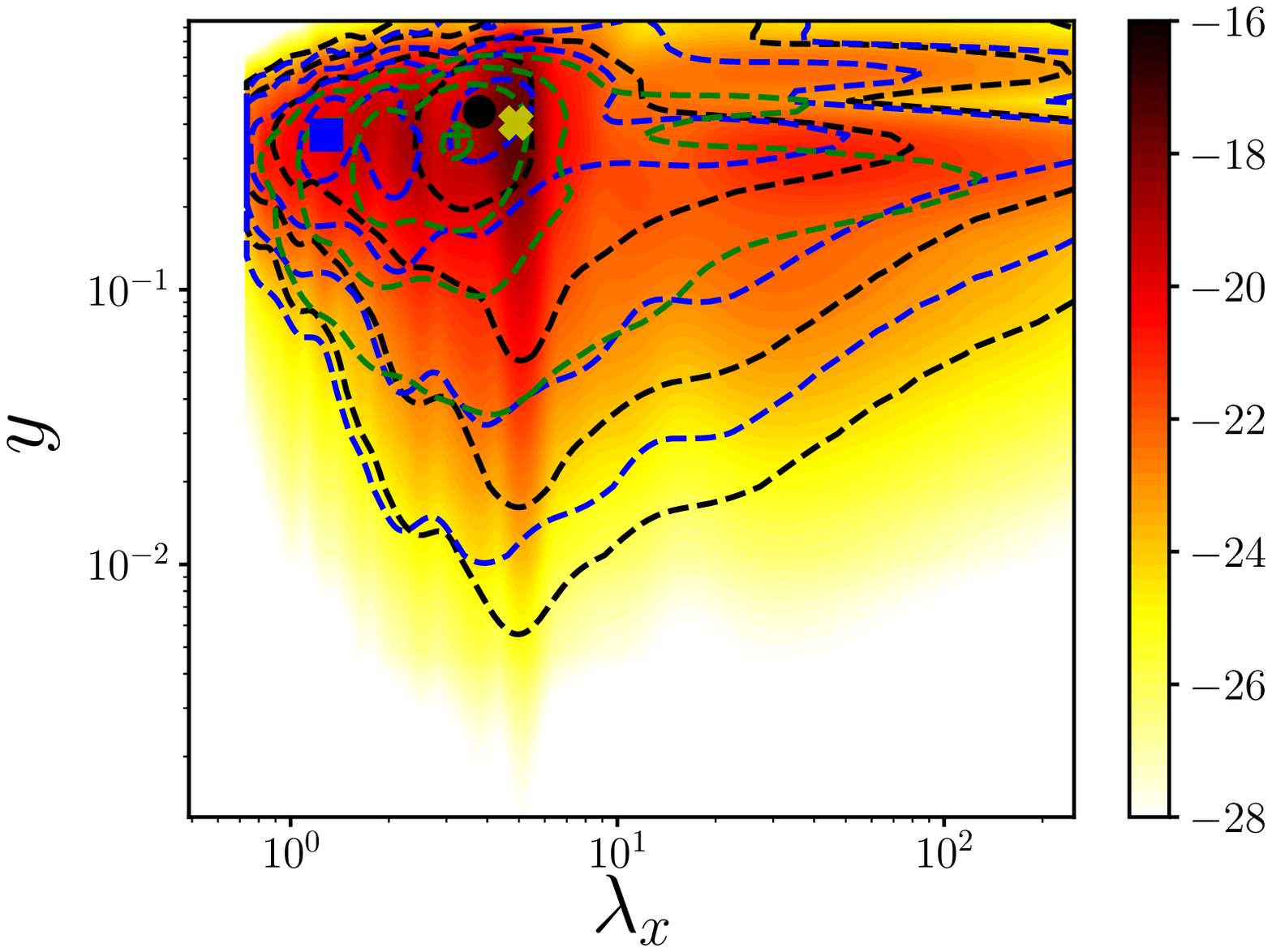}}
         \subfigure[$k_x E_{vv}(k_x)$]{\includegraphics[width=.325\columnwidth]{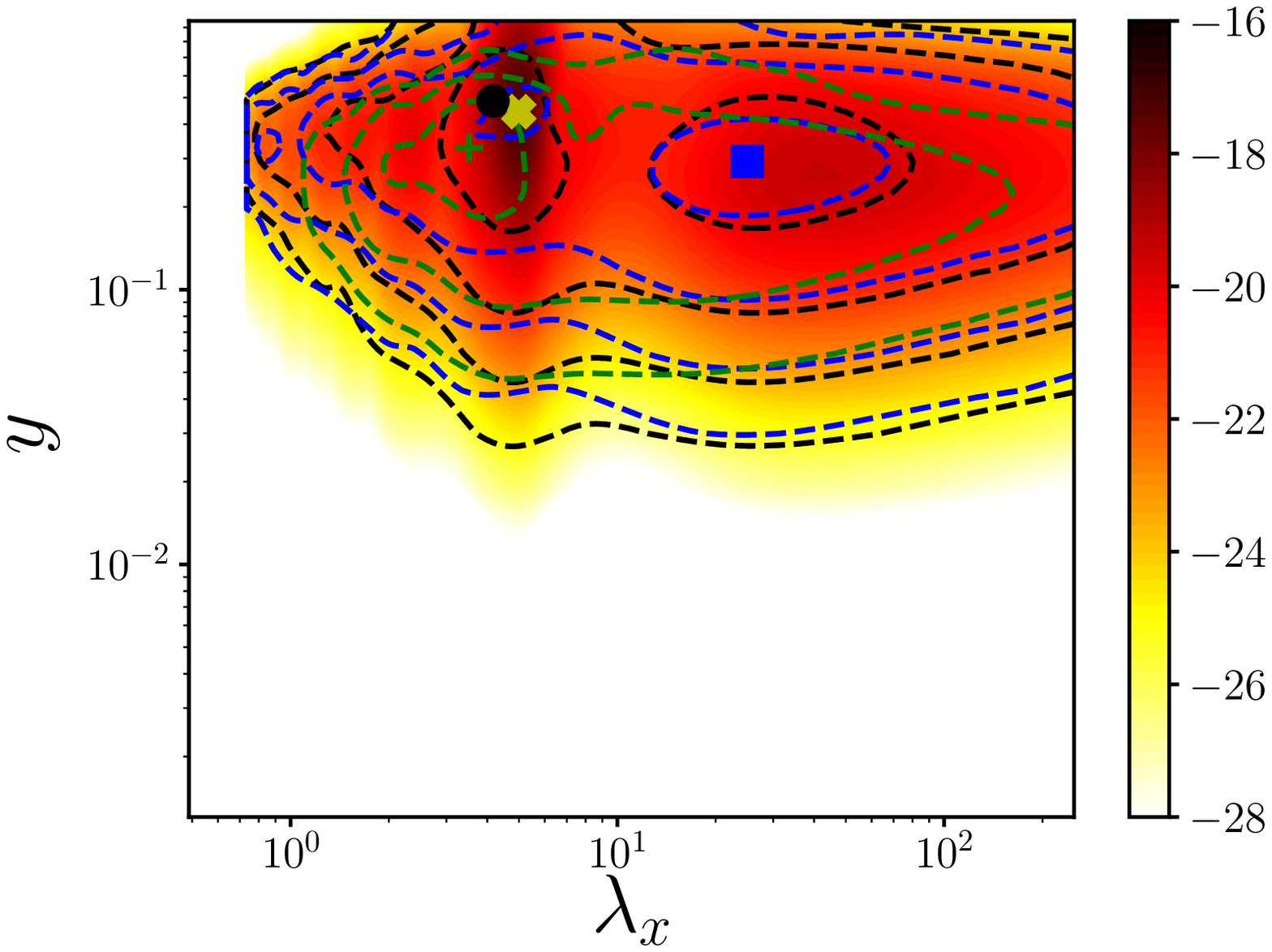}}
         \subfigure[$k_x E_{ww}(k_x)$]{\includegraphics[width=.325\columnwidth]{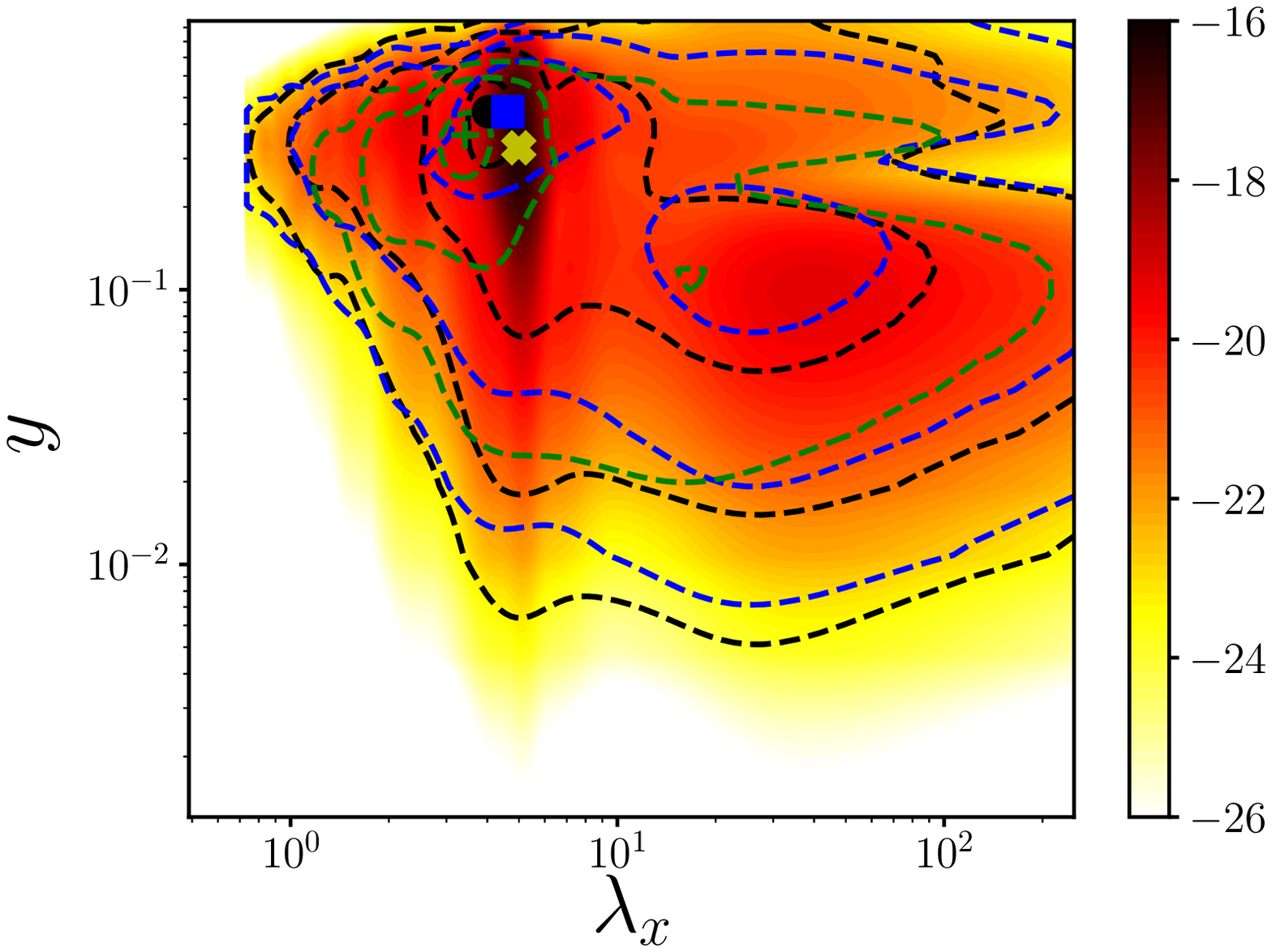}}
         \subfigure[$k_z E_{uu}(k_z)$]{\includegraphics[width=.325\columnwidth]{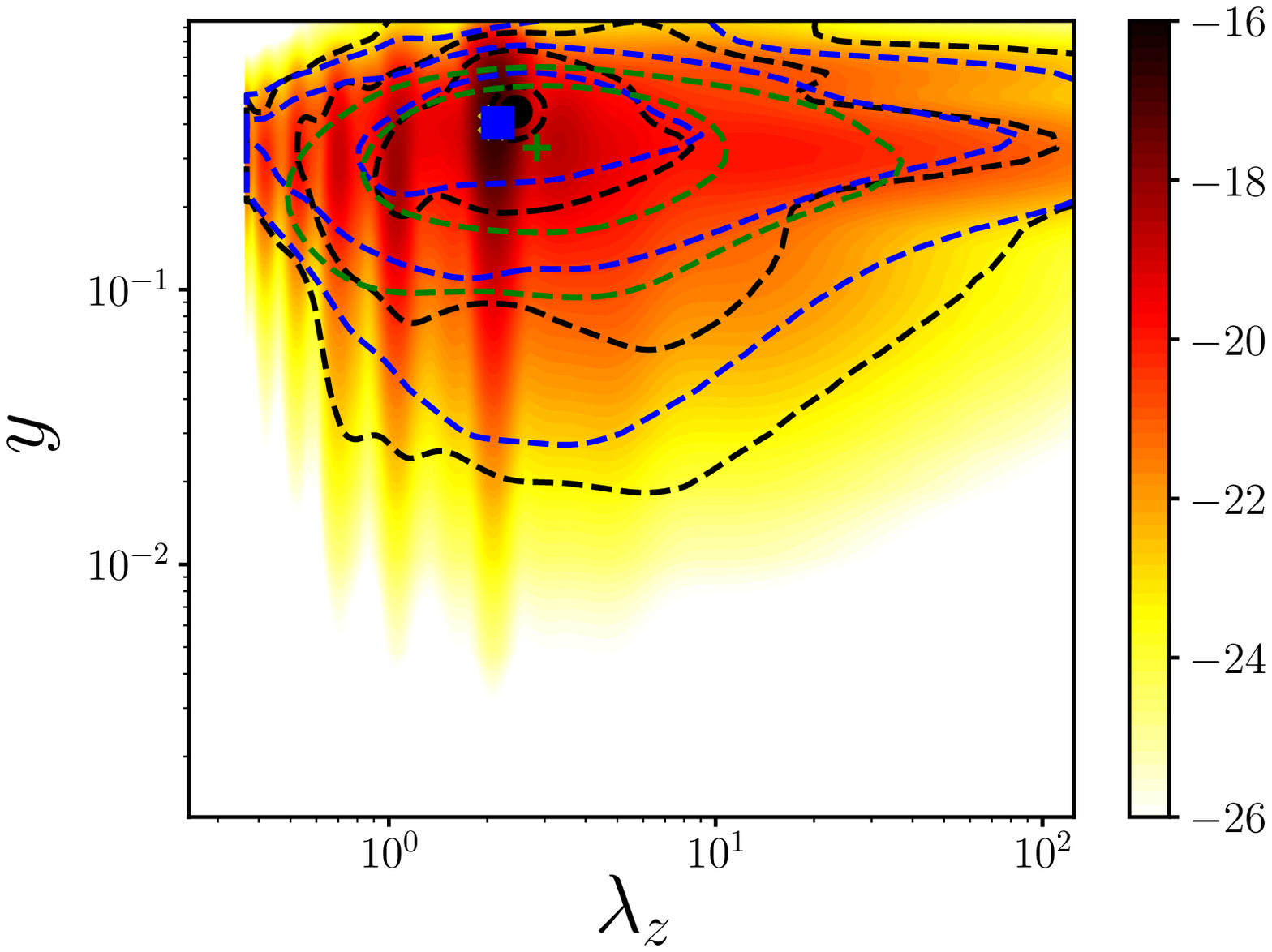}}
         \subfigure[$k_z E_{vv}(k_z)$]{\includegraphics[width=.325\columnwidth]{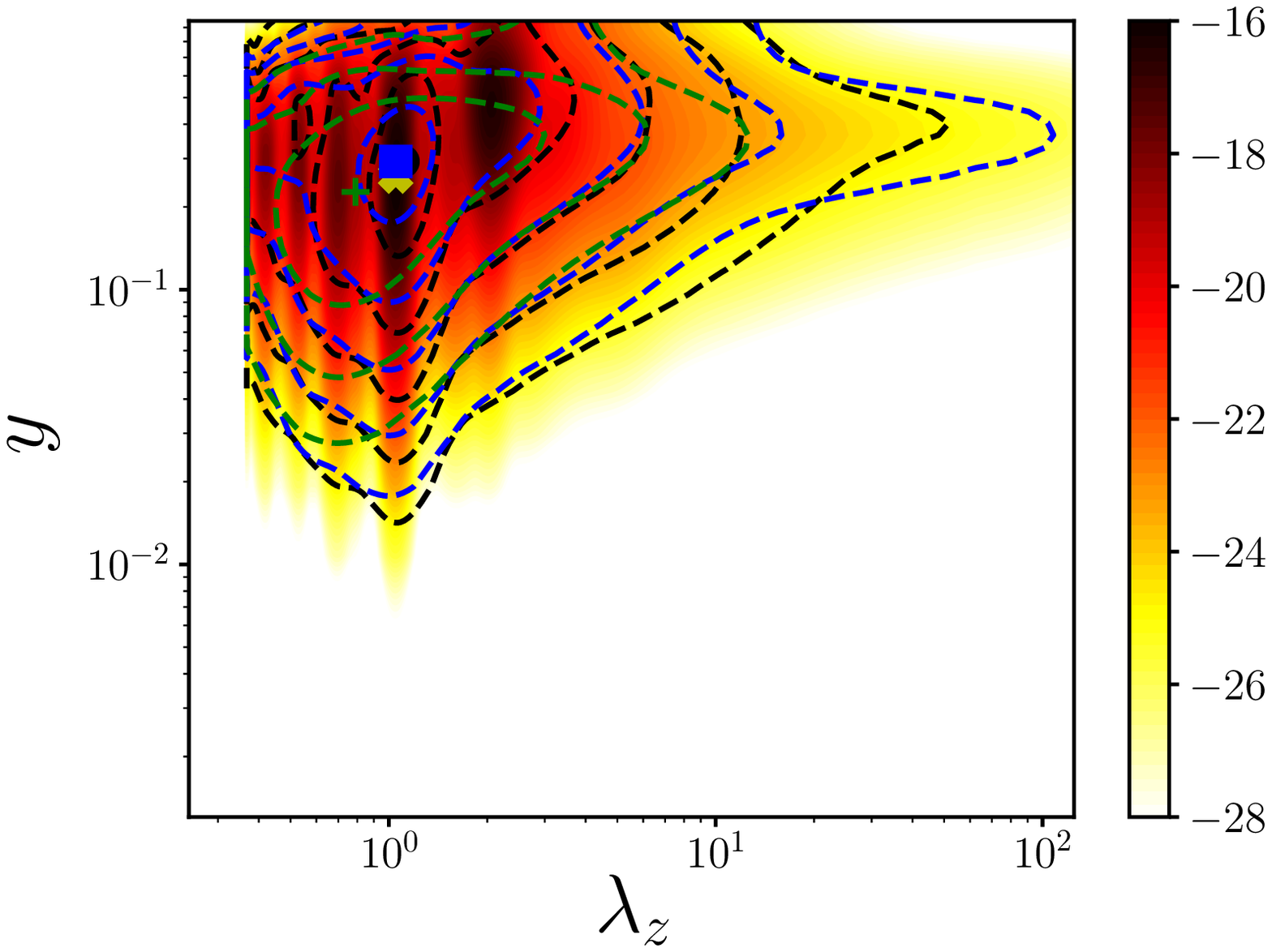}}
         \subfigure[$k_z E_{ww}(k_z)$]{\includegraphics[width=.325\columnwidth]{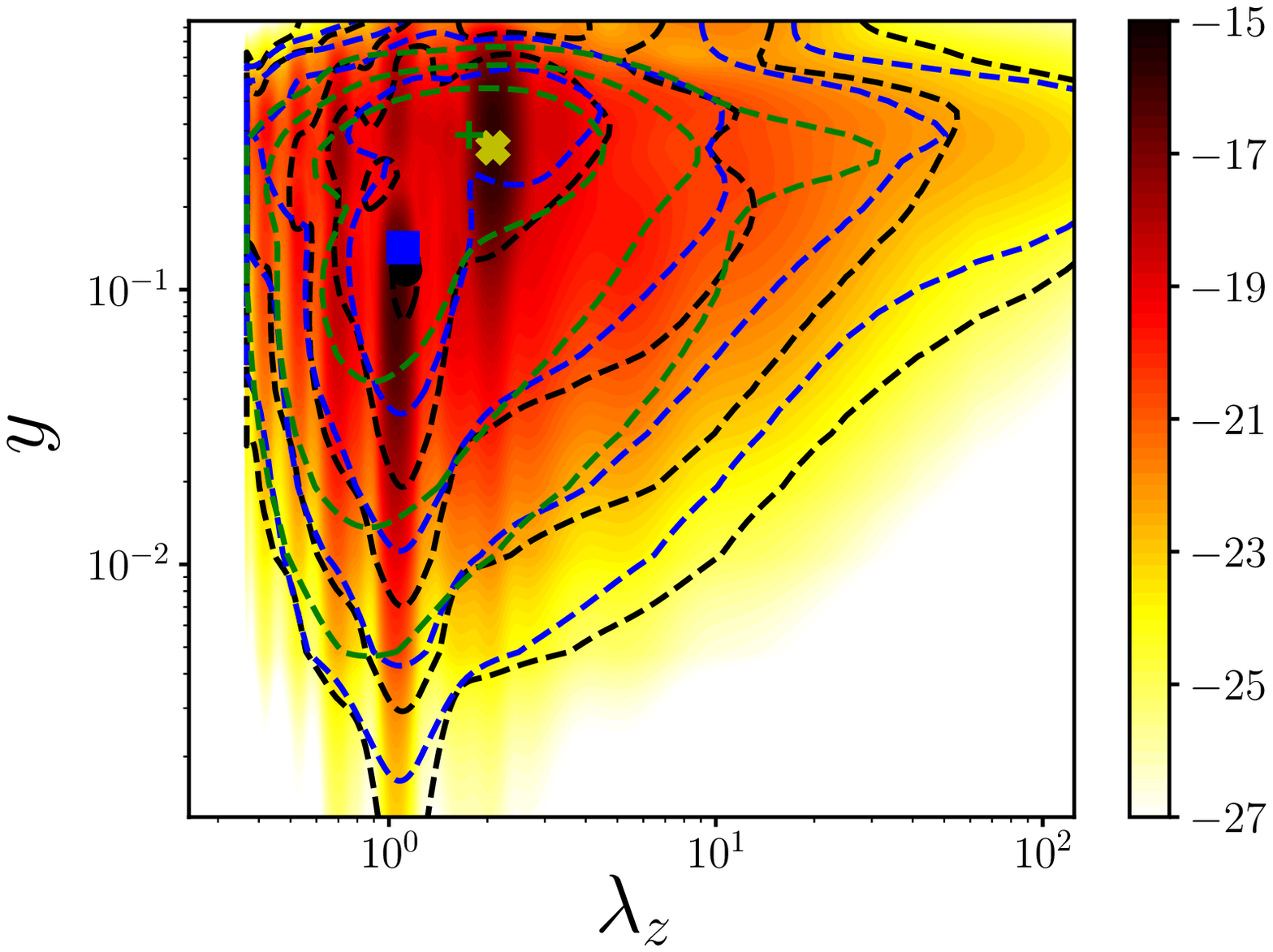}}
         \caption[]{Logarithm of the premultiplied spectral energy versus the wall-normal distance $y^+$ for the initial optimal solution at different Reynolds numbers at $T = 100$: $Re = 1000$ for $E_0 = 5.5 \times 10^{-7}$ (coloured contours), $Re = 1150$ for $E_0 = 4.7 \times 10^{-8}$ (black contours), $Re = 1250$ for $E_0 = 2.9 \times 10^{-8}$ (blue contours) and $Re = 1568$ for $E_0 = 5.8 \times 10^{-9}$ (green contours). The symbols indicate the peaks of the energy, also reported in table \ref{tab:kd}.}\label{energy_spectra_diffRe}
\end{figure}

\noindent The minimal seed for turbulent bands obtained  at $Re = 1150$ for $E_{0_{min}} = 4.7 \times 10^{-8}$ 
is reported in figure \ref{opt_sol_Re1150_E04p7e-8_T100} (a). Noticeably, it does not present significant differences compared to that at the same target time with a higher input energy, provided in figure \ref{opt_sol_Re1150_E01p1e-7_T100} (a). A three-dimensional visualization, provided in figure \ref{fig:3D-minRe1150} (b) shows the streamwise and spanwise alternance of finite-size streaks together with spanwise-inclined vortical structures closer to the wall.
As shown in figure \ref{opt_sol_Re1150_E04p7e-8_T100} (b-c), these finite-size streaks increase their streamwise length, and the initial wave packet turns at $t=25$ into four pairs of elongated streamwise streaks. At $t=50$, some wiggling of the streaks can be recovered in the center of the wave packet, which turns at target time into an arrow-shaped puff-like packet (figure \ref{opt_sol_Re1150_E04p7e-8_T100} (d)), as also shown in the three-dimensional view of figure \ref{fig:3D-minimalseed-T} (b).
Comparing figure \ref{opt_sol_Re1150_E01p1e-7_T100} (b) with figure \ref{opt_sol_Re1150_E04p7e-8_T100}(d), one can also notice that, at target time, the minimal perturbation differs from that computed for $E_0>{E_0}_{min}$, having completely lost symmetry along the $z$ axis. A clear overhang region is  present as well, and the structure in the $x-y$ plane recalls that experimentally observed in turbulent spots by \cite{lemoult2013}. Moreover, the large-scale flow, shown in figure \ref{2d_opt_sol_Re1150_E04p7e-8_T100}, maintains the previously observed quadrupolar structure, presenting large-scale vortices almost symmetric in the spanwise direction. The same large-scale quadrupolar structure surrounding the minimal-energy wakepacket is observed also for the other considered Reynolds numbers (not shown). Whereas, the small-scale minimal perturbations are found to considerably change with the Reynolds numbers, as shown in figure  \ref{fig:3D-minRe1150}. In particular, a further localization of the initial wavepacket is observed for increasing $Re$, leading to a minimal structure at $Re=1568$ which closely resembles that found for the plane Couette and the asymptotic suction boundary-layer flow  in small domains \citep{duguet2013,rabin2012triggering,CherubiniPoF2015}. Apart from the spatial localization, the main structures observed within the minimal-seed wavepackets are essentially the same, namely, finite-size streaks flanking upstream-tilted elongated vortices.  
As a result of the discrepancies recovered at initial time, the minimal seeds obtained at different $Re$ evolve differently in time, presenting an increasing localization for larger values of $Re$, as provided in figure \ref{fig:3D-minimalseed-T}. 
However, all of the wavepackets present an arrow-shaped structure, with the downstream region essentially characterized by low-speed streaks, and the core region filled with small-scale vortices together with some coherent streamwise streaks. Moreover, the large-scale quadrupolar structure observed at $t=0$ is maintained at target time for all the considered values of $Re$, as shown in figure \ref{2d_opt_sol_Re1150_E04p7e-8_T100} (b) for $Re=1150$. 
\noindent
\\
The premultiplied energy spectra of the minimal seed obtained at
$Re=1150$ %
are reported in figure \ref{energy_spectra_Re1150} by the coloured contours.
The streamwise perturbation energy peaks for $\lambda_z\approx 2.45$ and $\lambda_x\approx 3.78$, whereas the wall-normal and spanwise perturbation energy have largest amplitude at $\lambda_z\approx 1$, $\lambda_x\approx4$ with a secondary peak for $\lambda_z\approx2$ and $\lambda_x\approx 25$. 
The lowest of these wavelengths are consistent with those reported by \cite{lemoult2014} during the development of turbulent spots  at similar Reynolds numbers, and are linked to the finite size of the streaks.  Whereas, the largest wavelengths are directly linked to the streamwise and spanwise size of the wavepacket. 
Very similar spectra are found to characterize the nonlinear optimal perturbations at higher initial energy (solid contours in figure \ref{energy_spectra_Re1150}).
The premultiplied energy spectra of the minimal seeds obtained for the other considered values of $Re$
are reported in figure \ref{energy_spectra_diffRe} by the shaded contours ($Re=1000$) and the dashed lines (black for $Re=1150$, blue for $Re=1250$, green for $Re=1568$). 
 As a consequence of the increased spatial localization, the overall distribution of the energy spectra is displaced towards higher values of $\lambda_x,\lambda_z$ when $Re$ increases. This effect is coupled with a narrowing and displacement of the spectra towards higher values of $y$, which can be a consequence of an increased localization also in the wall-normal direction. The primary and secondary peak values, reported in table \ref{tab:kd} are also considerably influenced by the Reynolds number.

\begin{table}
\begin{center}
\begin{tabular}{|m{0.7cm}|m{1.8cm}|m{1.8cm}|m{1.8cm}|m{1.8cm}|m{1.8cm}|m{1.8cm}|}
  $Re$ & $(\lambda_x)_u$ & $(\lambda_z)_u$ & $(\lambda_x)_v$ & $(\lambda_z)_v$ & $(\lambda_x)_w$ & $(\lambda_z)_w$\\ [0.5ex] 
  $1000$ & $4.897 \ (2.390)$ & $2.117 \ (3.307)$ & $4.995 \ (38.45)$ & $1.049 \ (2.198)$ & $4.995 \ (36.34)$ & $2.080 \ (1.042)$\\
  $1150$ & $3.784 \ (1.618)$ & $2.449 \ (1.093)$ & $4.163 \ (25.56)$ & $1.105 \ (2.049)$ & $4.028 \ (25.56)$ & $1.125 \ (2.218)$\\
  $1250$ & $1.287 \ (2.236)$ & $2.153 \ (2.114)$ & $24.98 \ (4.991)$ & $1.049 \ (0.508)$ & $4.625 \ (24.16)$ & $1.105 \ (2.639)$\\
  $1568$ & $3.244 \ (1.916)$ & $2.838 \ (1.217)$ & $3.518 \ (18.75)$ & $0.790 \ (1.992)$ & $3.6421 \ (20.12)$ & $1.759 \ (0.958)$\\
  
\end{tabular}
\caption{Streamwise and spanwise wavelengths $\lambda_{x,z}$ associated with the primary and secondary (in brackets) peaks of the premultiplied energy spectra of $u,v,w$, shown in figure (\ref{energy_spectra_diffRe}) for different Reynolds numbers.}\label{tab:kd}
\end{center}
\end{table}

\subsection{Minimal seed evolution in time}\label{sec:evolution}
\begin{figure}
         \centering
       {\includegraphics[width=.5\columnwidth]{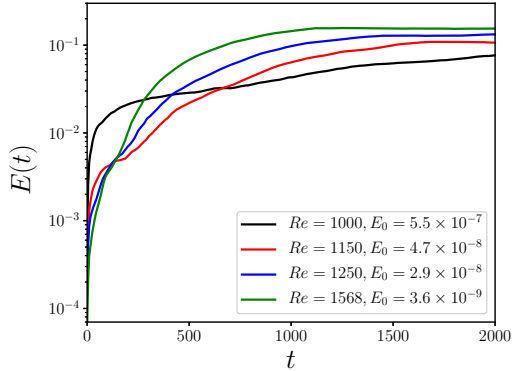}}
         \caption[]{Time evolution of the kinetic energy 
         for  the minimal seeds obtained for the  different considered Reynolds numbers.} 
         \label{kineticenergy_T100_comp}
\end{figure}

In this section, we analyse the time evolution of the minimal seeds towards the turbulent bands. 
\noindent In figure \ref{kineticenergy_T100_comp} (a), the time evolution of the kinetic energy obtained from direct numerical simulations initialised with the minimal seeds, is reported. In all cases, the kinetic energy strongly increases in time until saturating towards a statistically constant value. For the lowest considered Reynolds numbers,  we observe a rapid initial increase of the kinetic energy, followed by a slow phase of saturation of the energy. Whereas, for larger $Re$,  the initial growth is slower and leads to lower values of the kinetic energy at small time. One can notice once again that at $Re=1000$ the flow appears to behave rather differently from what observed at larger Reynolds numbers. 
 However, the minimal seeds at larger $Re$, despite having lower initial energy, tend towards higher values of the kinetic energy at large times, suggesting that for larger Reynolds numbers, turbulence eventually occupies a larger portion of the domain. 

\begin{figure}
        \centering
        \subfigure[$t = 200$]{\includegraphics[width=.325\columnwidth]{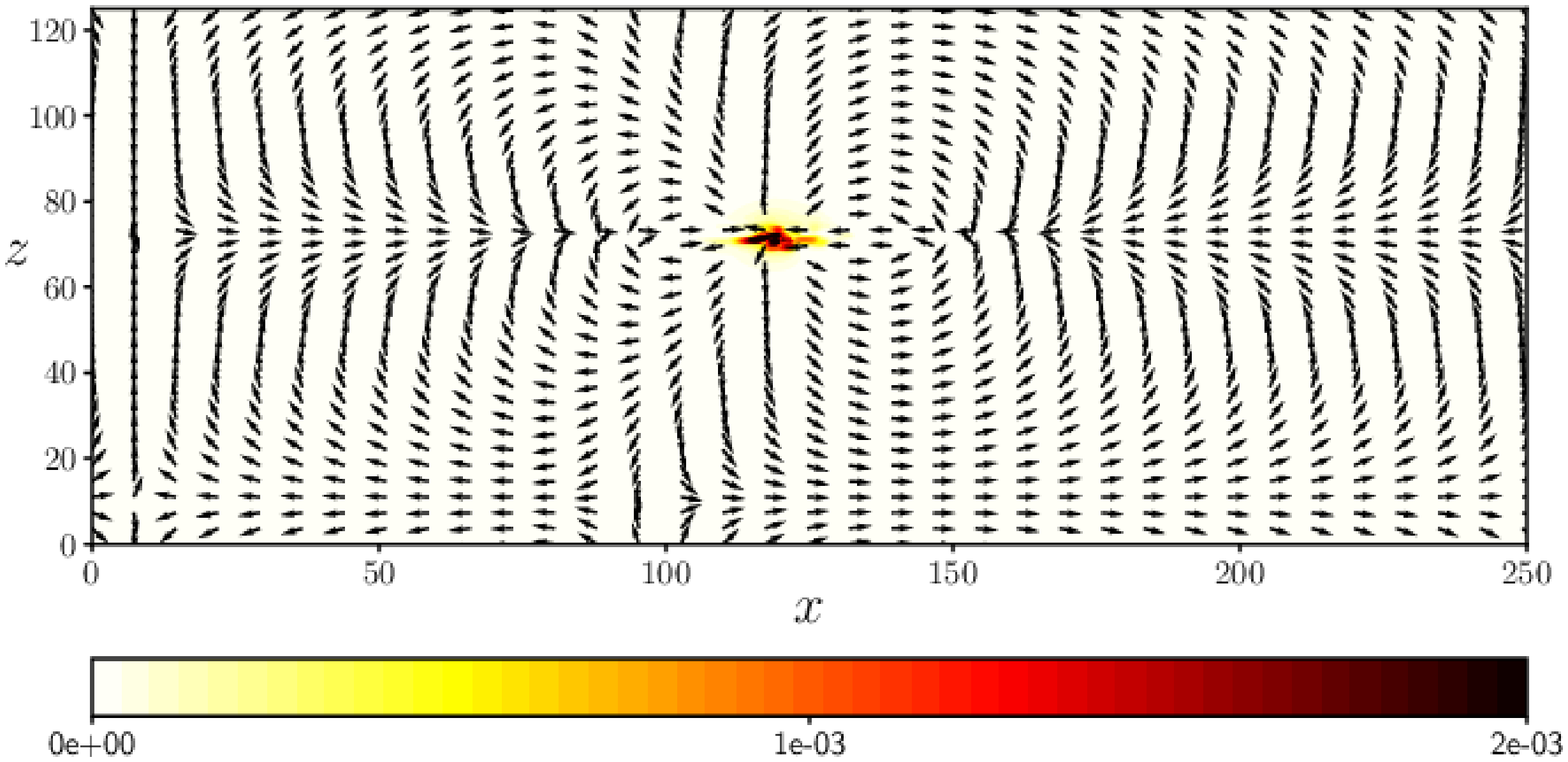}}
        \subfigure[$t = 500$]{\includegraphics[width=.325\columnwidth]{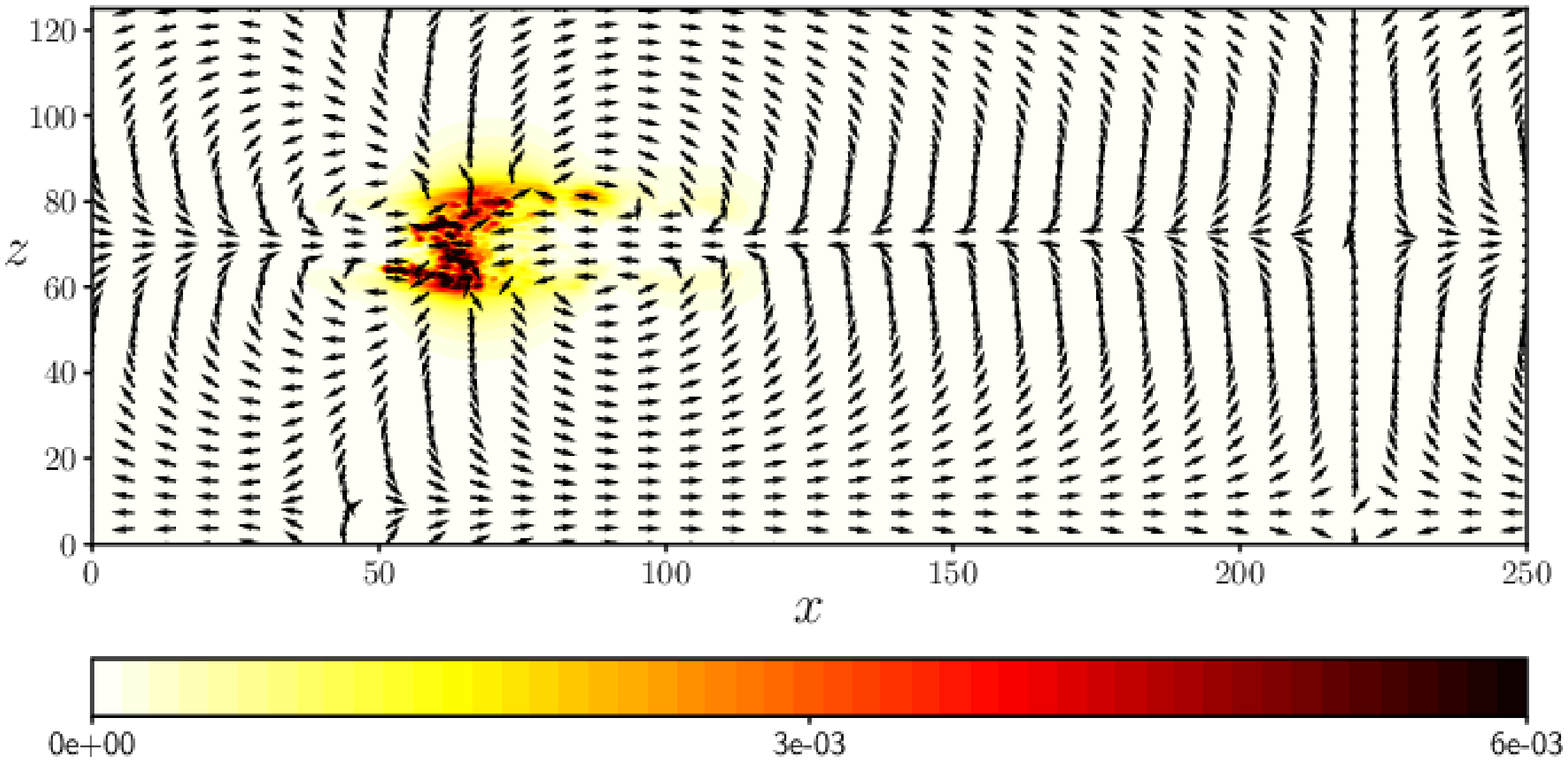}}
        \subfigure[$t = 900$]{\includegraphics[width=.325\columnwidth]{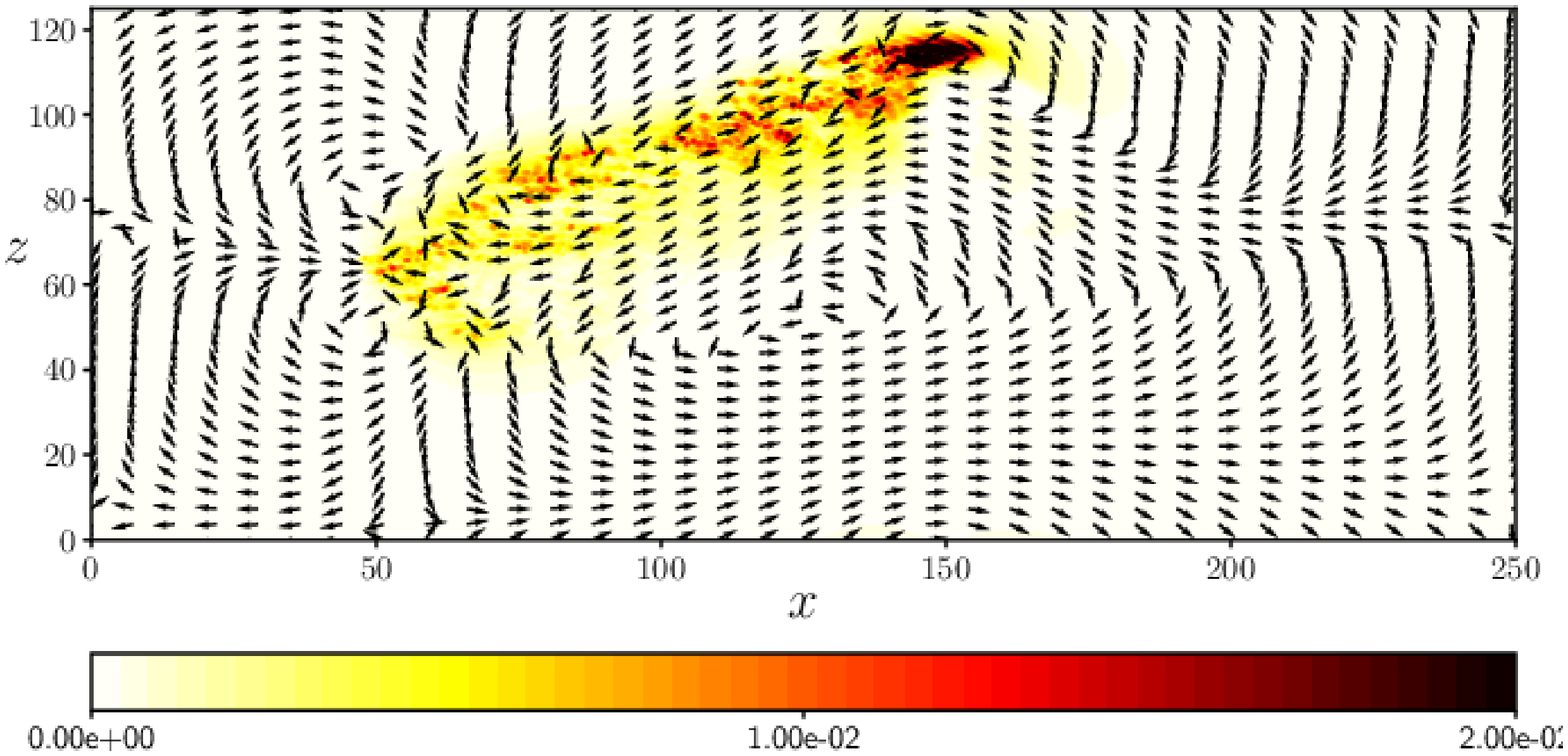}}
        \subfigure[$t = 1500$]{\includegraphics[width=.325\columnwidth]{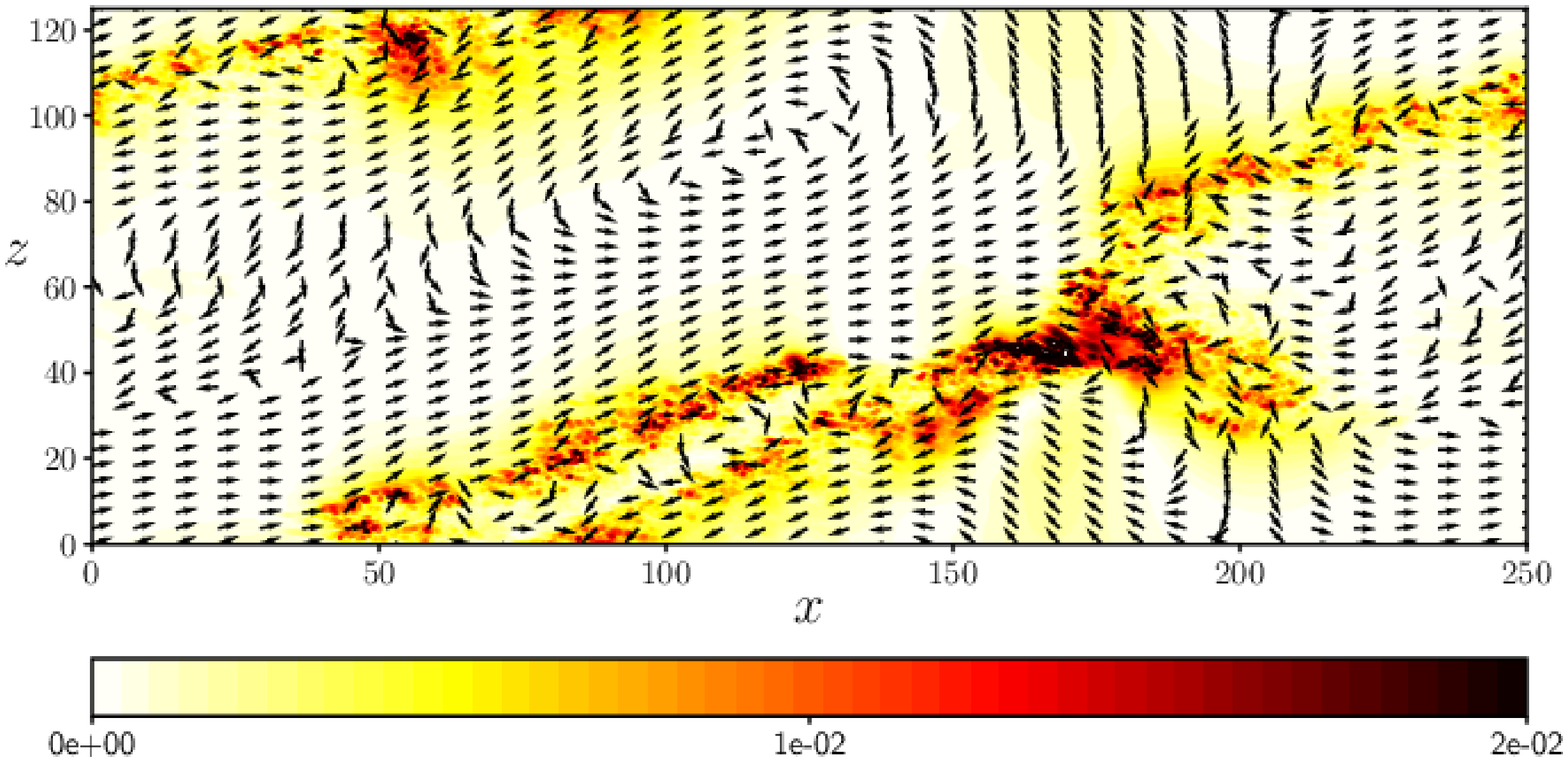}}
        \subfigure[$t = 2500$]{\includegraphics[width=.325\columnwidth]{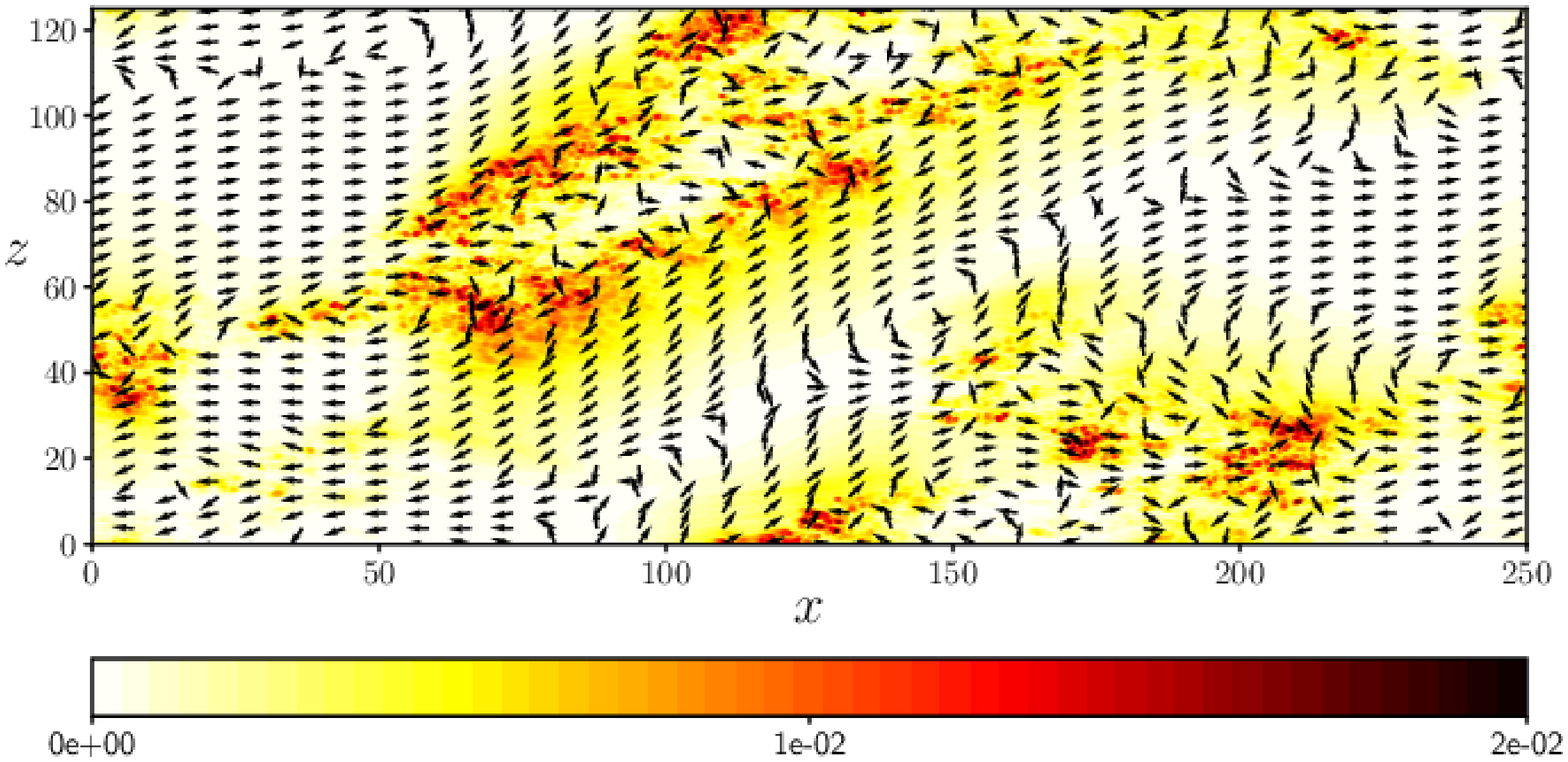}}
        \subfigure[$t = 4000$]{\includegraphics[width=.325\columnwidth]{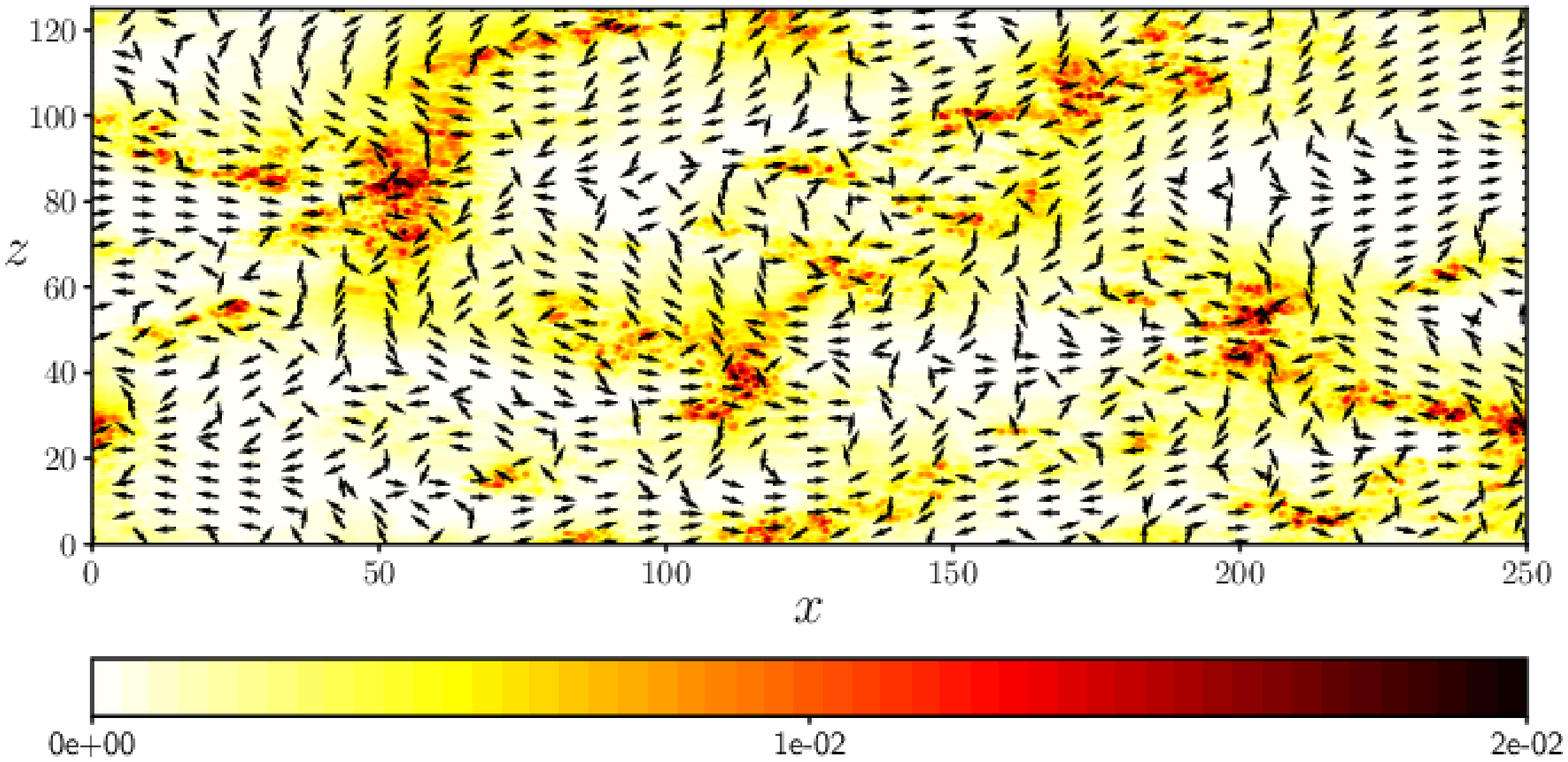}}
        \caption[]{Isocontours of the crossflow energy $E_{cf}$ together with the normalized $y$-integrated large scale flow (vectors) for several instantaneous fields ($Re = 1150$, $E_0 = 4.7 \times 10^{-8}$, $T = 100$).}
        \label{evolution_E04p7e-8} 
\end{figure}

\begin{figure}
        \centering
        \subfigure[$t = 150$]{\includegraphics[width=.325\columnwidth]{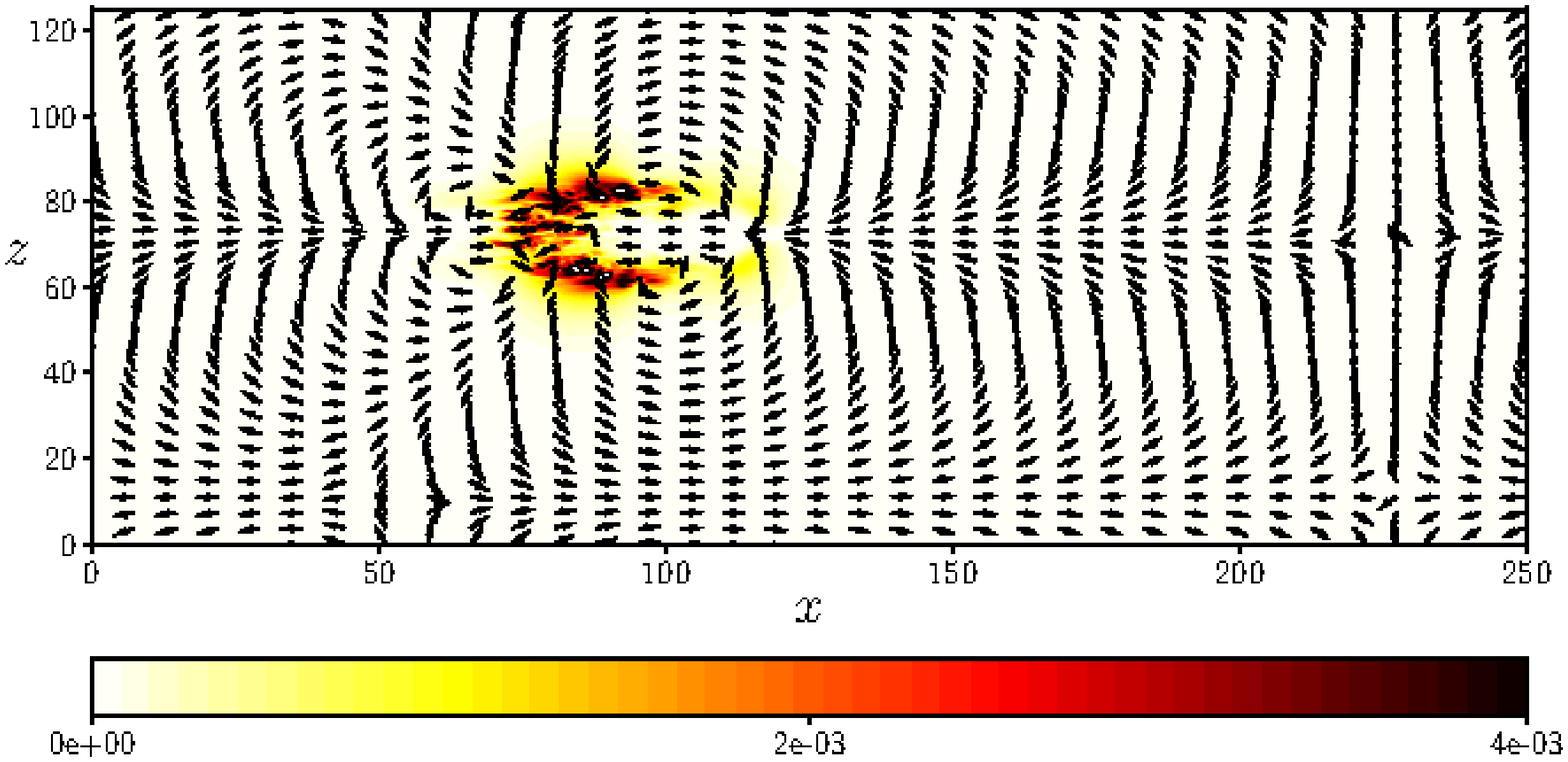}}
        \subfigure[$t = 500$]{\includegraphics[width=.325\columnwidth]{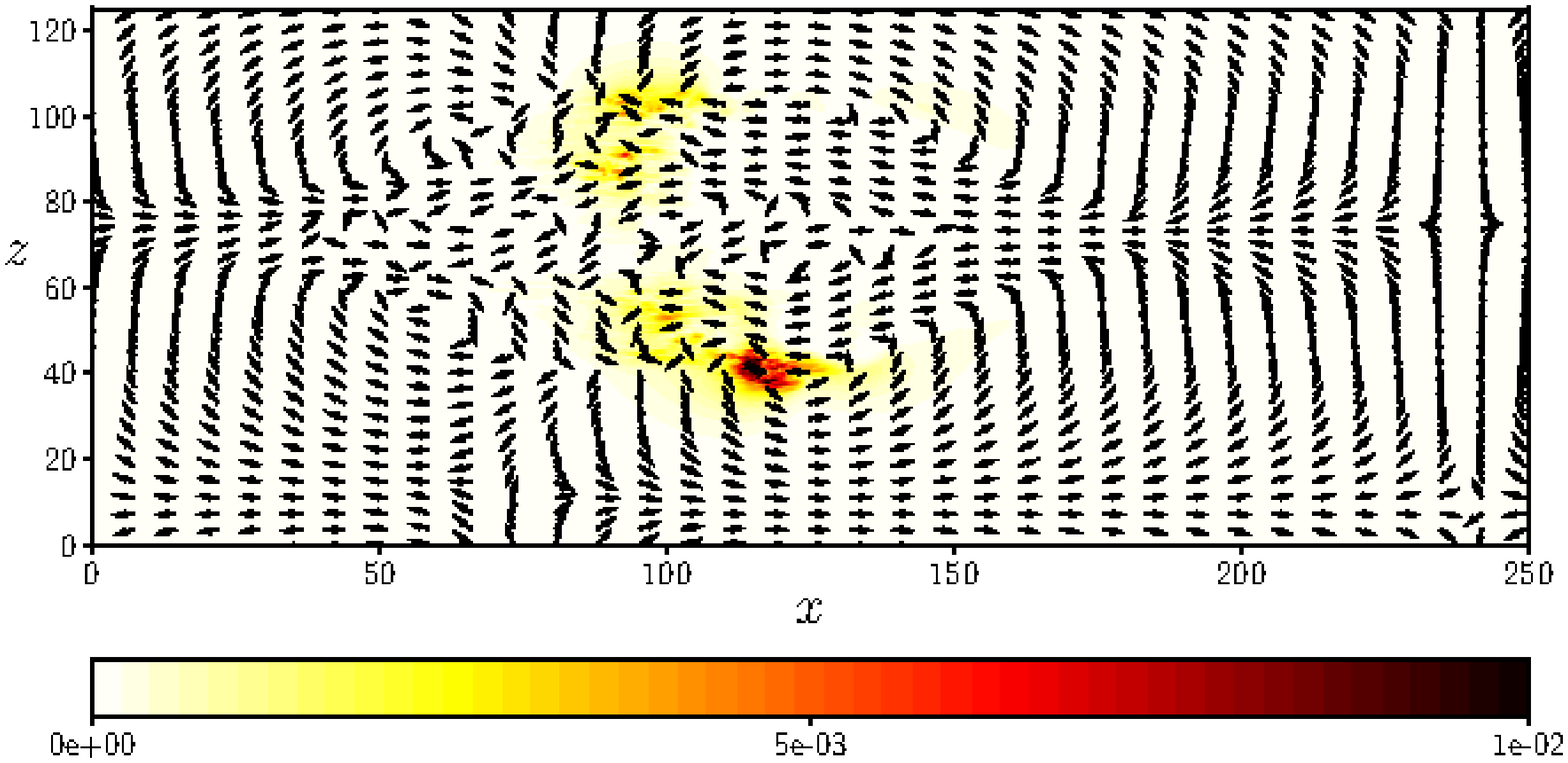}}
        \subfigure[$t = 900$]{\includegraphics[width=.325\columnwidth]{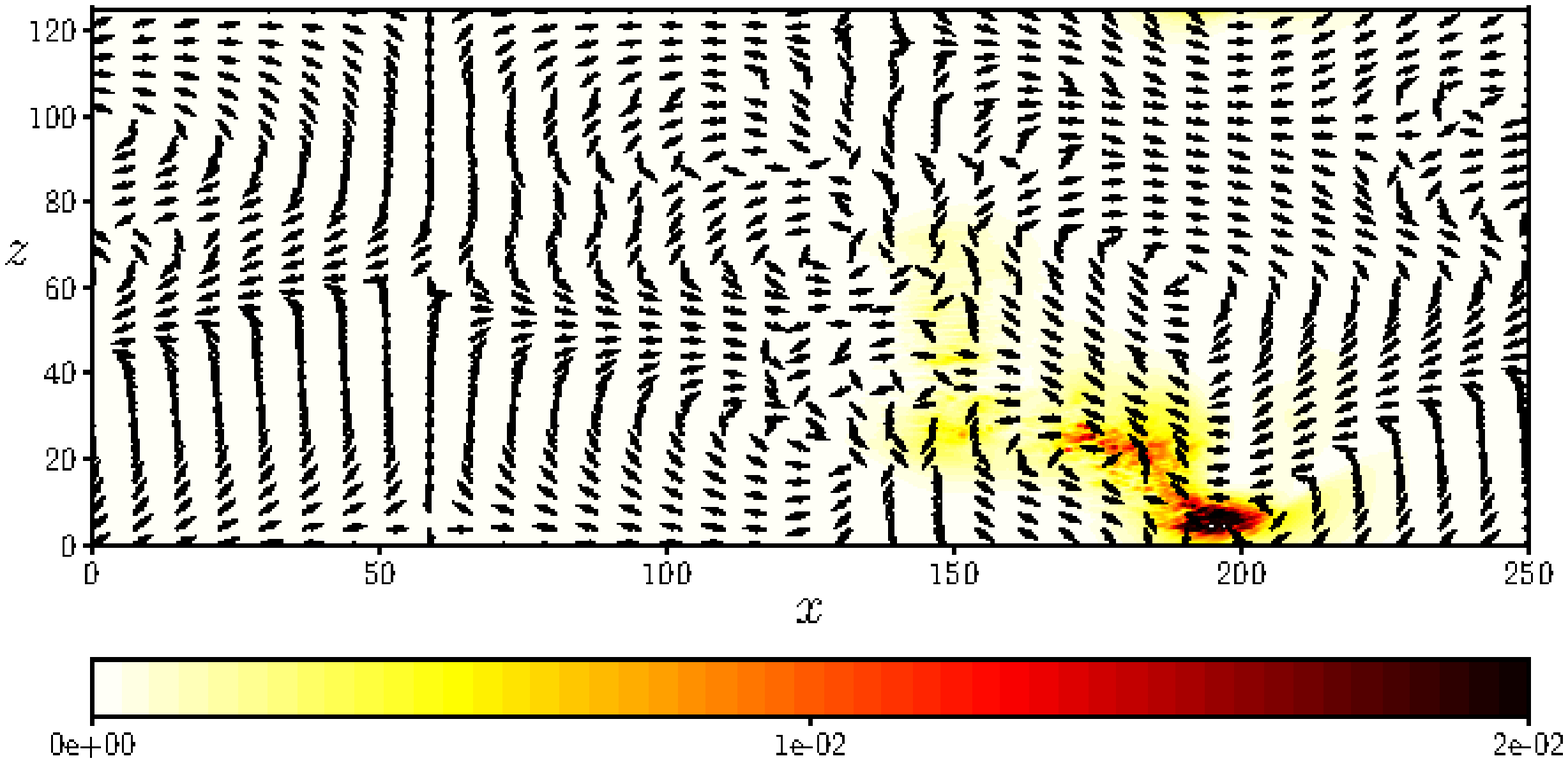}}
        \subfigure[$t = 1500$]{\includegraphics[width=.325\columnwidth]{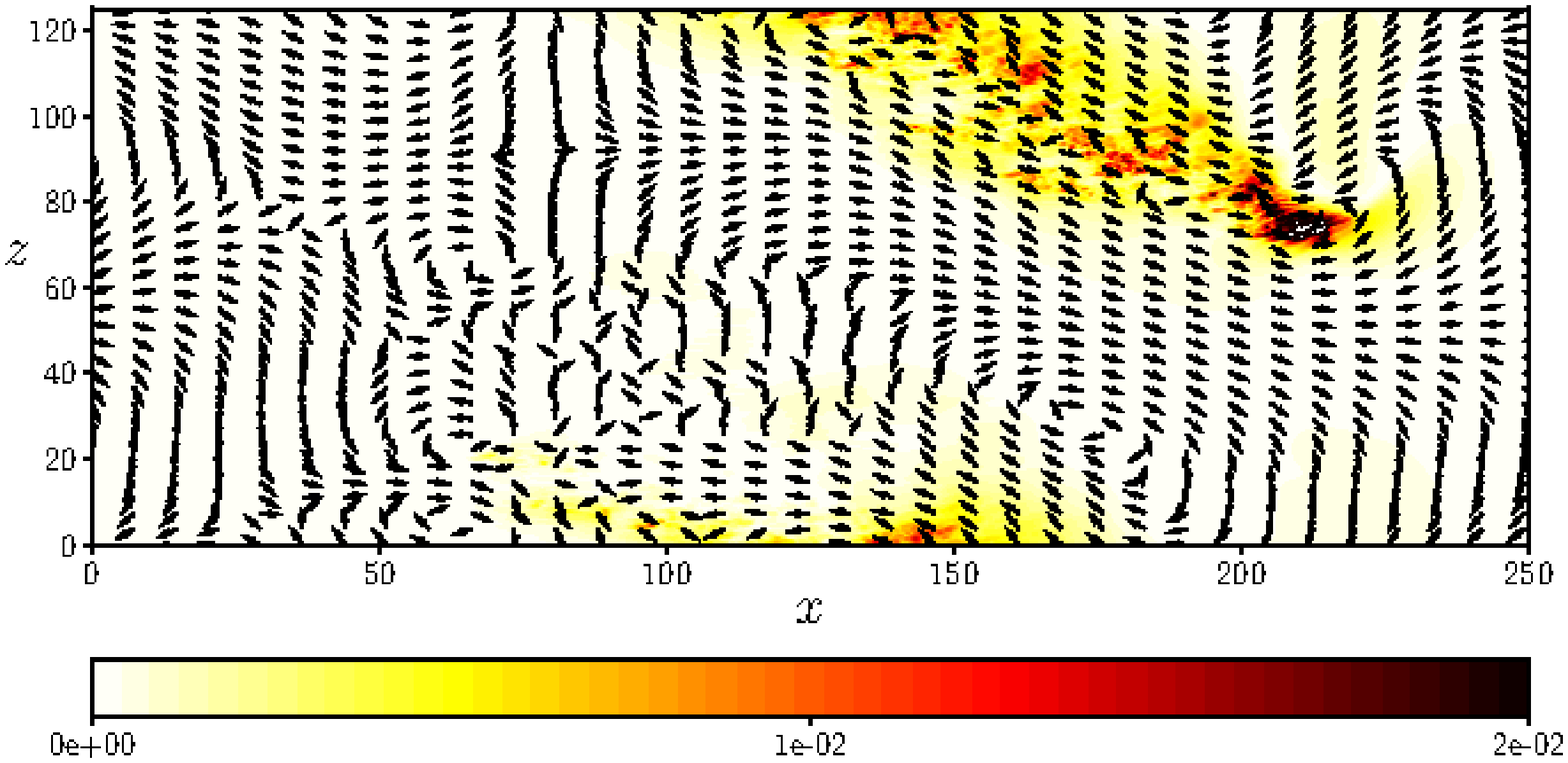}}
        \subfigure[$t = 2500$]{\includegraphics[width=.325\columnwidth]{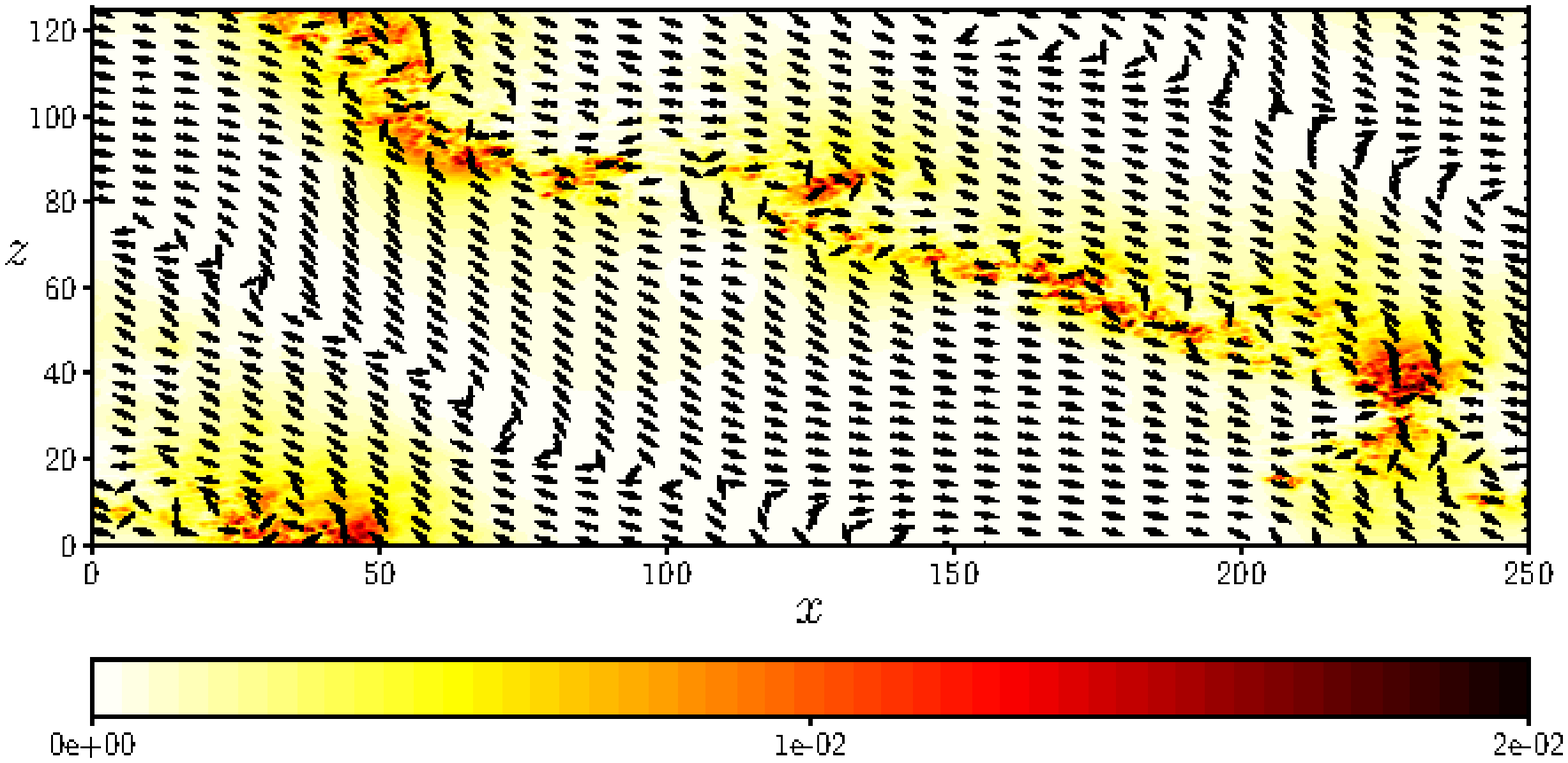}}
        \subfigure[$t = 3000$]{\includegraphics[width=.325\columnwidth]{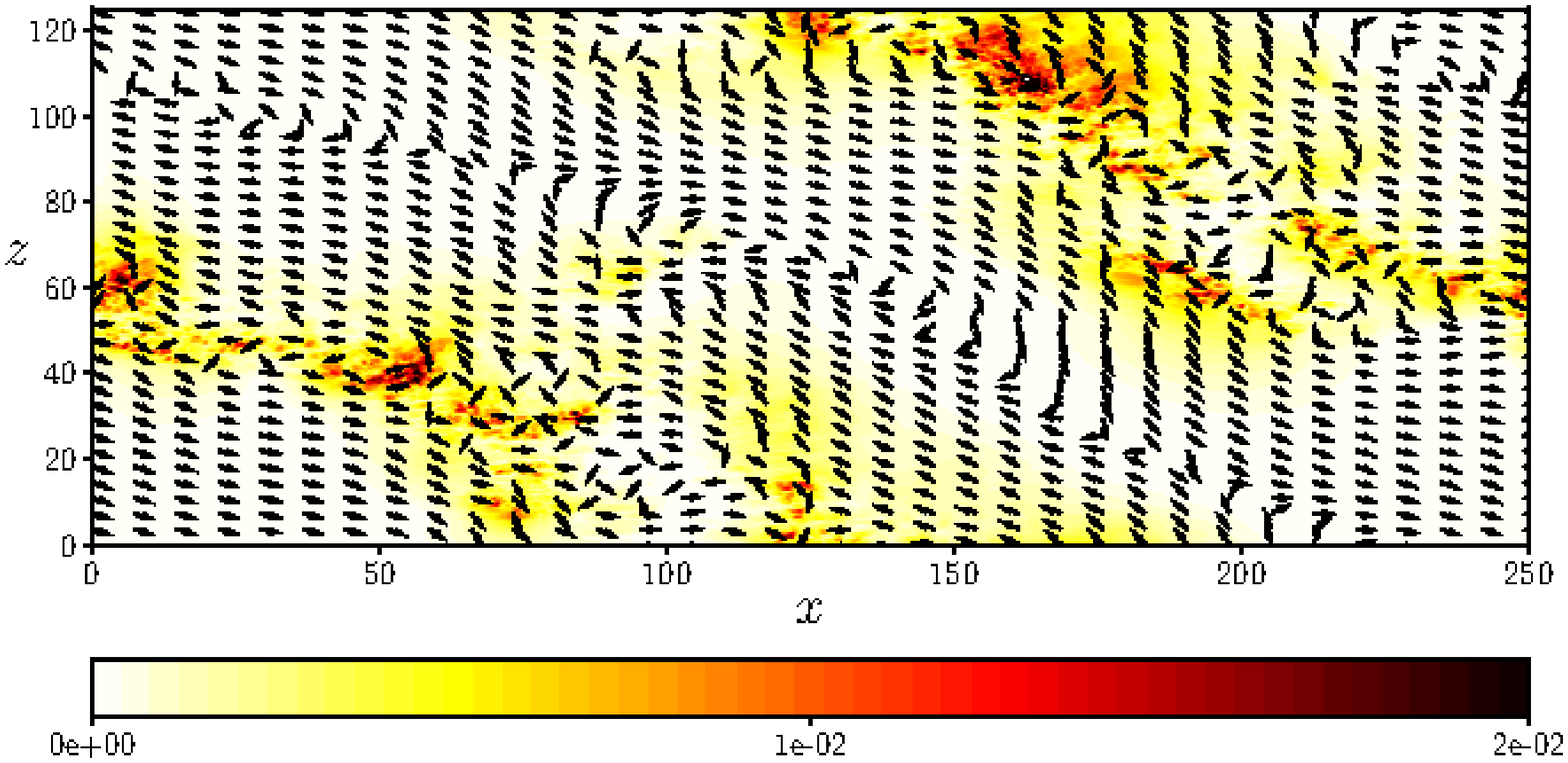}}
        \caption[]{Isocontour of the crossflow energy $E_{cf}$ together with the normalized $y$-integrated large scale flow for several instantaneous fields ($Re = 1000$, $E_0 = 5.5 \times 10^{-7}$, $T = 100$).}
        \label{evolution_Re1000} 
\end{figure}

\begin{figure}
        \centering
        \subfigure[$t = 200$]{\includegraphics[width=.325\columnwidth]{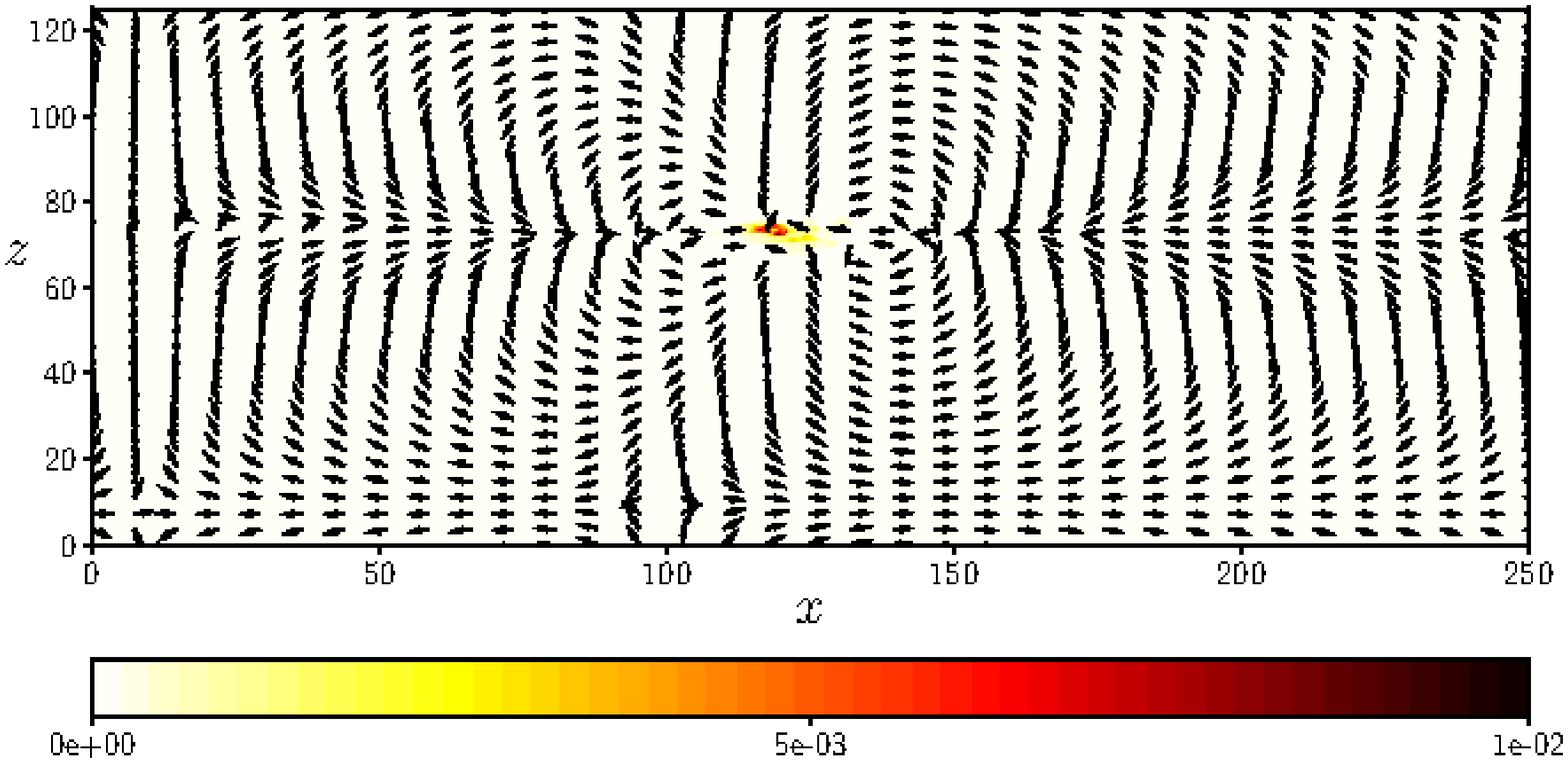}}
        \subfigure[$t = 500$]{\includegraphics[width=.325\columnwidth]{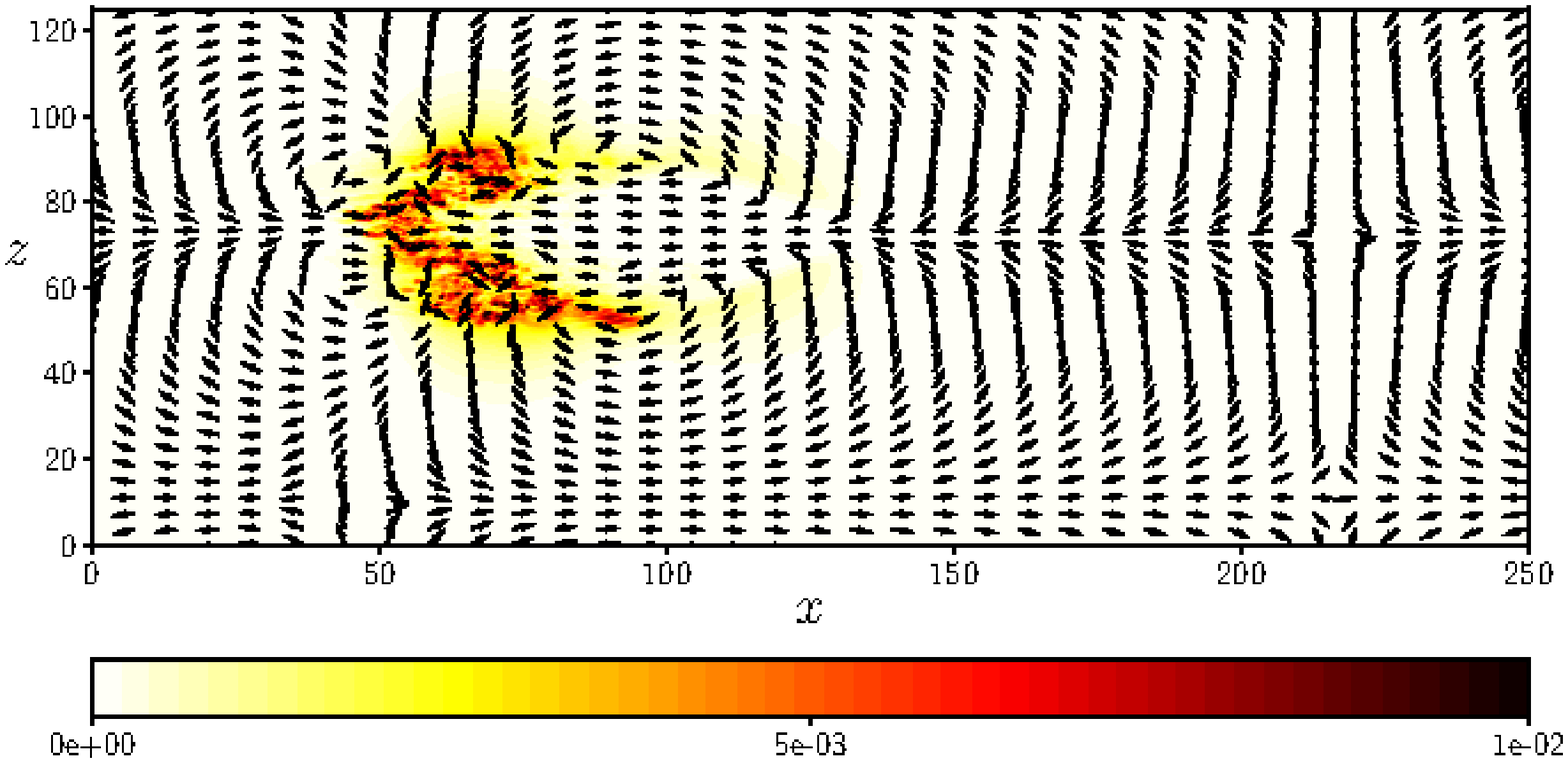}}
        \subfigure[$t = 900$]{\includegraphics[width=.325\columnwidth]{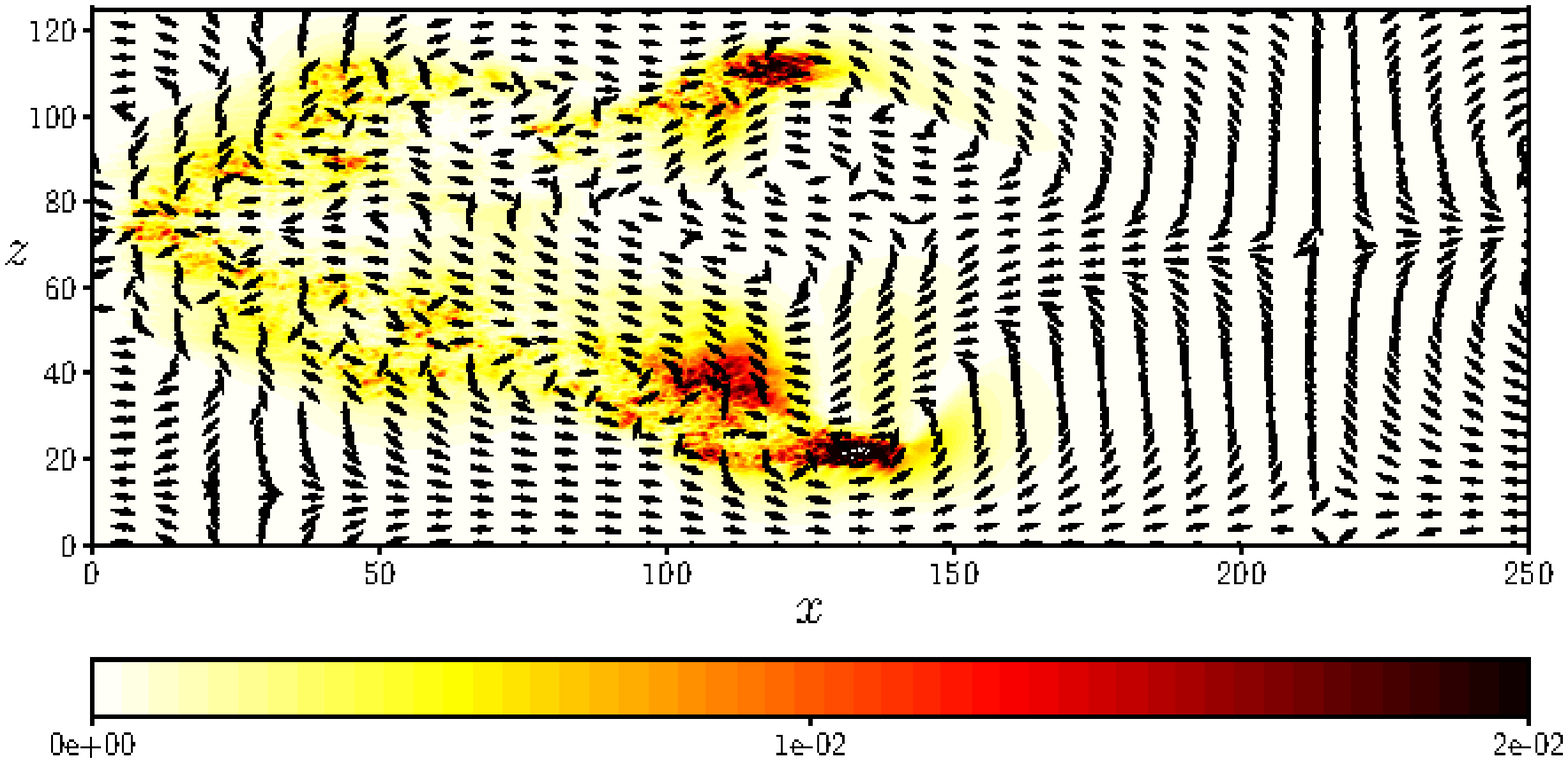}}
        \subfigure[$t = 1200$]{\includegraphics[width=.325\columnwidth]{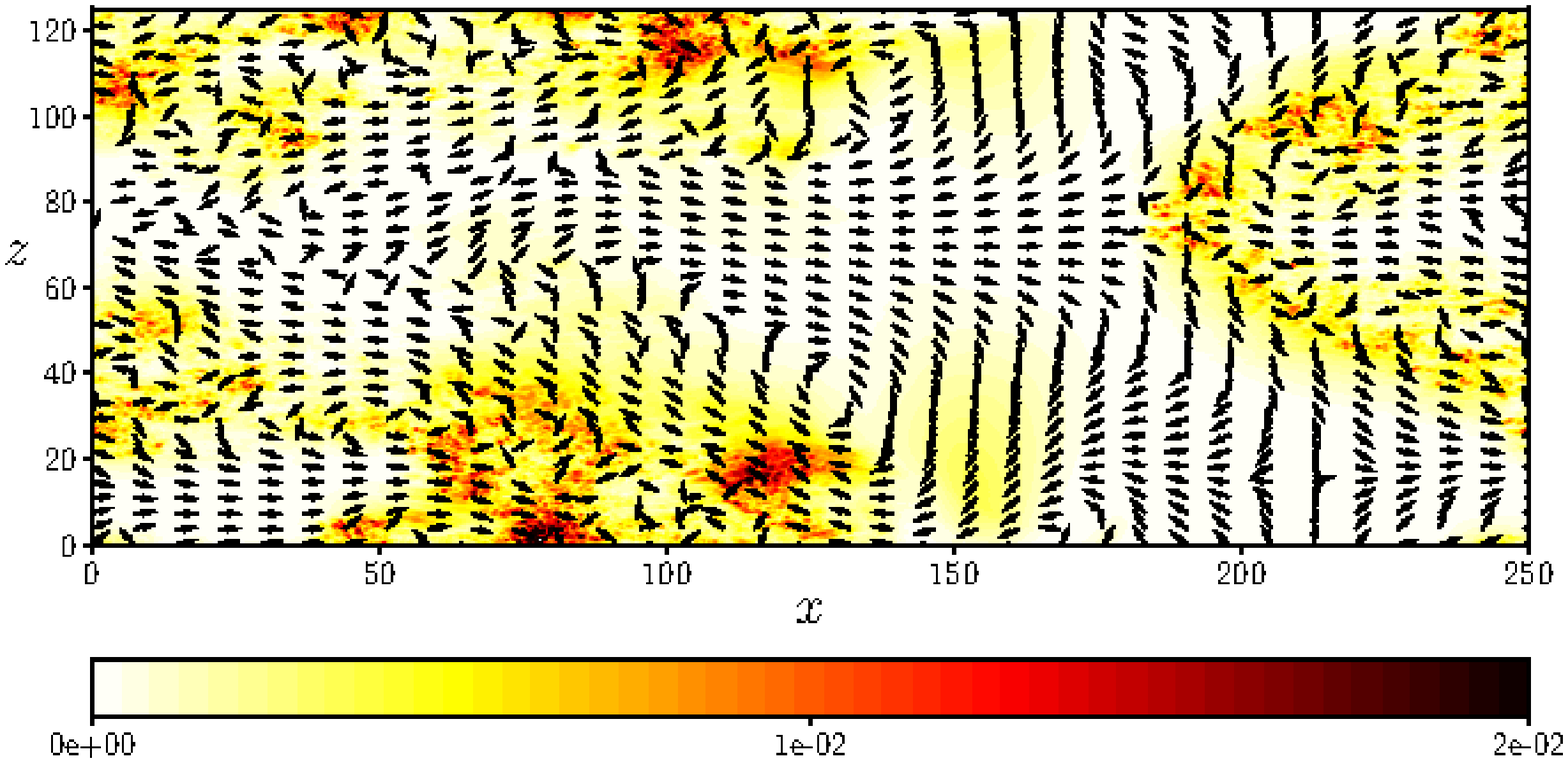}}
        \subfigure[$t = 1500$]{\includegraphics[width=.325\columnwidth]{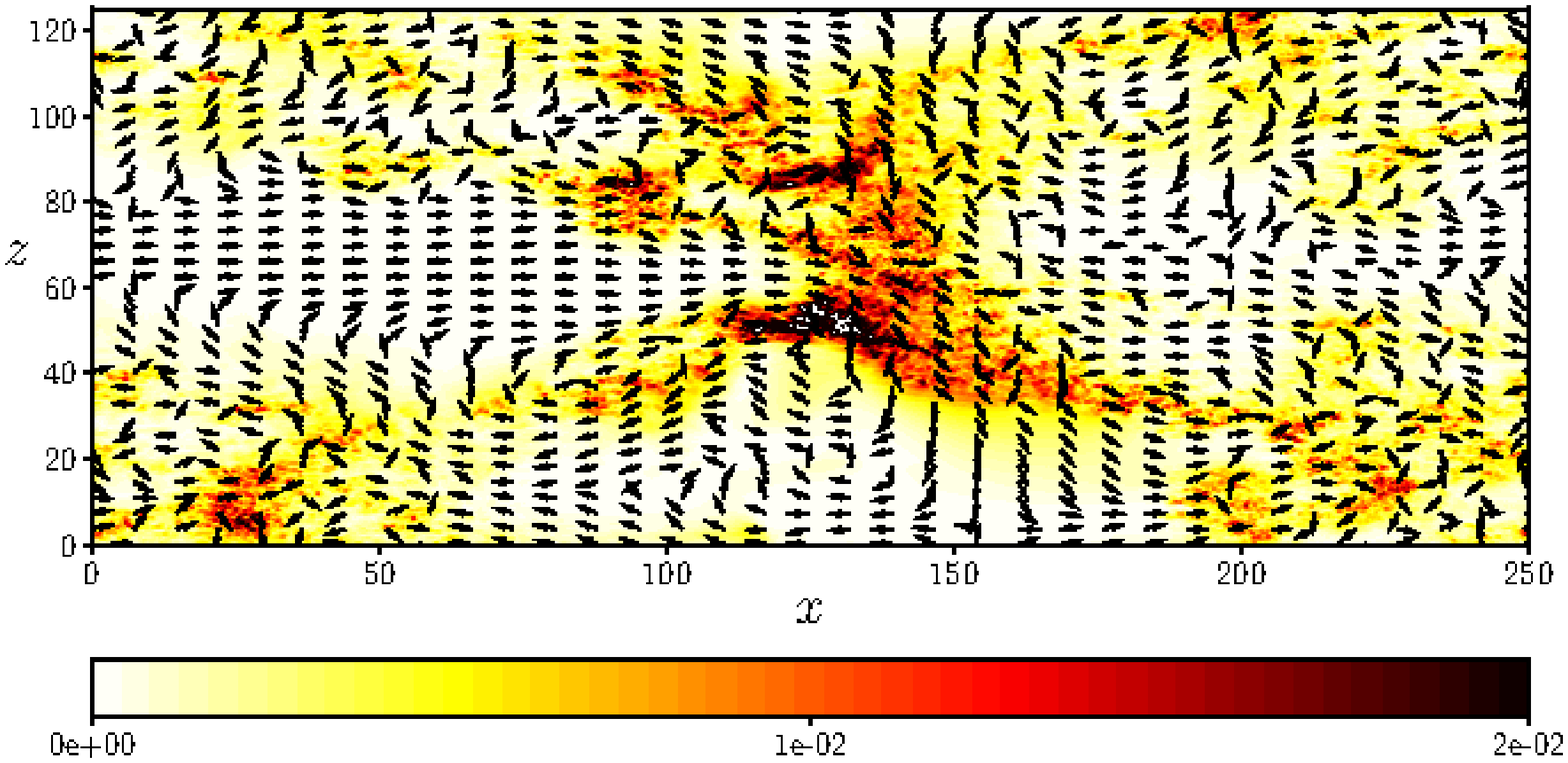}}
        \subfigure[$t = 3000$]{\includegraphics[width=.325\columnwidth]{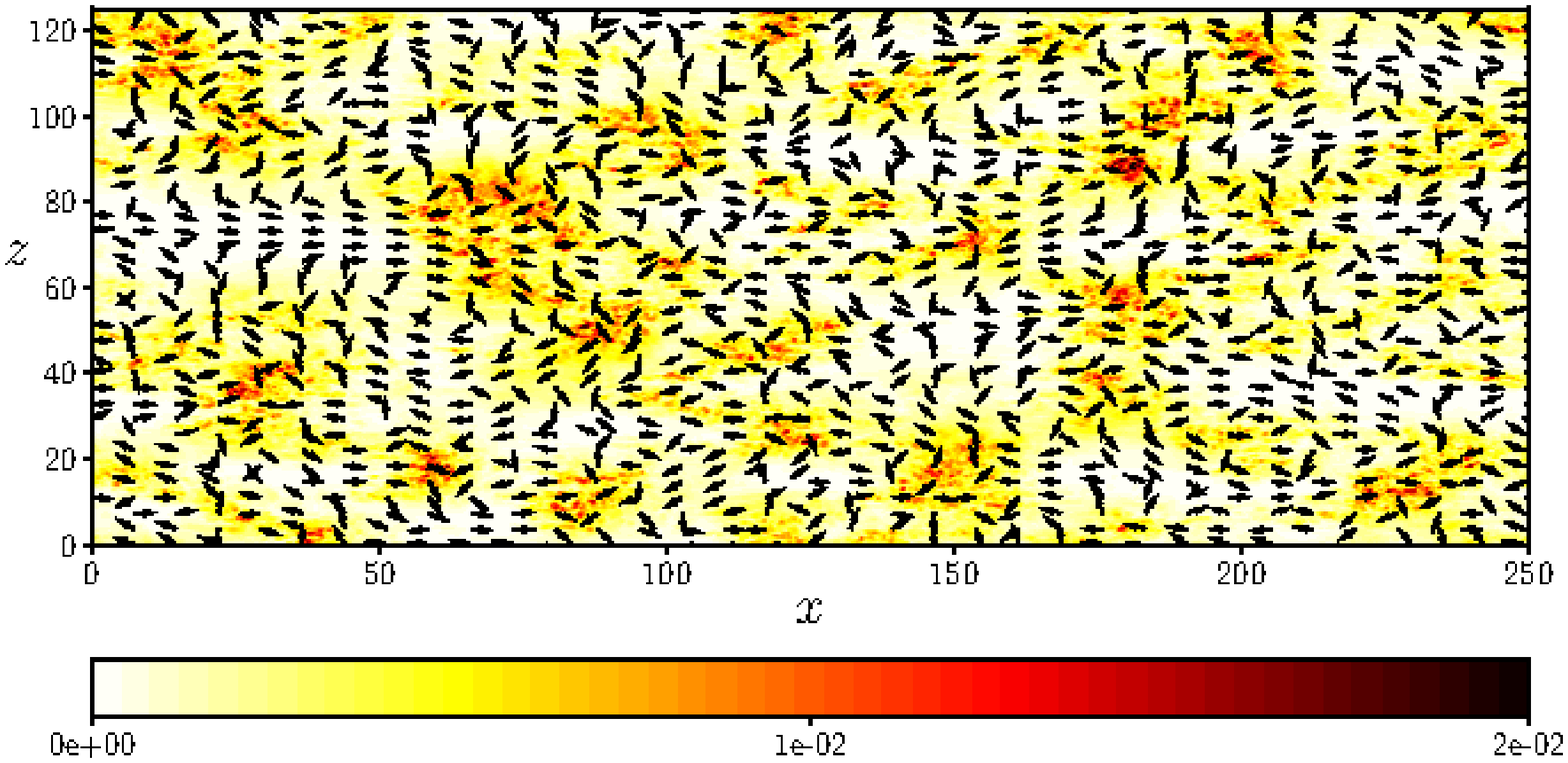}}
        \caption[]{Isocontour of the crossflow energy $E_{cf}$ together with the normalized $y$-integrated large scale flow for several instantaneous fields ($Re = 1250$, $E_0 = 2.9 \times 10^{-8}$, $T = 100$).}
        \label{evolution_E02p9e-8_Re1250} 
\end{figure}

\begin{figure}
        \centering
        \subfigure[$t = 200$]{\includegraphics[width=.325\columnwidth]{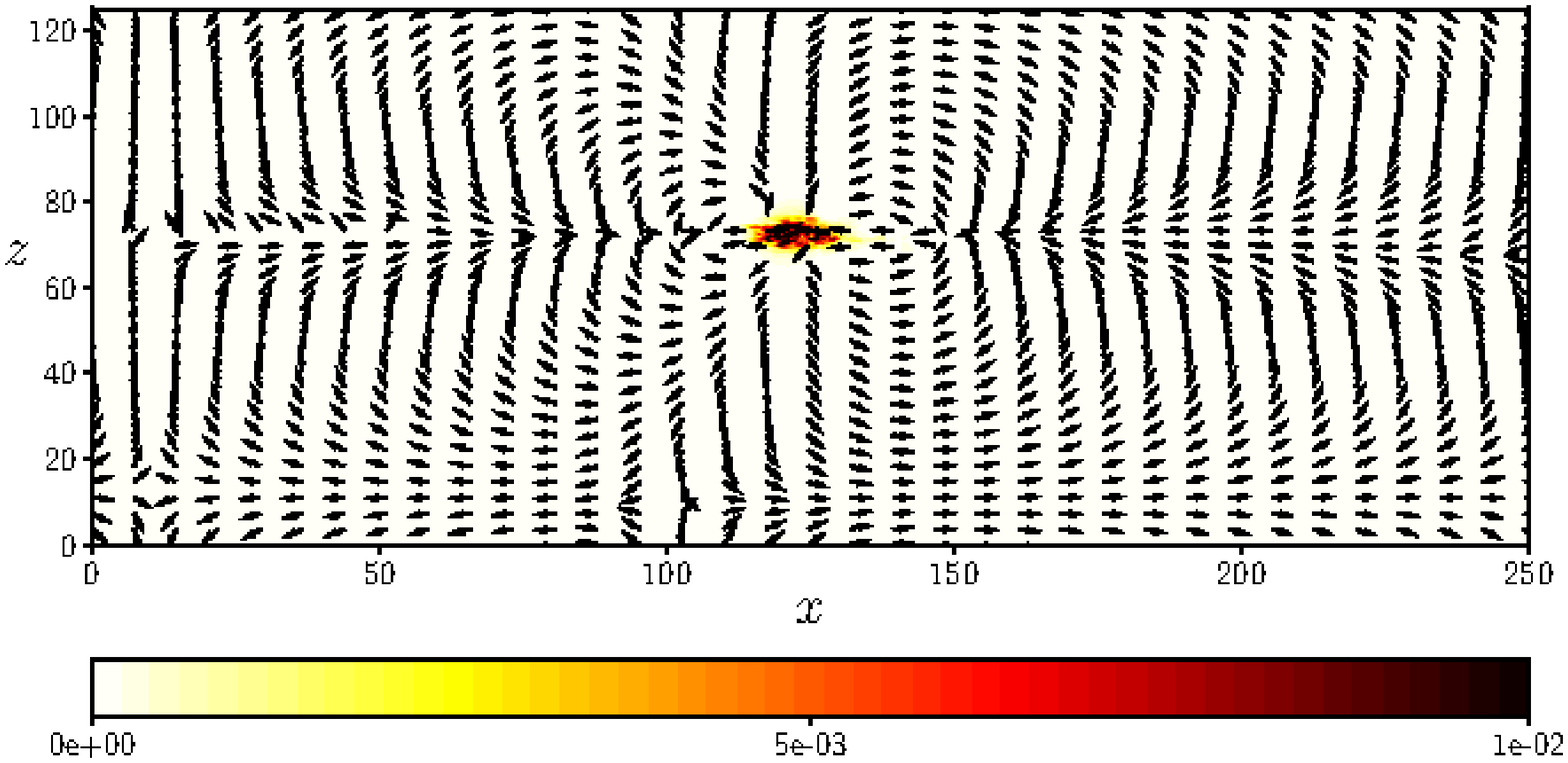}}
        \subfigure[$t = 300$]{\includegraphics[width=.325\columnwidth]{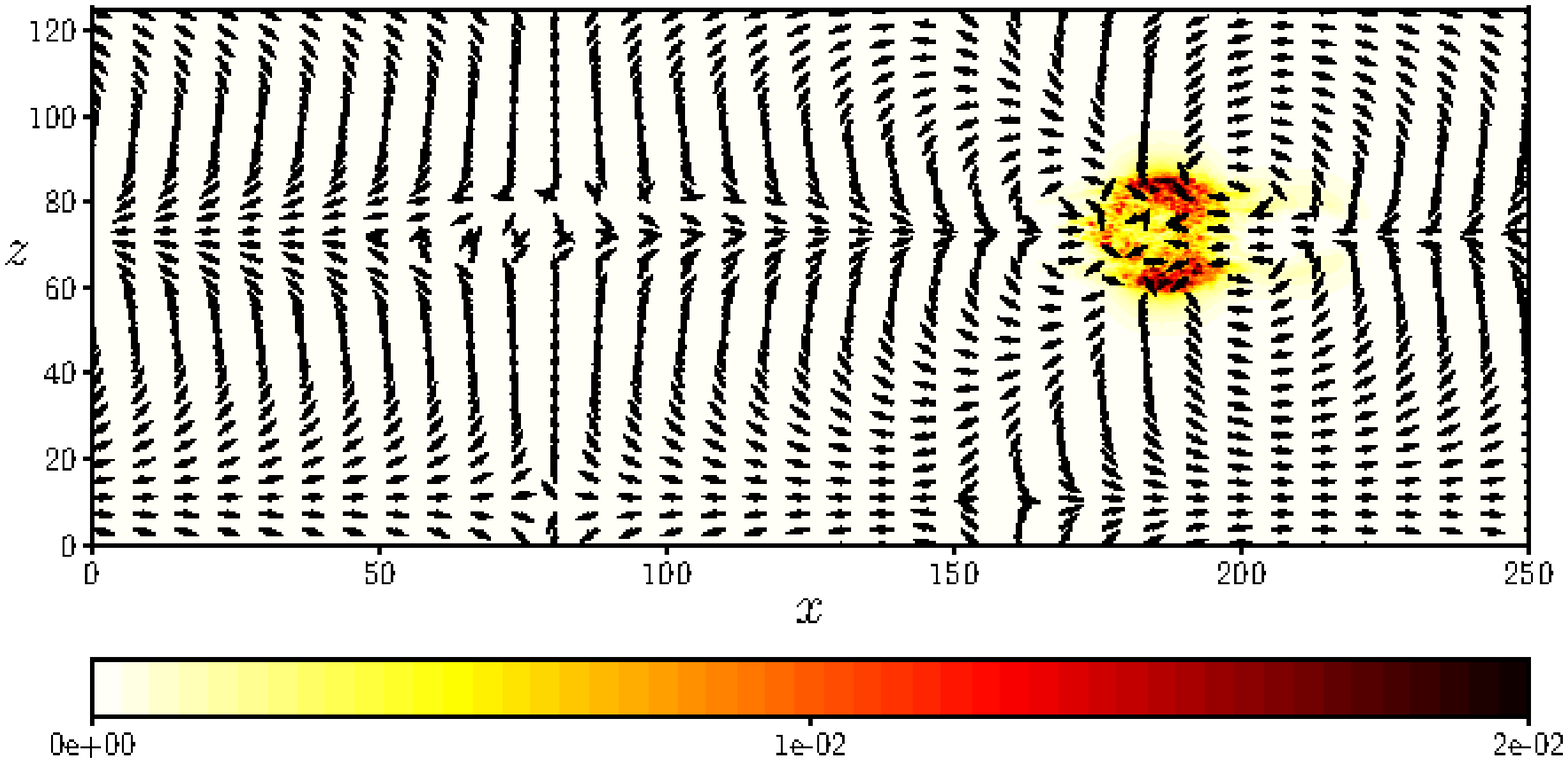}}
        \subfigure[$t = 350$]{\includegraphics[width=.325\columnwidth]{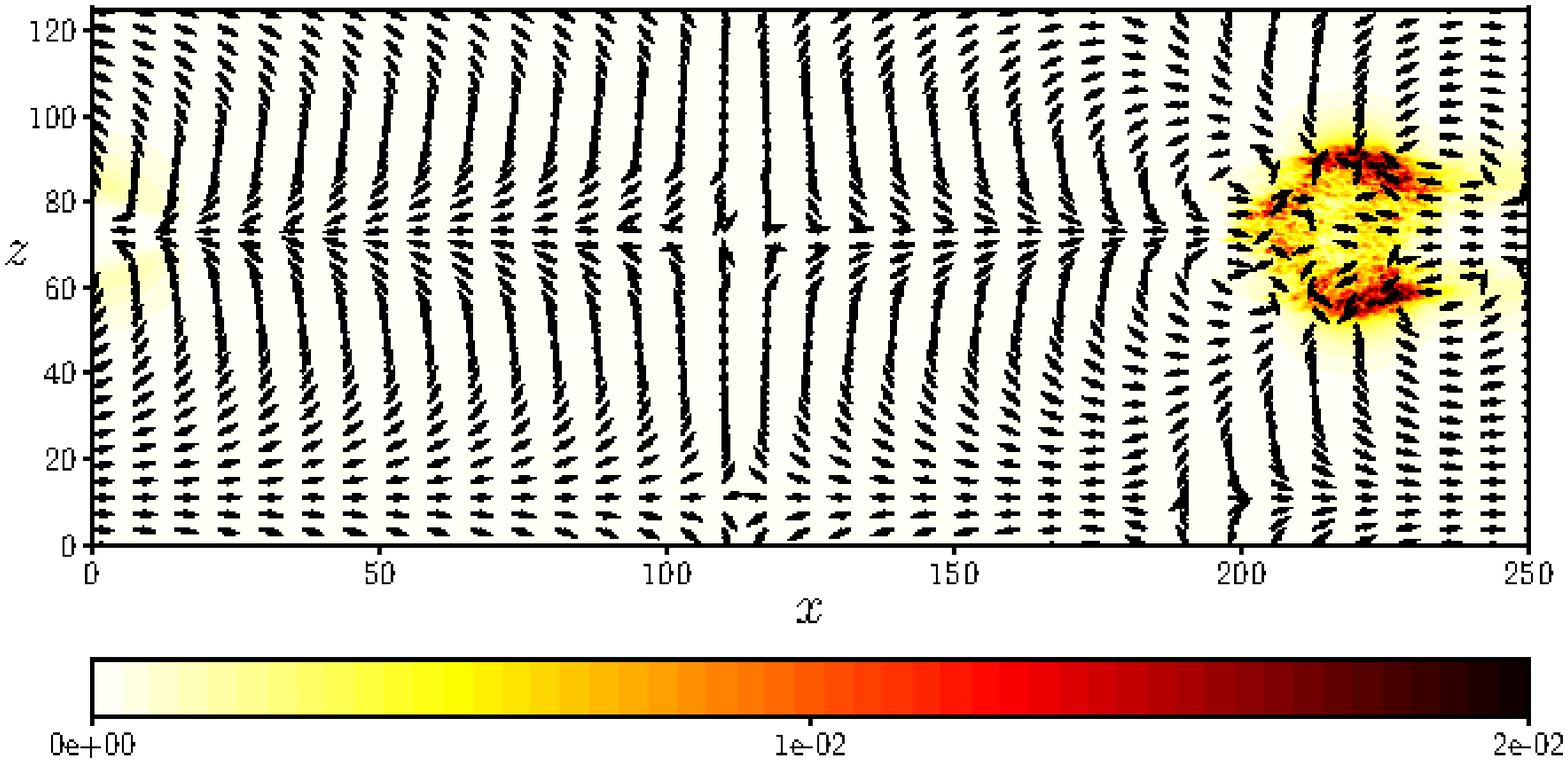}}
        \subfigure[$t = 500$]{\includegraphics[width=.325\columnwidth]{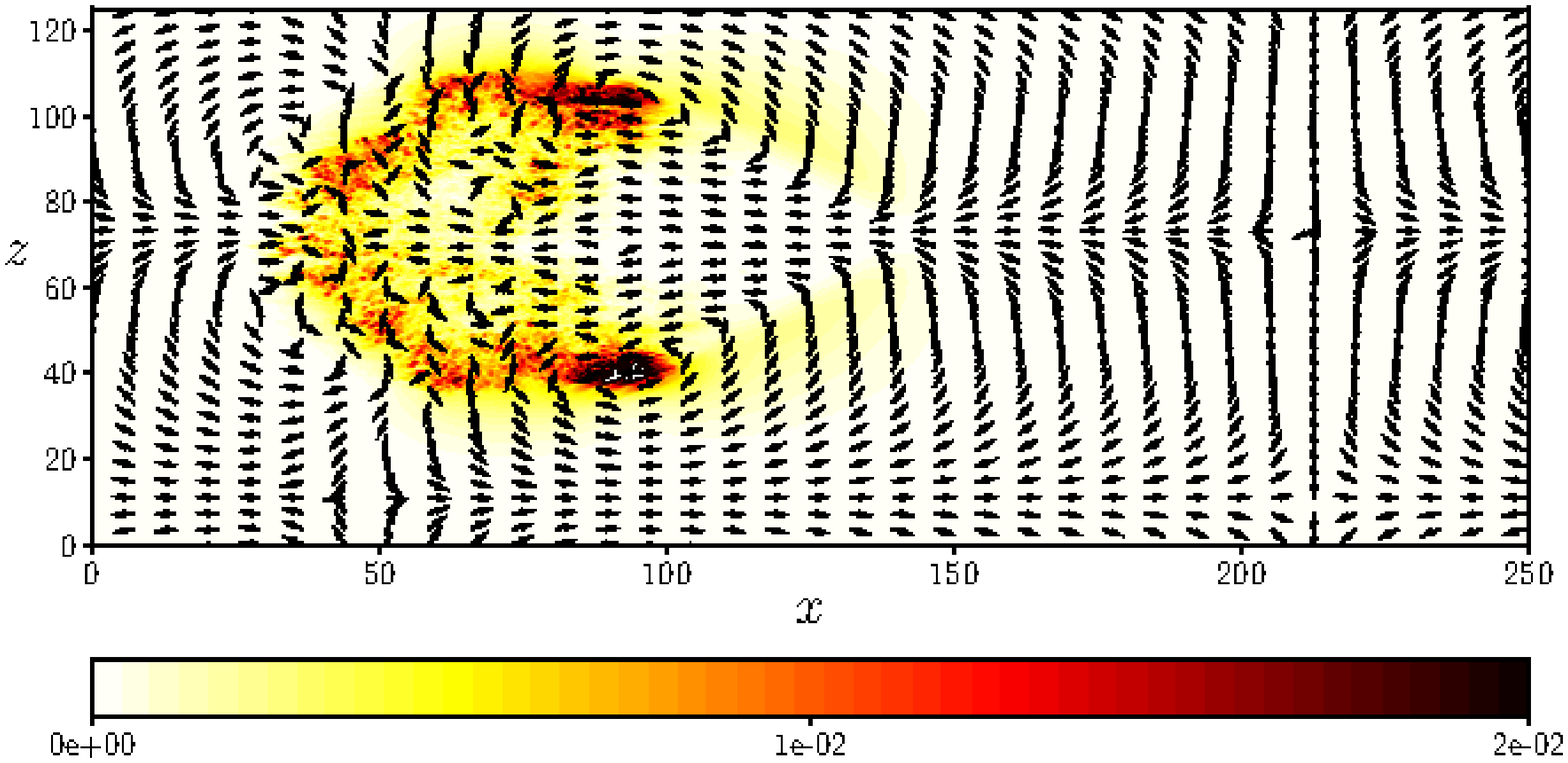}}
        \subfigure[$t = 600$]{\includegraphics[width=.325\columnwidth]{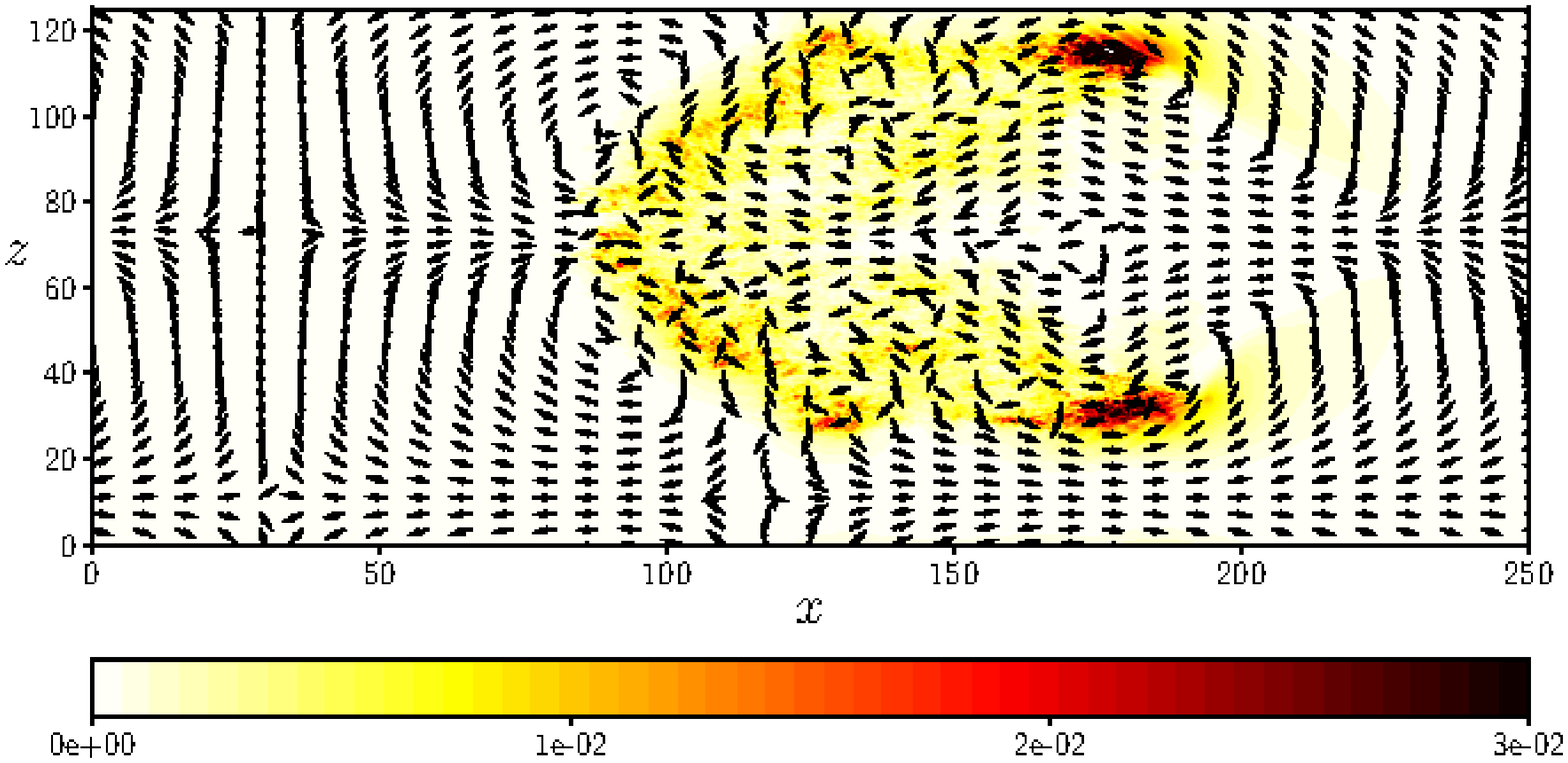}}
        \subfigure[$t = 900$]{\includegraphics[width=.325\columnwidth]{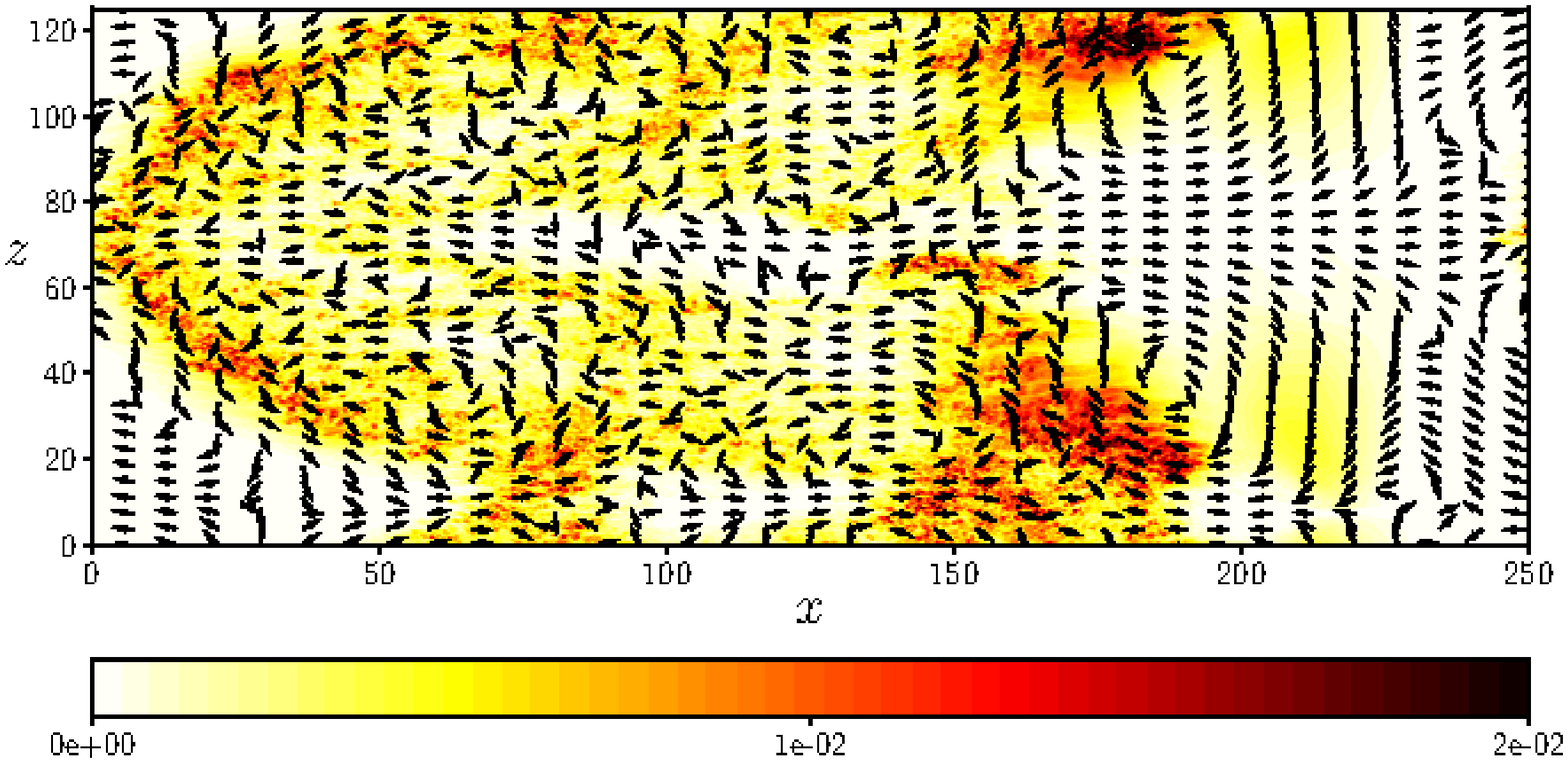}}
        \caption[]{Isocontour of the crossflow energy $E_{cf}$ together with the normalized $y$-integrated large scale flow for several instantaneous fields ($Re = 1568$, $E_0 = 3.6 \times 10^{-9}$, $T = 150$).}
        \label{evolution_E05p8e-9_Re1568} 
\end{figure}

This can be verified by analysing the time evolution of the crossflow energy and  $y$-averaged flow fields for the minimal seeds at different Reynolds numbers. 
For $Re=1150$, figure  \ref{evolution_E04p7e-8} shows that, 
as previously noticed (see also figure \ref{opt_sol_Re1150_E04p7e-8_T100}), the localised minimal solution breaks rapidly its symmetry along the spanwise direction, presenting a clearly asymmetric (but still spatially compact) structure at $t=500$, as shown in figure  \ref{evolution_E04p7e-8}  (b). This asymmetric wavepacket evolves via nucleation of new streaky structures (see \cite{ParenteRapid2021} concerning the mechanism of creation of the streaks) in the direction of the inclined laminar-turbulent interface, clearly forming a singular turbulent band, as it can be observed for $t = 900$. The newly-formed  turbulent band continues growing in an oblique direction with angle $\approx28^{\circ}$ until reaching the periodic boundaries, where it interacts with itself ($t=1500$). This triggers splitting of the previously isolated band ($t=2500$), which saturates reaching a  laminar-turbulent pattern filling the whole domain at $t = 4000$. The same behaviour has been observed by \cite{tao2013} and \cite{xiong2015} by injecting in a plane Poiseuille flow a "seed" of the turbulent bands, similar in shape to the instantaneous field at $t = 500$ in figure \ref{evolution_E04p7e-8}. 
 Analysing the large-scale flow, we can observe the formation of a small recirculation zone upstream of the spot during its evolution. Moreover, when the bands are formed, the large-scale flow is found to turn clockwise around bands with positive angle and anti-clockwise around bands with negative angle.  In fact, all the bands are formed in correspondence with the shear layer which divides the different vortices. 
A rather similar behaviour is observed at $Re=1000$, as shown in figure \ref{evolution_Re1000}. Despite at small times ($t=150$) the minimal seed has evolved into an almost symmetric V-shaped spot, one of its two legs weakens in time ($t=500$) and completely disappears at $t=900$, evolving into a single band as observed for $Re=1150$.
 Whereas, looking at the evolution in time for the cases at higher Reynolds number, the flow presents the same behaviour observed experimentally and numerically when turbulence is triggered by a spot (\cite{carlson1982},  \cite{henningson1991}, \cite{aida2010}, \cite{aida2011}). In fact, the localised perturbation initially evolves in the domain forming a turbulent spot (see figure \ref{opt_sol_Re1150_E01p1e-7_T100}),  turning into a V-shape at $t = 500$,  as shown in figure \ref{evolution_E02p9e-8_Re1250}. At this time, two distinct fronts of the spots can be observed, which evolve in two symmetric bands with angle $\approx\pm 45^{\circ}$ growing obliquely in the domain, as shown at  $t = 900$. At $t = 1200$, they start to interact which each other, forming a spatio-temporally complex final state composed by a coexistence of turbulent and laminar patterns ($t=3000$). Qualitatively the same behaviour is observed at $Re=1568$, as shown in figure \ref{evolution_E05p8e-9_Re1568}, although the spatial spreading of the  bands  appears to be more rapid than at lower $Re$, despite the initial energy of the perturbation is lower. Also in these cases, the bands are found to form right in the mixing layer between two large-scale counter-rotating vortices. 
 Notice also that the same quasi-symmetric behaviour can be observed at lower $Re$, for a larger initial energy. In fact,  the nonlinear optimal perturbation computed for $Re=1150$ and $E_0=4.7 \times 10^{-8}>{E_0}_{min}$ evolves in two distinct bands, showing a time evolution  corresponding to that of the minimal seed at largest $Re$ (not shown). \\
 An explanation of this behaviour can be attempted by recalling that, in the channel flow, turbulent stripes have a probability of decay that increases with time, and that decreases with the Reynolds number \citep{Paranjape2019}. Thus, as all minimal seeds present an almost spanwise-symmetric structure, two proto-bands begin to be created at the edges of the large-scale vortices characterising the minimal seed. However, the probability of decay of these bands is higher for low Reynolds number, and increases in time, so when $Re$ is sufficiently low one of these bands rapidly dies out, leading to the development of one isolated band.  Increasing $Re$, the probability of decay of an initial band is lower, while the probability of splitting increases. Thus, both oblique bands originated at the sides of the minimal seed  survive longer in time, until they split and  interact, rapidly leading to the establishment of a spatio-temporally complex final state. 
 Notice that injecting more initial energy at low value of $Re$ has the same effect of increasing $Re$. In fact, an optimal perturbation with $E_0>{E_0}_{min}$ is less spatially localized, and is able to reach a much larger kinetic energy at $T=100$, leading to more spatially-extended and energetic proto-bands, which allows their sustainment for a longer time. 
 \begin{figure}
        \centering
        \subfigure[$t = 0$]{\includegraphics[width=.24\columnwidth]{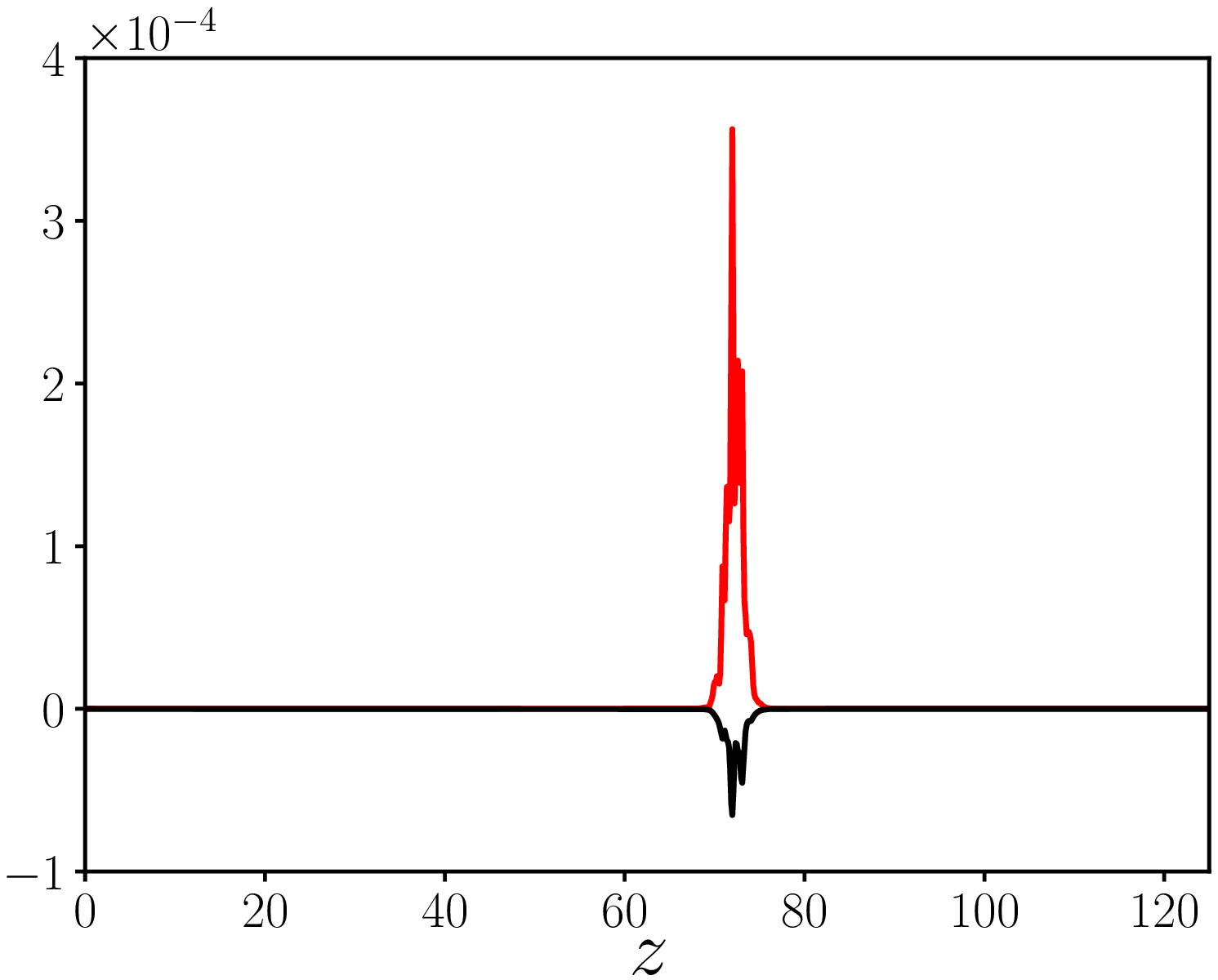}}
        \subfigure[$t = 100$]{\includegraphics[width=.24\columnwidth]{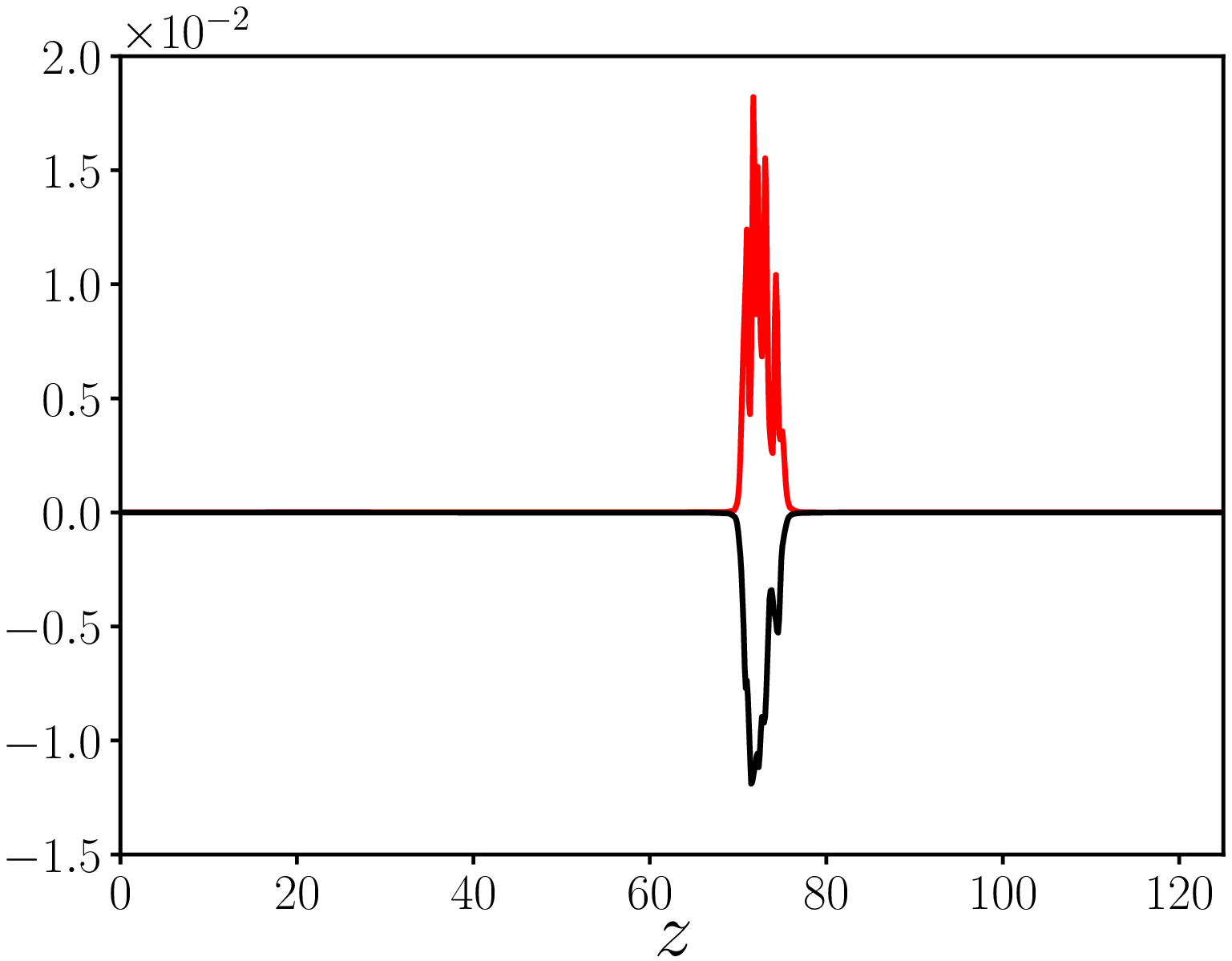}}
        \subfigure[$t = 500$]{\includegraphics[width=.24\columnwidth]{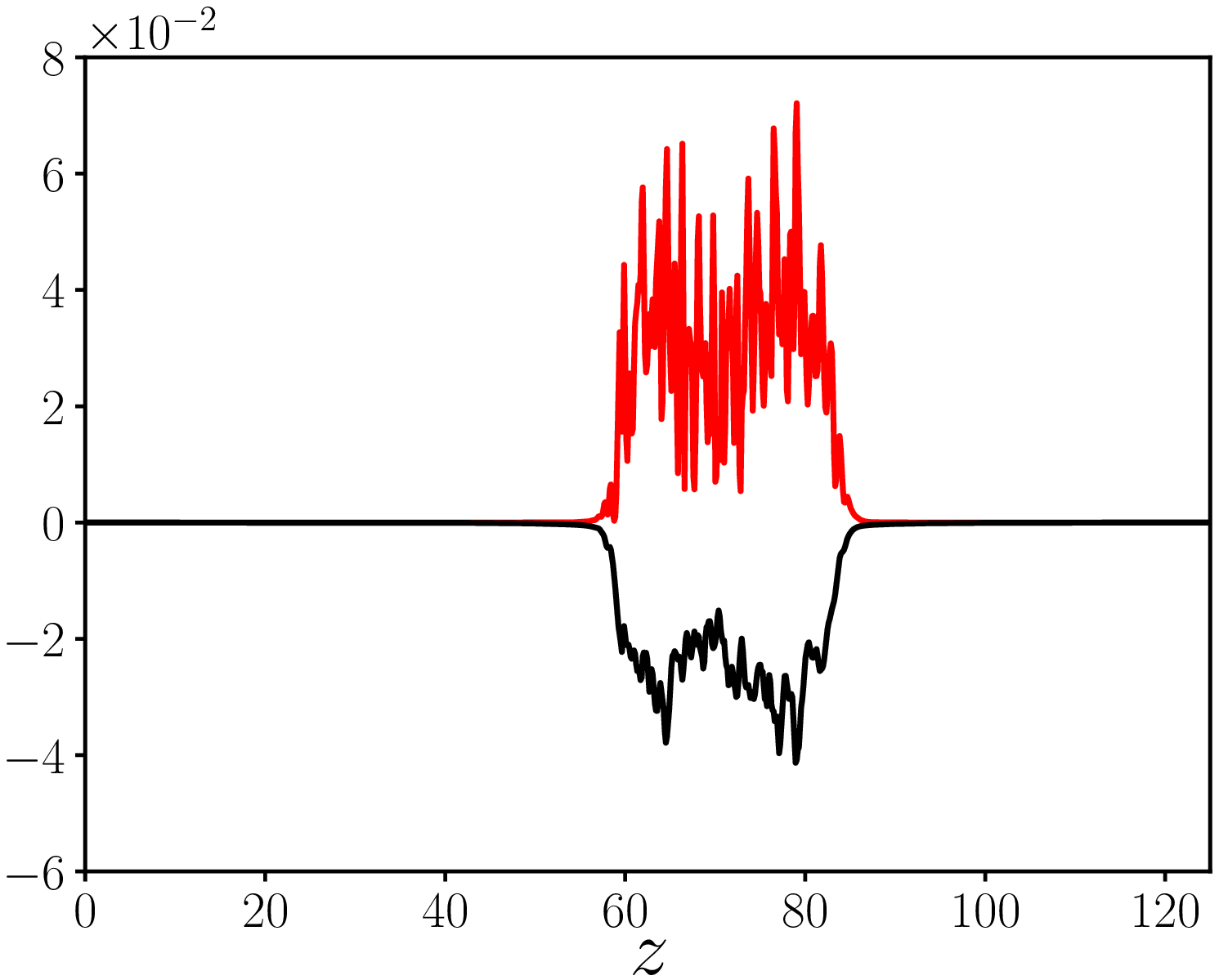}}
        \subfigure[$t = 900$]{\includegraphics[width=.24\columnwidth]{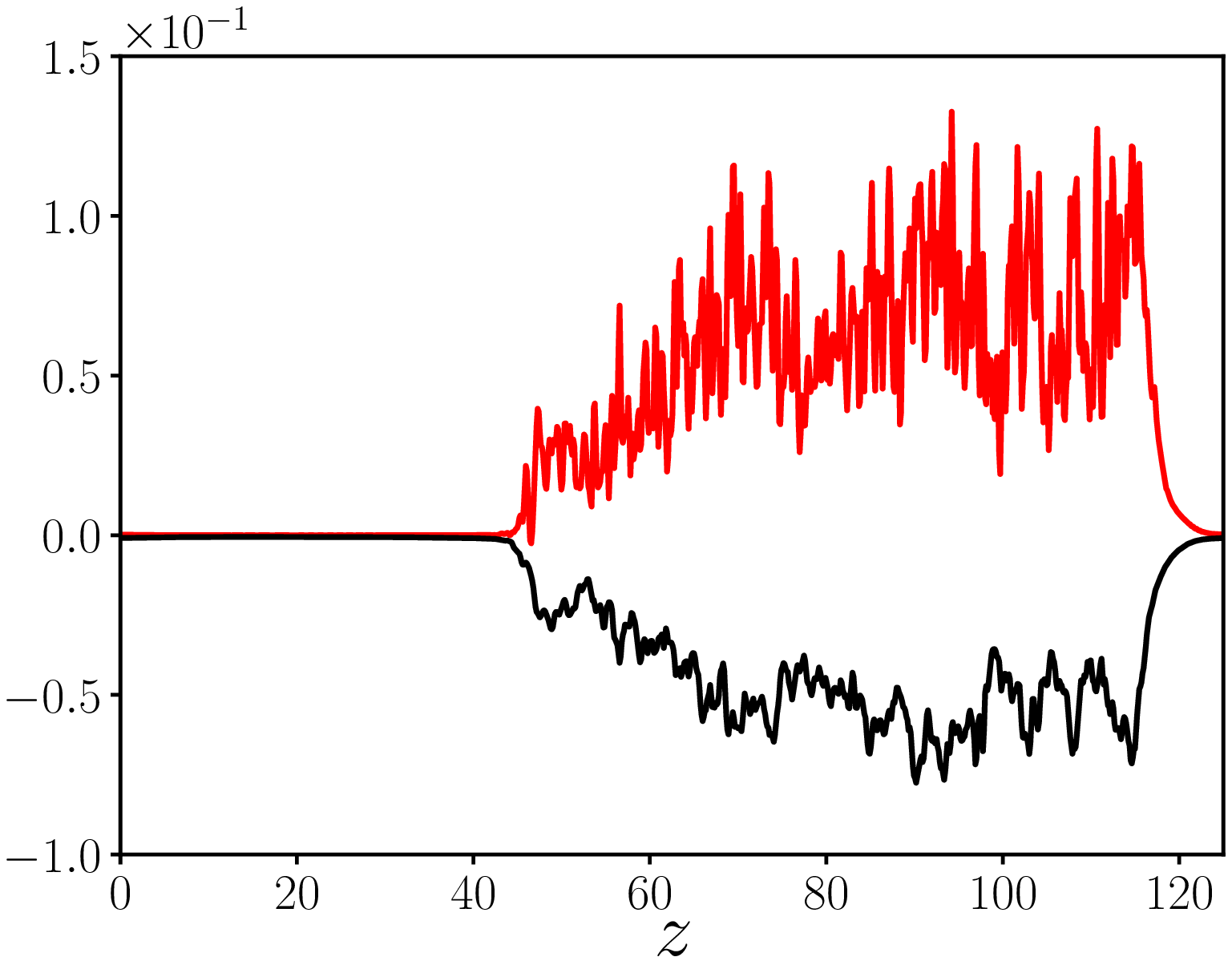}}
        \caption[]{Spanwise distribution of the production ($P$, red line) and dissipation ($-\epsilon$, black line) terms integrated in  $x-y$ planes for different instantaneous fields obtained evolving in time the minimal seed for $Re = 1150$, $E_0 = 4.7 \times 10^{-8}$, $T = 100$.}\label{fig:production-z-Re1000}
 \end{figure}
 
 \begin{figure}
        \centering
        \subfigure[$t = 0$]{\includegraphics[width=.24\columnwidth]{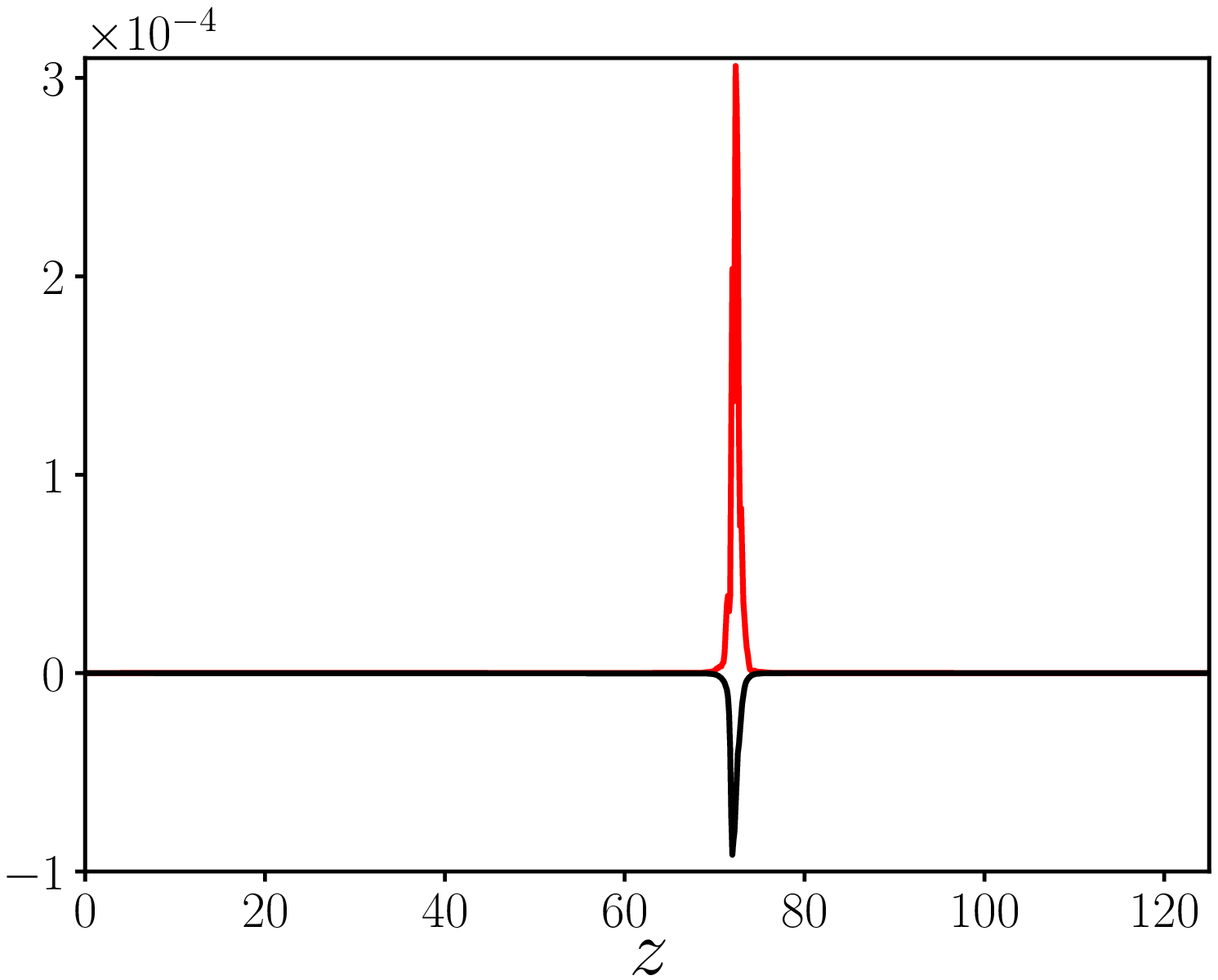}}
        \subfigure[$t = 100$]{\includegraphics[width=.24\columnwidth]{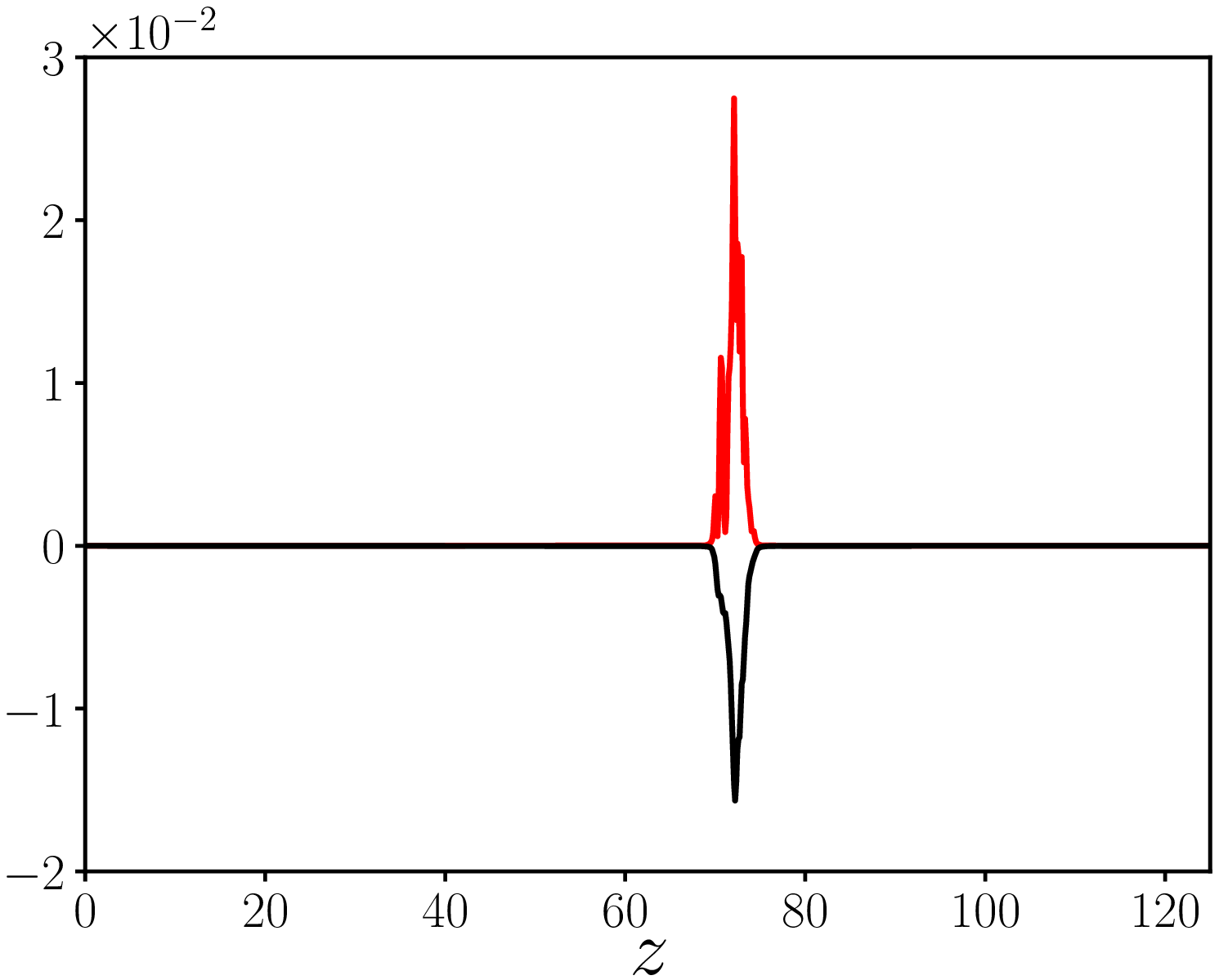}}
        \subfigure[$t = 500$]{\includegraphics[width=.24\columnwidth]{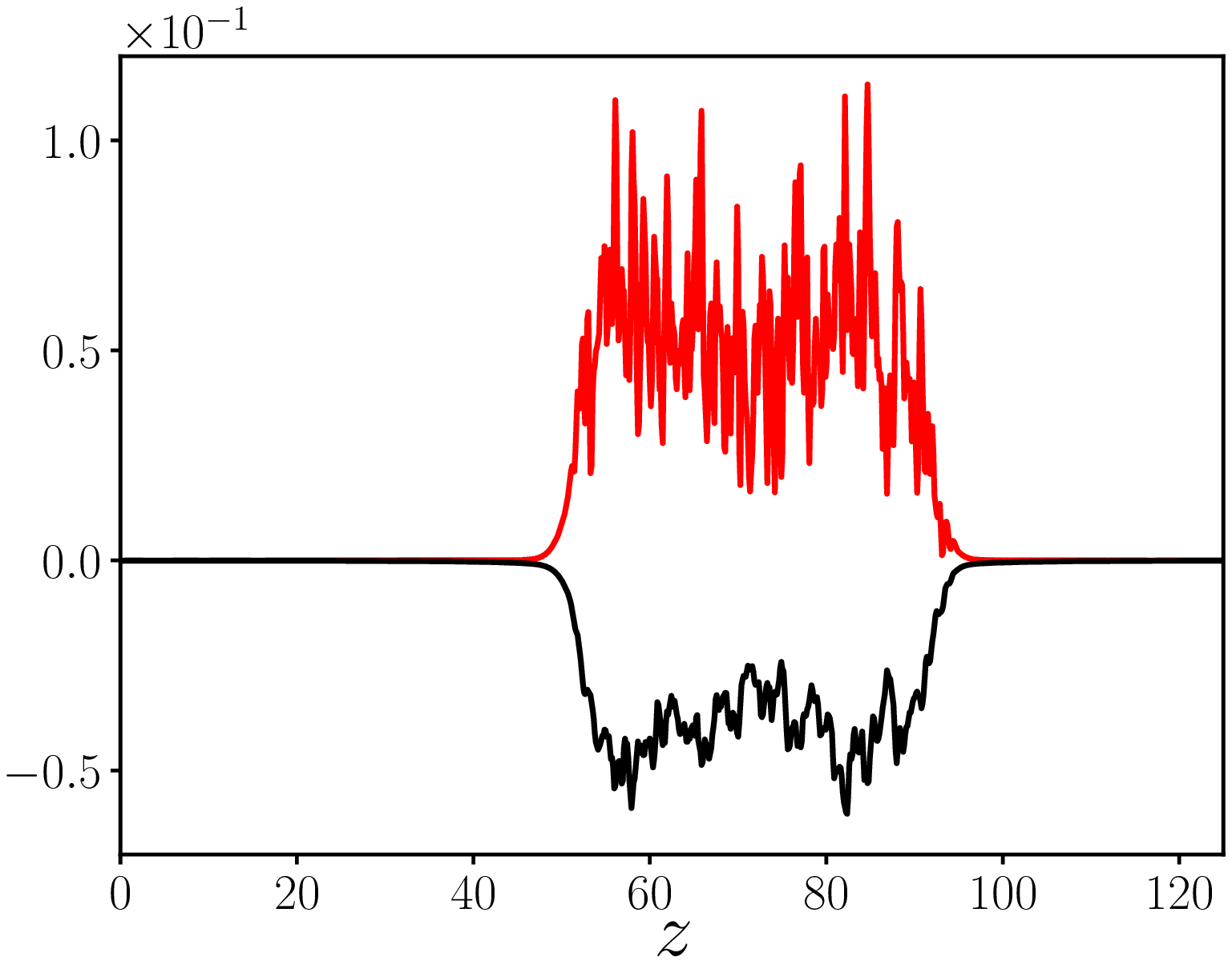}}
        \subfigure[$t = 900$]{\includegraphics[width=.24\columnwidth]{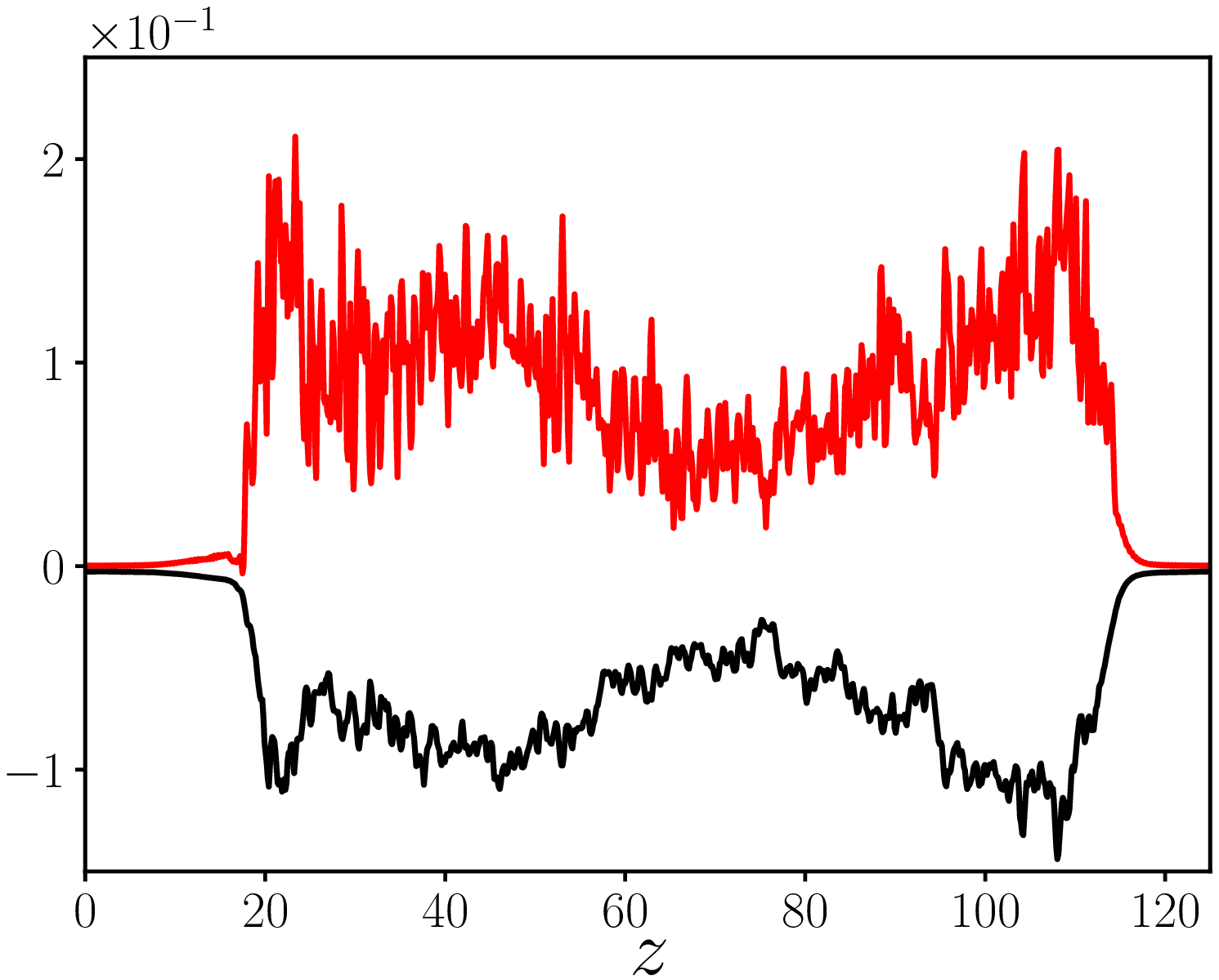}}
        \caption[]{Spanwise distribution of the production ($P$, red line) and dissipation ($-\epsilon$, black line) terms integrated in  $x-y$ planes for different instantaneous fields  obtained evolving in time the minimal seed for $Re = 1250$, $E_0 = 2.9 \times 10^{-8}$, $T = 100$.}\label{fig:production-z-Re1250}
 \end{figure}

To corroborate this conjecture, we make use the Reynolds-Orr equation to evaluate the 
 production and dissipation of kinetic energy as, respectively:
 \begin{equation}
 P=-{u}_i' {u}_j' \frac{\partial {U}_i}{\partial x_j},   \quad \quad \epsilon = \frac{2}{Re}{s_{ij}'s_{ij}'} \quad \text{with} \quad s_{ij}'=\frac{1}{2} \left(\frac{\partial {u}_i'}{\partial x_j}+\frac{\partial {u}_j'}{\partial x_i}\right),
 \end{equation}
 where the Einstein summation convention has been used. 
One can compare the time evolution of the production and dissipation terms, integrated in $x-y$ planes, for the minimal seeds at $Re=1150$ and $Re=1250$,  provided in  figure \ref{fig:production-z-Re1000} and \ref{fig:production-z-Re1250}. For both Reynolds numbers, at $t=0$ the production presents one single peak almost in the center of the spot ($z\approx70$), which is found to exceed dissipation of almost an order of magnitude. Production of kinetic energy leads to a slight increase of the spanwise size of the wavepacket ($t=100$), together with a further increase of the kinetic-energy production, probably due to the nucleation of new streaks which produce kinetic energy thanks to the lift-up effect. Due to the breakdown of the structures inside the spot, dissipation increases as well, reaching almost the same value than the production term. Notice also that a weak secondary peak begins to be visible in the production term. At $t=500$, the spot has strongly increased its size in the spanwise direction, presenting an almost symmetric shape with two peaks at $z\approx55-60$ and $z\approx 80-85$. However, at $t=900$, the evolution of the production and dissipation terms begins to strongly differ between the two considered Reynolds numbers. For $Re=1250$, the spanwise distribution of the dissipation and production terms remains almost spanwise symmetric, centered at $z\approx70$, with two distinct laminar-turbulent fronts at $z\approx20$ and $z\approx 110$ (see figure \ref{fig:production-z-Re1250}). Whereas, for $Re=1150$, the left-most part of the packet has almost faded away, while the right-most one has expanded up to $z\approx110$, as shown in figure \ref{fig:production-z-Re1000}.
The analysis of the production and dissipation terms clearly indicates that the minimal seed for turbulent bands leads to the generation of two almost symmetric regions of high production and dissipation, which can be seen as two distinct proto-bands. However, when the Reynolds number is lower, the weaker of these two proto-bands rapidly decays, leading to the development of an isolated band. Whereas, at larger Reynolds number, both bands survive for a sufficiently long amount of time to begin interacting between themselves. 

\section{Conclusion}\label{sec:conclusions}
In this work we have investigated the minimal-energy perturbations for the generation of turbulent bands in plane Poiseuille flow. A nonlinear optimization  maximising the kinetic energy at a given target time, coupled with initial energy bisection, has been used. The optimization was performed in very large domains, for a range of Reynolds number for which turbulent bands are sustained and lead to a spatio-temporally complex turbulent-laminar final state, namely 
$Re = 1000, 1150, 1250, 1568$ (the lowest value representing the threshold $Re$ for which bands splitting and turbulence spreading can be observed). 
The influence of the Reynolds number on the minimal energy threshold for generating turbulent bands ($E_{{0}_{min}}$), is analysed. In accordance with previous works carried out on other shear flows in small domains, the minimal seed has been found to scale with $Re$ following a power-law $E_{{0}_{min}} \propto Re^{-\gamma}$, although a sufficiently good fit is found only by restraining the analysis to $Re>1000$. However, the value of $\gamma$ recovered in the present work ($\approx 8.5$) is approximately $4$ time larger than the values  reported in  previous works ($\gamma \approx 2.7$ in \cite{duguet2013}, $\gamma\approx 2$ in \cite{CherubiniPoF2015}), 
probably due to the much larger size of the considered domain. 
\\
 For all the Reynolds numbers analysed, the minimal-energy perturbation able to generate turbulent bands 
is a spatially localised spot-like structure composed by finite-size streaks and elongated vortices. 
A more marked localization of the minimal seed is found when $Re$ increases. 
As previously reported for the channel flow in the presence of spatially-localised spots,  a large-scale flow having a quadrupolar structure has been found to surround the small-scale localised minimal perturbations.
These minimal perturbations have dominant wavelengths $\approx 4$ and $\approx 1$ in the streamwise and spanwise directions, respectively. 
Nonlinear optimal perturbations with energy higher than minimal, are characterized by similar shape and wavenumbers.
\\
The evolution of the minimal seeds towards the turbulent bands has been investigated. For $Re < 1250$, the minimal seeds  evolve in time creating an isolated oblique band. Whereas, for $Re \ge 1250$, it gives rise to two distinct bands which grow almost symmetrically in the spanwise direction. 
This almost symmetrical evolution is observed also at lower $Re$ for non-minimal optimal perturbations. 
An analysis of the production and dissipation of kinetic energy integrated in the streamwise and wall-normal directions shows that in all cases the initial spot-like perturbation evolves in an almost symmetric fashion, giving rise to two proto-bands at the edges of the large-scale flow characterizing the minimal seeds.
However, since the  probability  of  decay  of  bands  increases in time and is  higher  for  low  Reynolds  number, when $Re$ is sufficiently low one of these bands rapidly dies out, leading to the development of one isolated band. Whereas, for larger values of $Re$, the probability of decay of an initial band is lower, while the probability of splitting increases. Thus, both oblique bands originated at the sides of the minimal seed survive longer in time, until they split and interact, rapidly leading to the establishment of a spatio-temporally complex final state. Injecting more initial energy at low value of $Re$ has the same effect of increasing $Re$, since a more spatially-extended disturbance with higher kinetic energy is reached at a small time, leading to longer and more energetic proto-bands, able to be sustained for a longer time. \\
This work elucidates two (apparently distinct) minimal-energy mechanisms for the generation of turbulent bands in channel flow. It appears that both the initial and the final states are very sensitive to the energy and Reynolds numbers characterising the flow, highlighting the complexity of the laminar-turbulent patterned state and its initial seed. 
The selection of one of these two mechanisms appears to be affected by the probability of decay of the newly-created stripe, which increases with time, but decreases with the Reynolds number.
Future work will aim at extending the present investigation to other shear flows presenting spatially-patterned turbulence. 
\section*{Acknowledgements}

This work was granted access to the HPC resources of IDRIS under the allocation 2020-A0072A06362 and A0092A06362 made by GENCI.

\section*{Declaration of Interests}
The authors report no conflict of interest

\bibliographystyle{jfm}
\bibliography{biblio.bib}

\begin{thebibliography}{55}
\expandafter\ifx\csname natexlab\endcsname\relax\def\natexlab#1{#1}\fi
\def\au#1{#1} \def\ed#1{#1} \def\yr#1{#1}\def\at#1{#1}\def\jt#1{\textit{#1}}
  \def\bt#1{#1}\def\bvol#1{\textbf{#1}} \def\vol#1{#1} \def\pg#1{#1}
  \def\publ#1{#1}\def\arxiv#1{#1}\def\org#1{#1}\def\st#1{\textit{#1}}

\bibitem[Aida {\em et~al.\/}(2010)Aida, Tsukahara \& Kawaguchi]{aida2010}
{\sc \au{Aida, Hiroshi}, \au{Tsukahara, Takahiro} \& \au{Kawaguchi, Yasuo}}
  \yr{2010} Dns of turbulent spot developing into turbulent stripe in plane
  poiseuille flow.  \bt{In {\em Fluids Engineering Division Summer Meeting\/}},
  ,  \vol{vol. 49484},  \pg{pp. 2125--2130}.

\bibitem[Aida {\em et~al.\/}(2011)Aida, Tsukahara \& Kawaguchi]{aida2011}
{\sc \au{Aida, Hiroshi}, \au{Tsukahara, Takahiro} \& \au{Kawaguchi, Yasuo}}
  \yr{2011} Development of a turbulent spot into a stripe pattern in plane
  poiseuille flow.  \bt{In {\em Seventh International Symposium on Turbulence
  and Shear Flow Phenomena\/}}. Begel House Inc.

\bibitem[Avila {\em et~al.\/}(2011)Avila, Moxey, de~Lozar, Avila, Barkley \&
  Hof]{avila2011}
{\sc \au{Avila, Kerstin}, \au{Moxey, David}, \au{de~Lozar, Alberto}, \au{Avila,
  Marc}, \au{Barkley, Dwight} \& \au{Hof, Bj{\"o}rn}} \yr{2011}  \at{The onset
  of turbulence in pipe flow}.  \jt{Science}  \bvol{333}~(6039),
  \pg{192--196}.

\bibitem[Barkley \& Tuckerman(2005)]{barkley2005}
{\sc \au{Barkley, Dwight} \& \au{Tuckerman, Laurette~S}} \yr{2005}
  \at{Computational study of turbulent laminar patterns in couette flow}.
  \jt{Physical Review Letters}  \bvol{94}~(1),  \pg{014502}.

\bibitem[Carlson {\em et~al.\/}(1982)Carlson, Widnall \& Peeters]{carlson1982}
{\sc \au{Carlson, Dale~R}, \au{Widnall, Sheila~E} \& \au{Peeters, Martin~F}}
  \yr{1982}  \at{A flow-visualization study of transition in plane poiseuille
  flow}.  \jt{Journal of Fluid Mechanics}  \bvol{121},  \pg{487--505}.

\bibitem[Cherubini {\em et~al.\/}(2015)Cherubini, De~Palma \&
  Robinet]{CherubiniPoF2015}
{\sc \au{Cherubini, Stefania}, \au{De~Palma, Pietro} \& \au{Robinet,
  Jean-Christophe}} \yr{2015}  \at{Nonlinear optimals in the asymptotic suction
  boundary layer: Transition thresholds and symmetry breaking}.  \jt{Physics of
  Fluids}  \bvol{27}~(3),  \pg{034108}.

\bibitem[Cherubini {\em et~al.\/}(2010{\natexlab{{\em a\/}}})Cherubini,
  De~Palma, Robinet \& Bottaro]{cherubini2010}
{\sc \au{Cherubini, Stefania}, \au{De~Palma, Pietro}, \au{Robinet, J-Ch} \&
  \au{Bottaro, Alessandro}} \yr{2010{\natexlab{{\em a\/}}}}  \at{Rapid path to
  transition via nonlinear localized optimal perturbations in a boundary-layer
  flow}.  \jt{Physical Review E}  \bvol{82}~(6),  \pg{066302}.

\bibitem[Cherubini {\em et~al.\/}(2011)Cherubini, De~Palma, Robinet \&
  Bottaro]{cherubini2011}
{\sc \au{Cherubini, Stefania}, \au{De~Palma, Pietro}, \au{Robinet, J-C} \&
  \au{Bottaro, Alessandro}} \yr{2011}  \at{The minimal seed of turbulent
  transition in the boundary layer}.  \jt{Journal of Fluid Mechanics}
  \bvol{689},  \pg{221--253}.

\bibitem[Cherubini {\em et~al.\/}(2010{\natexlab{{\em b\/}}})Cherubini,
  Robinet, Bottaro \& De~Palma]{cherubiniJFMlin2010}
{\sc \au{Cherubini, S.}, \au{Robinet, J.-C.}, \au{Bottaro, A.} \& \au{De~Palma,
  P.}} \yr{2010{\natexlab{{\em b\/}}}}  \at{Optimal wave packets in a boundary
  layer and initial phases of a turbulent spot}.  \jt{Journal of Fluid
  Mechanics}  \bvol{656},  \pg{231–259}.

\bibitem[Duguet {\em et~al.\/}(2013)Duguet, Monokrousos, Brandt \&
  Henningson]{duguet2013}
{\sc \au{Duguet, Yohann}, \au{Monokrousos, Antonios}, \au{Brandt, Luca} \&
  \au{Henningson, Dan~S}} \yr{2013}  \at{Minimal transition thresholds in plane
  couette flow}.  \jt{Physics of Fluids}  \bvol{25}~(8),  \pg{084103}.

\bibitem[Duguet \& Schlatter(2013)]{DuguetPRL2013}
{\sc \au{Duguet, Y.} \& \au{Schlatter, P.}} \yr{2013}  \at{Oblique
  laminar-turbulent interfaces in plane shear flows}.  \jt{Physical Review
  Letters}  \bvol{110},  \pg{034502}.

\bibitem[Duguet {\em et~al.\/}(2010)Duguet, Schlatter \&
  Henningson]{duguet2010}
{\sc \au{Duguet, Yohann}, \au{Schlatter, Philipp} \& \au{Henningson, Dan~S}}
  \yr{2010}  \at{Formation of turbulent patterns near the onset of transition
  in plane couette flow}.  \jt{Journal of Fluid Mechanics}  \bvol{650},
  \pg{119}.

\bibitem[Eckhardt {\em et~al.\/}(2007)Eckhardt, Schneider, Hof \&
  Westerweel]{eckhardt2007}
{\sc \au{Eckhardt, Bruno}, \au{Schneider, Tobias~M}, \au{Hof, Bjorn} \&
  \au{Westerweel, Jerry}} \yr{2007}  \at{Turbulence transition in pipe flow}.
  \jt{Annual Review Fluid Mechanics}  \bvol{39},  \pg{447--468}.

\bibitem[Emmons(1951)]{emmons1951}
{\sc \au{Emmons, Howard~W}} \yr{1951}  \at{The laminar-turbulent transition in
  a boundary layer-part i}.  \jt{Journal of the Aeronautical Sciences}
  \bvol{18}~(7),  \pg{490--498}.

\bibitem[Farano {\em et~al.\/}(2016)Farano, Cherubini, Robinet \&
  De~Palma]{farano2016}
{\sc \au{Farano, Mirko}, \au{Cherubini, Stefania}, \au{Robinet,
  Jean-Christophe} \& \au{De~Palma, Pietro}} \yr{2016}  \at{Subcritical
  transition scenarios via linear and nonlinear localized optimal perturbations
  in plane poiseuille flow}.  \jt{Fluid Dynamics Research}  \bvol{48}~(6),
  \pg{061409}.

\bibitem[Farano {\em et~al.\/}(2017)Farano, Cherubini, Robinet \&
  De~Palma]{farano2017}
{\sc \au{Farano, Mirko}, \au{Cherubini, Stefania}, \au{Robinet,
  Jean-Christophe} \& \au{De~Palma, Pietro}} \yr{2017}  \at{Optimal bursts in
  turbulent channel flow}.  \jt{Journal of Fluid Mechanics}  \bvol{817},
  \pg{35–60}.

\bibitem[Foures {\em et~al.\/}(2013)Foures, Caulfield \& Schmid]{Foures2013}
{\sc \au{Foures, D. P.~G.}, \au{Caulfield, C.~P.} \& \au{Schmid, P.~J.}}
  \yr{2013}  \at{Localization of flow structures using $\infty $ -norm
  optimization}.  \jt{Journal of Fluid Mechanics}  \bvol{729},  \pg{672–701}.

\bibitem[Gibson {\em et~al.\/}(2021)Gibson, Reetz, Azimi, Ferraro, Kreilos,
  Schrobsdorff, Farano, Yesil, Schutz, Culpo \& Schneider]{channelflow}
{\sc \au{Gibson, J.~F.}, \au{Reetz, F.}, \au{Azimi, S.}, \au{Ferraro, A.},
  \au{Kreilos, T.}, \au{Schrobsdorff, H.}, \au{Farano, M.}, \au{Yesil, A.~F.},
  \au{Schutz, S.~S.}, \au{Culpo, M.} \& \au{Schneider, T.~M}} \yr{2021}
  \at{Channelflow 2.0}.  \jt{in preparation} .

\bibitem[Griewank \& Walther(2000)]{griewank2000}
{\sc \au{Griewank, Andreas} \& \au{Walther, Andrea}} \yr{2000}  \at{Algorithm
  799: revolve: an implementation of checkpointing for the reverse or adjoint
  mode of computational differentiation}.  \jt{ACM Transactions on Mathematical
  Software (TOMS)}  \bvol{26}~(1),  \pg{19--45}.

\bibitem[Henningson \& Kim(1991)]{henningson1991}
{\sc \au{Henningson, Dan~S} \& \au{Kim, John}} \yr{1991}  \at{On turbulent
  spots in plane poiseuille flow}.  \jt{Journal of Fluid mechanics}
  \bvol{228},  \pg{183--205}.

\bibitem[Hinze {\em et~al.\/}(2006)Hinze, Walther \& Sternberg]{hinze2006}
{\sc \au{Hinze, Michael}, \au{Walther, Andrea} \& \au{Sternberg, Julia}}
  \yr{2006}  \at{An optimal memory-reduced procedure for calculating adjoints
  of the instationary navier-stokes equations}.  \jt{Optimal Control
  Applications and Methods}  \bvol{27}~(1),  \pg{19--40}.

\bibitem[Kashyap {\em et~al.\/}(2020)Kashyap, Duguet \& Dauchot]{kashyap2020a}
{\sc \au{Kashyap, Pavan~V}, \au{Duguet, Yohann} \& \au{Dauchot, Olivier}}
  \yr{2020}  \at{Flow statistics in the transitional regime of plane channel
  flow}.  \jt{Entropy}  \bvol{22}~(9),  \pg{1001}.

\bibitem[Kerswell(2018)]{KerswellARFM}
{\sc \au{Kerswell, R.R.}} \yr{2018}  \at{Nonlinear nonmodal stability theory}.
  \jt{Annual Review of Fluid Mechanics}  \bvol{50}~(1),  \pg{319--345}.

\bibitem[Kerswell {\em et~al.\/}(2014)Kerswell, Pringle \&
  Willis]{Kerswell2014}
{\sc \au{Kerswell, R.~R.}, \au{Pringle, C. C.~T.} \& \au{Willis, A.~P.}}
  \yr{2014}  \at{{An optimisation approach for analysing nonlinear stability
  with transition to turbulence in fluids as an exemplar }}.  \jt{Rep. Prog.
  Phys.}  \bvol{77},  \pg{085901}.

\bibitem[Klingmann(1992)]{klingmann1992}
{\sc \au{Klingmann, Barbro~GB}} \yr{1992}  \at{On transition due to
  three-dimensional disturbances in plane poiseuille flow}.  \jt{Journal of
  Fluid Mechanics}  \bvol{240},  \pg{167--195}.

\bibitem[Lagha \& Manneville(2007)]{lagha2007}
{\sc \au{Lagha, Maher} \& \au{Manneville, Paul}} \yr{2007}  \at{Modeling of
  plane couette flow. i. large scale flow around turbulent spots}.  \jt{Physics
  of Fluids}  \bvol{19}~(9),  \pg{094105}.

\bibitem[Lemoult {\em et~al.\/}(2013)Lemoult, Aider \& Wesfreid]{lemoult2013}
{\sc \au{Lemoult, Gr{\'e}goire}, \au{Aider, Jean-Luc} \& \au{Wesfreid,
  Jos{\'e}~Eduardo}} \yr{2013}  \at{Turbulent spots in a channel: large-scale
  flow and self-sustainability}.  \jt{Journal of Fluid Mechanics}  \bvol{731},
  \pg{R1}.

\bibitem[Lemoult {\em et~al.\/}(2014)Lemoult, Gumowski, Aider \&
  Wesfreid]{lemoult2014}
{\sc \au{Lemoult, Gr{\'e}goire}, \au{Gumowski, Konrad}, \au{Aider, Jean-Luc} \&
  \au{Wesfreid, Jos{\'e}~Eduardo}} \yr{2014}  \at{Turbulent spots in channel
  flow: an experimental study}.  \jt{The European Physical Journal E}
  \bvol{37}~(4),  \pg{1--11}.

\bibitem[Marensi {\em et~al.\/}(2019)Marensi, Willis \& Kerswell]{marensi2019}
{\sc \au{Marensi, Elena}, \au{Willis, Ashley~P.} \& \au{Kerswell, Rich~R.}}
  \yr{2019}  \at{Stabilisation and drag reduction of pipe flows by flattening
  the base profile}.  \jt{Journal of Fluid Mechanics}  \bvol{863},
  \pg{850–875}.

\bibitem[Monokrousos {\em et~al.\/}(2011)Monokrousos, Bottaro, Brandt, Di~Vita
  \& Henningson]{monokrousos2011}
{\sc \au{Monokrousos, Antonios}, \au{Bottaro, Alessandro}, \au{Brandt, Luca},
  \au{Di~Vita, Andrea} \& \au{Henningson, Dan~S}} \yr{2011}  \at{Nonequilibrium
  thermodynamics and the optimal path to turbulence in shear flows}.
  \jt{Physical Review Letters}  \bvol{106}~(13),  \pg{134502}.

\bibitem[Orr(1907)]{orr1907}
{\sc \au{Orr, William~M'F}} \yr{1907} The stability or instability of the
  steady motions of a perfect liquid and of a viscous liquid. part ii: A
  viscous liquid.  \bt{In {\em Proceedings of the Royal Irish Academy. Section
  A: Mathematical and Physical Sciences\/}}, ,  \vol{vol.~27},  \pg{pp.
  69--138}. JSTOR.

\bibitem[Paranjape(2019)]{Paranjape2019}
{\sc \au{Paranjape, C.}} \yr{2019}  \at{Onset of turbulence in plane poiseuille
  flow}.  \jt{PhD thesis}  \bvol{IST Austria}.

\bibitem[Parente {\em et~al.\/}(2021)Parente, Robinet, De~Palma \&
  Cherubini]{ParenteRapid2021}
{\sc \au{Parente, Enza}, \au{Robinet, J-Ch}, \au{De~Palma, Pietro} \&
  \au{Cherubini, Stefania}} \yr{2021}  \at{Linear and nonlinear optimal growth
  mechanisms for generating turbulent bands}.  \jt{Journal of Fluid Mechanics}
  \bvol{submitted}.

\bibitem[Prigent {\em et~al.\/}(2002)Prigent, Gr{\'e}goire, Chat{\'e}, Dauchot
  \& van Saarloos]{prigent2002}
{\sc \au{Prigent, Arnaud}, \au{Gr{\'e}goire, Guillaume}, \au{Chat{\'e},
  Hugues}, \au{Dauchot, Olivier} \& \au{van Saarloos, Wim}} \yr{2002}
  \at{Large-scale finite-wavelength modulation within turbulent shear flows}.
  \jt{Physical Review Letters}  \bvol{89}~(1),  \pg{014501}.

\bibitem[Pringle \& Kerswell(2010)]{pringle2010}
{\sc \au{Pringle, Chris~CT} \& \au{Kerswell, Rich~R}} \yr{2010}  \at{Using
  nonlinear transient growth to construct the minimal seed for shear flow
  turbulence}.  \jt{Physical Review Letters}  \bvol{105}~(15),  \pg{154502}.

\bibitem[Pringle {\em et~al.\/}(2012)Pringle, Willis \& Kerswell]{pringle2012}
{\sc \au{Pringle, Chris~CT}, \au{Willis, Ashley~P} \& \au{Kerswell, Rich~R}}
  \yr{2012}  \at{Minimal seeds for shear flow turbulence: using nonlinear
  transient growth to touch the edge of chaos}.  \jt{Journal of Fluid
  Mechanics}  \bvol{702},  \pg{415--443}.

\bibitem[Rabin {\em et~al.\/}(2012)Rabin, Caulfield \&
  Kerswell]{rabin2012triggering}
{\sc \au{Rabin, SME}, \au{Caulfield, CP} \& \au{Kerswell, RR}} \yr{2012}
  \at{Triggering turbulence efficiently in plane couette flow}.  \jt{Journal of
  Fluid Mechanics}  \bvol{712},  \pg{244--272}.

\bibitem[Rabin {\em et~al.\/}(2014)Rabin, Caulfield \& Kerswell]{Rabin2014}
{\sc \au{Rabin, SME}, \au{Caulfield, CP} \& \au{Kerswell, RR}} \yr{2014}
  \at{Designing a more nonlinearly stable laminar flow via boundary
  manipulation}.  \jt{Journal of Fluid Mechanics}  \bvol{738}.

\bibitem[Reynolds(1883)]{reynolds1883}
{\sc \au{Reynolds, Osborne}} \yr{1883}  \at{Iii. an experimental investigation
  of the circumstances which determine whether the motion of water shall be
  direct or sinuous, and of the law of resistance in parallel channels}.
  \jt{Proceedings of the royal society of London}  \bvol{35}~(224-226),
  \pg{84--99}.

\bibitem[Schumacher \& Eckhardt(2001)]{schumacher2001}
{\sc \au{Schumacher, J{\"o}rg} \& \au{Eckhardt, Bruno}} \yr{2001}
  \at{Evolution of turbulent spots in a parallel shear flow}.  \jt{Physical
  Review E}  \bvol{63}~(4),  \pg{046307}.

\bibitem[Shimizu \& Manneville(2019)]{shimizu2019}
{\sc \au{Shimizu, Masaki} \& \au{Manneville, Paul}} \yr{2019}  \at{Bifurcations
  to turbulence in transitional channel flow}.  \jt{Physical Review Fluids}
  \bvol{4}~(11),  \pg{113903}.

\bibitem[Song \& Xiao(2020)]{song2020}
{\sc \au{Song, Baofang} \& \au{Xiao, Xiangkai}} \yr{2020}  \at{Trigger
  turbulent bands directly at low reynolds numbers in channel flow using a
  moving-force technique}.  \jt{Journal of Fluid Mechanics}  \bvol{903},
  \pg{A43}.

\bibitem[Tao {\em et~al.\/}(2018)Tao, Eckhardt \& Xiong]{tao2018}
{\sc \au{Tao, JJ}, \au{Eckhardt, Bruno} \& \au{Xiong, XM}} \yr{2018}
  \at{Extended localized structures and the onset of turbulence in channel
  flow}.  \jt{Physical Review Fluids}  \bvol{3}~(1),  \pg{011902}.

\bibitem[Tao \& Xiong(2013)]{tao2013}
{\sc \au{Tao, Jianjun} \& \au{Xiong, Xiangming}} \yr{2013}  \at{The unified
  transition stages in linearly stable shear flows}.  \jt{Fourteenth Asia
  Congress of Fluid Mechanics, Hanoi and Halong, Oct. 15-19} .

\bibitem[Tao \& Xiong(2017)]{tao2017}
{\sc \au{Tao, Jianjun} \& \au{Xiong, Xiangming}} \yr{2017}  \at{The unified
  transition stages in linearly stable shear flows}.  \jt{arXiv preprint
  arXiv:1710.02258} .

\bibitem[Tsukahara {\em et~al.\/}(2014)Tsukahara, Kawaguchi \&
  Kawamura]{tsukahara2014}
{\sc \au{Tsukahara, Takahiro}, \au{Kawaguchi, Yasuo} \& \au{Kawamura, Hiroshi}}
  \yr{2014}  \at{An experimental study on turbulent-stripe structure in
  transitional channel flow}.  \jt{arXiv preprint arXiv:1406.1378} .

\bibitem[Tsukahara {\em et~al.\/}(2005)Tsukahara, Seki, Kawamura \&
  Tochio]{tsukahara2005}
{\sc \au{Tsukahara, Takahiro}, \au{Seki, Yohji}, \au{Kawamura, Hiroshi} \&
  \au{Tochio, Daisuke}} \yr{2005} Dns of turbulent channel flow at very low
  reynolds numbers.  \bt{In {\em Fourth International Symposium on Turbulence
  and Shear Flow Phenomena\/}}. Begel House Inc.

\bibitem[Tuckerman \& Barkley(2011)]{tuckerman2011}
{\sc \au{Tuckerman, Laurette~S} \& \au{Barkley, Dwight}} \yr{2011}
  \at{Patterns and dynamics in transitional plane couette flow}.  \jt{Physics
  of Fluids}  \bvol{23}~(4),  \pg{041301}.

\bibitem[Tuckerman {\em et~al.\/}(2020)Tuckerman, Chantry \&
  Barkley]{tuckerman2020}
{\sc \au{Tuckerman, Laurette~S}, \au{Chantry, Matthew} \& \au{Barkley, Dwight}}
  \yr{2020}  \at{Patterns in wall-bounded shear flows}.  \jt{Annual Review of
  Fluid Mechanics}  \bvol{52}.

\bibitem[Tuckerman {\em et~al.\/}(2014)Tuckerman, Kreilos, Schrobsdorff,
  Schneider \& Gibson]{tuckerman2014}
{\sc \au{Tuckerman, Laurette~S}, \au{Kreilos, Tobias}, \au{Schrobsdorff,
  Hecke}, \au{Schneider, Tobias~M} \& \au{Gibson, John~F}} \yr{2014}
  \at{Turbulent-laminar patterns in plane poiseuille flow}.  \jt{Physics of
  Fluids}  \bvol{26}~(11),  \pg{114103}.

\bibitem[Vavaliaris {\em et~al.\/}(2020)Vavaliaris, Beneitez \&
  Henningson]{vavaliaris2020}
{\sc \au{Vavaliaris, Chris}, \au{Beneitez, Miguel} \& \au{Henningson, Dan~S}}
  \yr{2020}  \at{Optimal perturbations and transition energy thresholds in
  boundary layer shear flows}.  \jt{Physical Review Fluids}  \bvol{5}~(6),
  \pg{062401}.

\bibitem[Wang {\em et~al.\/}(2020)Wang, Guet, Monchaux, Duguet \&
  Eckhardt]{wang_duguet2020}
{\sc \au{Wang, Zhe}, \au{Guet, Claude}, \au{Monchaux, Romain}, \au{Duguet,
  Yohann} \& \au{Eckhardt, Bruno}} \yr{2020}  \at{Quadrupolar flows around
  spots in internal shear flows}.  \jt{Journal of Fluid Mechanics}  \bvol{892},
   \pg{A27}.

\bibitem[Xiao \& Song(2020)]{xiao2020}
{\sc \au{Xiao, Xiangkai} \& \au{Song, Baofang}} \yr{2020}  \at{The growth
  mechanism of turbulent bands in channel flow at low reynolds numbers}.
  \jt{Journal of Fluid Mechanics}  \bvol{883}.

\bibitem[Xiong {\em et~al.\/}(2015)Xiong, Tao, Chen \& Brandt]{xiong2015}
{\sc \au{Xiong, Xiangming}, \au{Tao, Jianjun}, \au{Chen, Shiyi} \& \au{Brandt,
  Luca}} \yr{2015}  \at{Turbulent bands in plane-poiseuille flow at moderate
  reynolds numbers}.  \jt{Physics of Fluids}  \bvol{27}~(4),  \pg{041702}.

\bibitem[Zuccher {\em et~al.\/}(2004)Zuccher, Luchini \& Bottaro]{zuccher2004}
{\sc \au{Zuccher, Simone}, \au{Luchini, Paolo} \& \au{Bottaro, Alessandro}}
  \yr{2004}  \at{Algebraic growth in a blasius boundary layer: optimal and
  robust control by mean suction in the nonlinear regime}.  \jt{Journal of
  Fluid Mechanics}  \bvol{513},  \pg{135}.

\end{thebibliography}

\end{document}